\documentclass[usenatbib]{mn2e}
\usepackage{epsfig}
\usepackage{amssymb}
\usepackage{lscape,graphicx}
\usepackage{longtable}

\def\850{$850\,\mathrm{\mu m}$}

\usepackage{graphicx}
\usepackage{epstopdf}
\def\lsim{\mathrel{\lower2.5pt\vbox{\lineskip=0pt\baselineskip=0pt
           \hbox{$<$}\hbox{$\sim$}}}}

\def\gsim{\mathrel{\lower2.5pt\vbox{\lineskip=0pt\baselineskip=0pt
           \hbox{$>$}\hbox{$\sim$}}}}

\newcommand{\mum}{$\,\mu$m}

\newcommand{\dmB}{\ifmmode \Delta m_{15}(B) \else $\Delta m_{15}(B)$\fi}
\newcommand{\msun}{\ifmmode \mathrm{M}_{\odot} \else $\mathrm{M}_{\odot}$\fi}
\newcommand{\msunyr}{\ifmmode \mathrm{M}_{\odot}\mathrm{yr}^{-1} \else $\mathrm{M}_{\odot}\mathrm{yr}^{-1}$\fi}
\newcommand{\lsun}{\ifmmode \mathrm{L}_{\odot} \else $\mathrm{L}_{\odot}$\fi}

\begin{document}\newcommand{\isnote}[1]{}

\title[SHADES submillimetre maps, catalogue and number counts] {The SCUBA HAlf Degree Extragalactic Survey (SHADES) -- II. Submillimetre maps, catalogue and number counts}
\author[Coppin et al.]{ 
\parbox[t]{\textwidth}{
K. Coppin$^{1}$, E.L. Chapin$^{1,2}$, A.M.J. Mortier$^{3}$, S.E. Scott$^{4}$, C. Borys$^{5}$, J.S. Dunlop$^{4}$, M. Halpern$^{1}$, D.H. Hughes$^{2}$, A. Pope$^{1}$, D. Scott$^{1}$, S. Serjeant$^{3}$, J. Wagg$^{2,6}$, D.M. Alexander$^{7}$,O. Almaini$^{8}$, I. Aretxaga$^{2}$, T. Babbedge$^{9}$,  P.N. Best$^{4}$, A. Blain$^{5}$, S. Chapman$^{5}$, D.L. Clements$^{9}$, M. Crawford$^{4}$, L. Dunne$^{4,8}$, S.A. Eales$^{13}$, A.C. Edge$^{10}$, D. Farrah$^{11}$, E. Gazta\~naga$^{2,12}$, W.K. Gear$^{13}$, G.L. Granato$^{14}$, T.R. Greve$^{5}$, M. Fox$^{9}$, R.J. Ivison$^{4, 15}$, M.J. Jarvis$^{16}$, T. Jenness$^{17}$,  C. Lacey$^{10}$, K. Lepage$^{1}$, R.G. Mann$^{4}$, G. Marsden$^{1}$, A. Martinez-Sansigre$^{16}$, S. Oliver$^{18}$, M.J. Page$^{19}$, J.A. Peacock$^{4}$, C.P. Pearson$^{20}$, W.J. Percival$^{21}$, R.S. Priddey$^{22}$, S. Rawlings$^{16}$, M. Rowan-Robinson$^{9}$, R.S. Savage$^{18}$, M. Seigar$^{23,17}$, K. Sekiguchi$^{24}$, L. Silva$^{25}$, C. Simpson$^{26,27}$, I. Smail$^{10}$, J.A. Stevens$^{22}$, T. Takagi$^{3}$, M. Vaccari$^{9,28}$, E. van Kampen$^{4,29}$, C.J. Willott$^{30}$
}
\\
$^{1}$ Department of Physics \& Astronomy, University of British Columbia, 6224 Agricultural Road, Vancouver, B.C., V6T 1Z1, Canada\\
$^{2}$ Instituto Nacional de Astrof\'{\i}sica, \'{O}ptica y Electr\'{o}nica, Apartado Postal 51 y 216, 72000 Puebla, Pue., Mexico\\
$^{3}$ Centre for Astrophysics and Planetary Science, School of Physical Sciences, University of Kent, Canterbury, Kent CT2 7NR, UK\\
$^{4}$ Institute for Astronomy, University of Edinburgh, Royal Observatory, Blackford Hill, Edinburgh EH9 3HJ, UK\\
$^{5}$ Caltech, 1200 E. California Blvd, Pasadena, CA 91125-0001, USA\\
$^{6}$ Harvard-Smithsonian Center for Astrophysics, Cambridge, MA 02138, USA\\
$^{7}$ Institute of Astronomy, University of Cambridge, Madingley Road, Cambridge CB3 0HA, UK\\
$^{8}$ The School of Physics and Astronomy, University of Nottingham, University Park, Nottingham NG7 2RD, UK \\
$^{9}$ Astrophysics Group, Blackett Laboratory, Imperial College, Prince Consort Rd., London SW7 2BW, UK\\
$^{10}$ Institute for Computational Cosmology, University of Durham, South Rd, Durham DH1 3LE, UK\\
$^{11}$ Department of Astronomy, Cornell University, Space Sciences Building, Ithaca, NY 14853, USA\\
$^{12}$ Institut d'Estudis Espacials de Catalunya, IEEC/CSIC, c/ Gran Capita 2-4, 08034, Barcelona, Spain\\
$^{13}$ School of Physics and Astronomy, Cardiff University, 5, The Parade, Cardiff CF24 3YB, UK\\ 
$^{14}$ Osservatorio Astronomico di Padova, Vicolo dell'Osservatorio, 5, I-35122, Padova, Italy\\
$^{15}$ UK ATC, Royal Observatory, Blackford Hill, Edinburgh EH9 3HJ, UK\\
$^{16}$ Department of Astrophysics, Denys Wilkinson Building, Keble Road, Oxford OX1 3RH, UK\\
$^{17}$ Joint Astronomy Centre, 660 N.\ A`oh\={o}k\={u} Place, University Park, Hilo, HI 96720, USA \\
$^{18}$ Astronomy Centre, University of Sussex, Falmer, Brighton BN1 9QH, UK\\
$^{19}$ Mullard Space Science Laboratory (MSSL), University College London, Holmbury St. Mary, Dorking, Surrey RH5 6NT, UK\\
$^{20}$ Institute of Space and Astronautical Science (ISAS), Yoshinodai 3-1-1, Sagamihara, Kanagawa 229 8510, Japan\\
$^{21}$ Institute of Cosmology and Gravitation, University of Portsmouth, 
Portsmouth P01 2EG, UK\\
$^{22}$ Centre for Astrophysics Research, Science and Technology Research Institute, University of Hertfordshire, College Lane, Hatfield, Hertfordshire AL10 9AB, UK\\
$^{23}$ Center for Cosmology, University of California Irvine, 4129 Frederick Reines Hall, Irvine, CA 92697-4575, USA\\
$^{24}$ Subaru Telescope, National Astronomical Observatory of Japan, 650 N.\ A`oh\={o}k\={u} Place, Hilo, HI 96720, USA\\
$^{25}$ Osservatorio Astronomico di Trieste, Via Tiepolo 11, I-34131, Trieste, Italy\\
$^{26}$ Astrophysics Research Institute, Liverpool John Moores University, Twelve Quays House, Egerton Wharf, Birkenhead CH41 1LD, UK\\
$^{27}$ Department of Physics, University of Durham, South Road, Durham DH1 3LE, UK\\
$^{28}$ Department of Astronomy, University of Padova, Vicolo dell'Osservatorio 2, 35122, Italy\\
$^{29}$ Institute for Astrophysics, University of Innsbruck, Technikerstr. 25, A-6020 Innsbruck, Austria\\
$^{30}$ Herzberg Institute of Astrophysics, National Research Council, 5071 West Saanich Rd, Victoria, B.C. V9E 2E7, Canada \\
\\
\\
\\
\\
\\
\\
}

\date{Accepted August 17 2006 by MNRAS}

\maketitle

\begin{abstract}
We present maps, source catalogue and number counts of the largest, most complete and unbiased extragalactic submillimetre survey:  the $850\,\mu\mathrm{m}$ SCUBA HAlf Degree Extragalactic Survey (SHADES).  Using the Submillimetre Common-User Bolometer Array (SCUBA) on the James Clerk Maxwell Telescope (JCMT), SHADES mapped two separate regions of sky:  the Subaru/\textit{XMM-Newton} Deep Field (SXDF) and the Lockman Hole East (LH).  Encompassing 93 per cent of the overall acquired data (i.e.~data taken up to 2004 February 1), these SCUBA maps cover $720\,\mathrm{arcmin}^{2}$ with an RMS noise level of about $2\,\mathrm{mJy}$ and have uncovered $>100$ submillimetre galaxies.  In order to ensure the utmost robustness of the resulting source catalogue, data reduction was independently carried out by four sub-groups within the SHADES team, providing an unprecedented degree of reliability with respect to other SCUBA catalogues available from the literature.  Individual source lists from the four groups were combined to produce a robust 120-object SHADES catalogue; an invaluable resource for follow-up campaigns aiming to study the properties of a complete and consistent sample of submillimetre galaxies.  For the first time, we present deboosted flux densities for each submillimetre galaxy found in a large survey.  Extensive simulations and tests were performed separately by each group in order to confirm the robustness of the source candidates and to evaluate the effects of false detections, completeness, and flux density boosting.  Corrections for these effects were then applied to the data to derive the submillimetre galaxy source counts.  SHADES has a high enough number of detected sources that meaningful differential counts can be estimated, unlike most submillimetre surveys which have to consider integral counts.  We present differential and integral source number counts and find that the differential counts are better fit with a broken power-law or a Schechter function than with a single power-law; the SHADES data alone significantly show that a break is required at several mJy, although the precise position of the break is not well constrained.  We also find that an $850\,\mathrm{\mu m}$ survey complete down to $2\,\mathrm{mJy}$ would resolve 20--30 per cent of the Far-IR background into point sources.

\end{abstract}


\section{Introduction}\label{intro}

Deep blank-field surveys conducted with the Submillimetre Common-User Bolometer Array (SCUBA) on the James Clerk Maxwell Telescope (JCMT) have resolved as much as 50 per cent (depending on the survey depth) of the Far-Infrared Background (FIB) into discrete, high redshift-sources with flux density levels of $S_{850}\gsim2\,\mathrm{mJy}$ (\citealt{Smail97}, \citealt{Hughes98}, \citealt{Barger98}, \citealt{Blain99}, \citealt{Barger99}, \citealt{Eales}, \citealt{Cowie}, \citealt{Scott}, \citealt{Webb}, \citealt{Borys2003}). Intensive campaigns to study these submillimetre galaxies (SMGs) and millimetre galaxies at other wavelengths have shown that they are primarily powered by star-formation, although many do harbour an Active Galactic Nucleus (AGN; \citealt{alexander05}).  

The implied star-formation rates are extremely high (100--$1000\,\mathrm{M}_{\odot}\,\mathrm{yr}^{-1}$), and since the well-known local radio/Far-Infrared correlation \citep{condon1992} apparently holds for these higher redshift sources (e.g~\citealt{kovacs2006}), the high submillimetre luminosities result in correspondingly high radio luminosities, which are detected in deep 1.4\,GHz images.  The resolution afforded by radio-interferometers results in precise optical identifications of $\gsim50\,\mathrm{per cent}$ of all known SMGs, and in this manner \cite{Chapman2005} derive a median redshift of $\left\langle z \right\rangle\sim2.2$ for the population using deep Keck spectroscopy.  While much of what is known about SMGs is based on the radio detected subset \citep{Ivison}, other studies (e.g.~\citealt{Pope}) find no significant differences in the radio-undetected population.

Many investigations (e.g.~\citealt{Smail97}; \citealt{Barger}, \citealt{Ivison}; \citealt{Smail2002}; \citealt{Webb}; \citealt{Borys2004}; \citealt{Wang}; \citealt{Greve}) have suggested that SMGs are likely to be associated with an early-phase in the formation of massive galaxies. The intensity of their starbursts, the resulting high metallicity, along with their large dynamical masses, high gas fractions and inferred strong clustering (\citealt{gravy}; \citealt{2004ApJ...617...64S}; \citealt{BCSI04}) are all suggestive of a close link to the formation phase of the most massive spheroids (e.g.~\citealt{2004ApJ...616...71S}).

Despite this progress, the samples being used in any given study are typically quite small, since SCUBA could only detect roughly one SMG per night in good weather conditions.  And while the total number of SMGs detected over the lifetime of the instrument\footnote{SCUBA was operational between 1997 and 2005} is now $\sim400$, these are mainly drawn from small ($\lsim100$ square arcminute) fields spread all over the sky, each observed and reduced by different groups using different techniques and source identification criteria.  The desire to obtain a well characterised sample of hundreds of SMGs in a large, contiguous area was the motivation for the SCUBA HAlf-Degree Extragalactic Survey (SHADES; \citealt{Mortier}, \citealt{vanKampen}).

SHADES is an ambitious wide extragalactic submillimetre survey, split evenly between 2 separate regions of sky:  the Subaru/\textit{XMM-Newton} Deep Field (SXDF) and the Lockman Hole East (LH).  The aim was to map $0.5\,\mathrm{deg}^{2}$ ($\simeq7$ times the area of its predecessor, the SCUBA 8-mJy Survey; \citealt{Scott}) to a comparable depth of $\sigma \simeq 2\,\mathrm{mJy}$ at $850\,\mu\mathrm{m}$, which is roughly three times the $1\,\sigma$ confusion limit imposed by the underlying sea of fainter unresolved sources in the coarse SCUBA beam (\citealt{Hogg01}; \citealt{Hughes98}; \citealt{Cowie}).  The survey began in late 2002, and finished when SCUBA was decomissioned in late 2005.  SHADES is the largest survey in terms of observing time carried out with SCUBA.

Many scientifically powerful results from SHADES will come from comparisons of the properties of sources found in the SHADES catalogue with observations at other wavelengths.  Some of these include radio identifications (Ivison et al. in preparation), FIR-radio photometric redshift estimates (Aretxaga et al. in preparation), and submillimetre-Spitzer-based SEDs and photometric redshifts (Clements et al., Eales et al. and Serjeant et al. in preparation).  In this paper we present new constraints on the numbers of submillimetre sources as a function of $850\,\mathrm{\mu m}$ flux density based on the SHADES catalogues.  A complementary analysis effort to constrain the numbers of faint submillimetre sources directly from our maps without the intermediate step of making a catalogue, a so-called $P(D)$ approach (e.g.~\citealt{Condon}, \citealt{Scheuer}), will be reported separately.

The main results presented in this paper are the SHADES catalogue and 850\mum\ number counts.  The observations are summarised in Section~\ref{obsdr}.  Section~\ref{dr} provides details about the data reduction.  For the first time in a submillimetre survey the data are processed by four independent data reduction pipelines in order to increase the robustness of the results.  In Section~\ref{compareall}, the amalgamation of the four source lists into a master $850\,\mathrm{\mu m}$ SHADES catalogue is described.  In Section~\ref{complete} two different approaches to derive the number counts are described, compared and contrasted.  The SHADES differential count measurements are provided here.  In Section~\ref{numcts}, models are fit to the differential counts, and the cumulative source counts are computed in order to compare them with previous data.  Concluding remarks regarding what was learned about data analysis by comparing the different reductions, as well as advice for future surveys, are given in Section~\ref{conc}.

\section{Observations}\label{obsdr}

The Subaru/\textit{XMM-Newton} Deep Field (SXDF) and Lockman Hole East (LH) regions are centred at (J2000) RA=\(2^{\mathrm{h}}17^{\mathrm{m}}57{\fs}5\), Dec=\(-5^{\circ}00'18{\farcs}5\) and RA=\(10^{\mathrm{h}}52^{\mathrm{m}}26{\fs}7\), Dec=\(57^{\circ}24'12{\farcs}6\), respectively.  They were observed with a resolution of 14.8 arcsec and 7.5 arcsec at 850 and $450\,\mu\mathrm{m}$, respectively, with SCUBA (\citealt{Holland}) on the 15-m JCMT atop Mauna Kea in Hawaii.  A detailed account of the survey design and observing strategy is given in \citet{Mortier} but is summarised briefly here.  The observing strategy consists of making three overlapping jiggle maps for each of six different chop throws (30, 44 and $68\,\mathrm{arcsec}$) and chop position angle (PA, 0 and $90\,\mathrm{deg}$ in RA/Dec coordinates) combinations, motivated by \citet{Emerson}.  Each set of six observations spans a range of airmass values, while the observations are limited to a $225\,\mathrm{GHz}$ atmospheric opacity range of $0.05<\tau_{\mathrm{CSO}}<0.1$, so as to maintain uniform coverage over the entire survey area.  The SHADES survey data collection took place between 1998 March and 2005 June (including the epoch of existing data from the LH portion of the SCUBA 8-mJy Survey, \citealt{Scott}, \citealt{Fox}).  

After a series of technical problems, SCUBA was officially decommissioned in 2005 July, with only a modest amount of SHADES data taken in its last year.  Only SCUBA data taken up until 2004 February 1, covering a total area of $720\,\mathrm{arcmin}^{2}$ down to an RMS noise level of $\simeq 2.2\,\mathrm{mJy}$, are included here.  The sky coverage is approximately evenly distributed between the two SHADES fields.  Complete SCUBA SHADES maps will contain only an additional roughly 7 per cent of the planned area, although these will be complemented with $1.1\,\mathrm{mm}$ AzTEC data \citep{aztec} which will be presented in a future paper.

\section{Data Reduction:  Four independent SHADES pipelines}\label{dr}  
In order to ensure the utmost robustness of the resulting $850\,\mathrm{\mu m}$ source catalogue, data reduction was independently carried out by four sub-groups drawn from within the SHADES consortium; we refer to these as Reductions A, B, C and D.  This provides an unprecedented degree of reliability with respect to other SCUBA source catalogues available from the literature.  The reduction of the $450\,\mathrm{\mu m}$ data, which are only of limited use, is discussed in Appendix~\ref{450umfluxes}.  A major strength of the SHADES analysis comes from the concordance of these reductions which have \textit{not} been modified to bring them into agreement.

All groups perform the basic reduction steps of combining 
the nods (i.e.~re-switching), 
flat-fielding, extinction correcting, 
despiking, removing the sky signal, calibrating, rebinning 
the data into pixel grids, 
and finally extracting sources.  All reduction groups also 
correct the astrometry of the SXDF 
observations prior to December 26, 2002 by $-6.7$ arcsec in 
Declination, in order to account for a JCMT 
pointing catalogue error for oCeti.  For all data taken on 
or after 2002 December 6 the known 
dead bolometer `G9' was manually removed from further analysis.  
Groups dealt with the presence 
of a 16 sample noise spike in the data in different ways, as first noted by 
\citet{Borys2004} and \citet{Webb_1stlook} 
(see also \citealt{Mortier}); this is further discussed in Appendix~\ref{noisespike}. 

\subsection{Description of the four independent reductions}

Reduction steps were performed by each group independently using some combination of SURF (SCUBA User Reduction Facility; \citealt{Jenness}) and their own locally-developed codes.  The basic steps of the analysis procedure are described in \citet{Mortier}.  Each reduction group has made different choices with respect to various detailed steps, which are described in Table \ref{procedures}.  Several of
the choices merit attention and are discussed below.

The teams each produce signal-to-noise ratio (S/N) 
maps convolved with the telescope point spread function. This procedure is
mathematically equivalent to fitting the point spread function (PSF) to the
data centred on each pixel and is the minimum variance estimator of
the brightness of isolated point sources if the background noise is
white (see e.g.~\citealt{vonH}).  In all four reductions, flux density 
measured when bolometers in the `off-beam chop' of the telescope
point at the region in question has essentially been folded in to the 
likelihood maps in order to increase the sensitivity of the maps.  
Two approaches have been used here: (1) fitting each pixel in the map to the multi-beam PSF; or (2) folding in the flux from the off-beams in the timestream data and then fitting each pixel in the map to a single Gaussian.  After calibrating, the final map noise RMS values for the LH and SXDF 
fields are 2.1 and $2.0\,\mathrm{mJy}\,\mathrm{beam}^{-1}$, respectively. 

Finally, the presence of the 16 sample noise spike does not warrant any major pipeline alterations, since its effect on these data was found to be almost negligible; an account of the investigation into this issue is presented in Appendix~\ref{noisespike}.  Several different checks were performed to test the Gaussianity of the noise and to check individual source robustness by examining spatial and temporal fits to point sources in the maps and this analysis is presented in Appendix~\ref{tests}.

\subsection{Opacity} The dominant cause of variation in the multiplicative 
factor between detector voltage and inferred source flux density is temporal
variation in the atmospheric opacity.  In all four reductions detector
voltages are divided by the atmospheric opacity inferred, 
using coefficients in \citet{arch01}, from either
the $225\,\mathrm{GHz}$ $\tau_{\mathrm{CSO}}$ measurements or 
the $183\,\mathrm{GHz}$ line of sight water vapour monitor (WVM; \citealt{Wiedner}) 
at the JCMT.  The differences in strategy described in 
Table~\ref{procedures} do not
lead to measurable differences in the data.  Largely because of
SHADES, there is a substantial body of data correlating opacity measured
via secant scans (`sky dips') for various chop-throws with the opacity 
inferred from measurements of the $\tau_{\mathrm{CSO}}$ or the JCMT WVM. Opacity monitors 
provide a lower noise measurement of opacity than is provided by any single sky-dip and 
do not cost any observing time (and are therefore preferred).
 
\subsection{Calibration}  The flux conversion factor (FCF) 
converts opacity-corrected bolometer voltages to flux densities.  It is
essentially the receiver responsivity.  The FCF is fairly stable on
both short and long time scales, but might depart from the long term
average as the telescope mirror is distorted by thermal gradients near
to sunset and sunrise \citep{Jenness_cal}.  Because the FCF 
at $850\,\mathrm{\mu m}$ is so stable, 
the long term average might be a better estimator of the FCF at a given 
moment than any
single measurement taken during stable conditions (monthly average FCF values 
have been tabulated on the JCMT 
website\footnote{http://www.jach.hawaii.edu/JCMT/continuum/calibration/sens/gains.html}, 
and were used to calibrate the data in Reduction D).  Over-reliance on individual FCF 
measurements could inject noise into
the analysis. However, Reduction B, which makes the most use 
of measured FCF values, produces maps whose noise is as low as 
the other reductions.  

\subsection{Pixel size}  Reductions A and B make `zero-footprint' maps using an optimal noise-weighted drizzling algorithm (essentially the limiting form of the method of \citealt{Fruchter02}) with a pixel size of $1\,\mathrm{arcsec}^{2}$ (see \citealt{Mortier}).  Reductions C and D use $3\times3\,\mathrm{arcsec}$ pixels.
The pressure to use larger pixels is driven by the observation of
\citet{Borys2003} that statistical analysis becomes unpredictable when
there are too few observations per pixel.  The concern that large
pixels might lead to larger uncertainty in the positions of the detected
sources is not supported in the data (as discussed in Section~\ref{compareposition}).  

\begin{table*}
\caption{Data Reduction Procedures.  `Secondary Extinction Monitor' refers to the source of the estimation of the sky opacity when the `Primary Extinction Monitor' was unavailable.}

\begin{center}
\begin{tabular}{ | p{1.5cm} | p{3.5cm} | p{3.5cm}| p{3.5cm} |p{3.5cm}|}
	\hline
	Step & Reduction A& Reduction B & Reduction C & Reduction D \\ \hline

Reduction Code Used&
	IDL-based routines used in the SCUBA 8-$\mathrm{mJy}$ Survey \citep{Scott}.  See \citet{SerjeantDR}. &
	Reduction A's IDL pipeline altered at the extinction correction, calibration and source extraction phases.  See \citet{Mortier}.&
	SURF scripts for reswitching and flat-fielding and locally developed code from \citet{Chapinthesis} thereafter.&
        SURF scripts for reswitching and flatfielding and locally developed C code thereafter \citep{Borys2003}. \\ \hline

Primary Extinction Monitor&
	Polynomial fit to $\tau_{\mathrm{CSO}}$. &
	Interpolated WVM readings. &
	Same as Reduction A. &
	Nearest WVM reading. \\ \hline

Secondary Extinction Monitor &
          Linear fit to neighbouring sky-dips and interpolation. &
          $\tau_{\mathrm{CSO}}$, sky-dip values. &
          Interpolated sky-dips. &
          Nearest $\tau_{\mathrm{CSO}}$ reading, linear fit to neighbouring sky-dips and interpolation.\\ \hline	
          
 Cosmic Rays&
           Iterative $3\,\sigma$ cuts until no signal is removed. &
           $\simeq 32$ successive $3\,\sigma$ cuts. &
           One $3\,\sigma$ cut. &
           Iterative $4.5\,\sigma$ cuts until no signal is removed. \\ \hline
           
 Baseline Subtraction &
            Subtract the temporal modal sky level obtained from a fit to all bolometers in the array iteratively with cosmic ray removal. &
            Subtract mode signal iteratively with cosmic rays. Handle bolometers exhibiting excess 16 sample noise separately. &
            Subtract array median. &
            Subtract mean of non-noisy bolometers from all bolometers iteratively with cosmic ray removal. \\ \hline
           
 Calibration &
            Calculate an FCF for each half night and apply to the data. &
            Linearly interpolate between all stable FCF measurements during a night, selected by eye. &
            Mean of measured FCF before and after sunrise/set (i.e.~two averages per shift).  Each of 3 chop throw \textit{amplitudes} handled separately. &
            Monthly average FCFs (single value of $219\,\mathrm{Jy}\,\mathrm{V^{-1}}\,\mathrm{beam^{-1}}$ for data before January 1999). \\ \hline
            
Flux Density Maps &
	  Inverse variance weighted flux density summed in $1\,\mathrm{arcsec}$ pixels. Each chop throw mapped separately (six flux density maps) and then combined to produce a single maximum likelihood flux density map.&
	  Same as Reduction A. &
	  Same as Reduction A, but with $3\,\mathrm{arcsec}$ pixels. &
	  Inverse variance weighted flux density summed in $3\,\mathrm{arcsec}$ pixels to produce a single flux density map.  Negative flux density from off-chop positions also summed in at the timestream level(see \citealt{Borys2003}). \\ \hline

Convolution &
       Form maximum likelihood point-source flux density and noise maps (and S/N maps) via noise-weighted convolution with the differential PSF that combines $14.5\,\mathrm{arcsec}$ Gaussians with the chop pattern. &

       Same as Reduction A. &

       Same as Reduction A, but with $14.7\,\mathrm{arcsec}$ Gaussians. &

       Form maximum likelihood point-source flux density and noise maps (and hence S/N maps) using noise-weighted convolution with a single $14.7\,\mathrm{arcsec}$ Gaussian (i.e.~without the chop pattern). \\ \hline

Cut \nobreak{Pixels} &
           No pixel cuts were made in flux density map.  Ignore pixels with S/N deviating more than $4\,\sigma$ from convolved map. &
           No pixel cuts were made in flux density map.  Ignore pixels with $\sigma>10$\,mJy in convolved map.  &
           Remove pixels with $\leq 10$ `hits' in flux density map. Ignore pixels with $\sigma>5$\,mJy in convolved map.&
           Remove pixels with $\leq 10$ `hits' in flux density map.  No pixels were ignored in convolved map.\\ \hline
          
Source Extraction &
	Positive peaks identified in the convolved signal map above a threshold.  A model was constructed by centering a normalised beam-map at the positions of the peaks in the convolved map.  Normalisation coefficients were calculated simultaneously, providing a minimum noise-weighted $\chi^{2}$ fit to the unconvolved signal map. &
	Peaks identified in the maximum likelihood S/N maps above a $2.5\,\sigma$ threshold. Flux densities identified at those positions in the maximum likelihood point source flux density maps. &
	Same as Reduction B. &
	Same as Reduction B. \\ \hline

		\hline
	\end{tabular}
\end{center}
\label{procedures}
\end{table*}
\normalsize

\begin{landscape}
\begin{figure}
\psfig{file=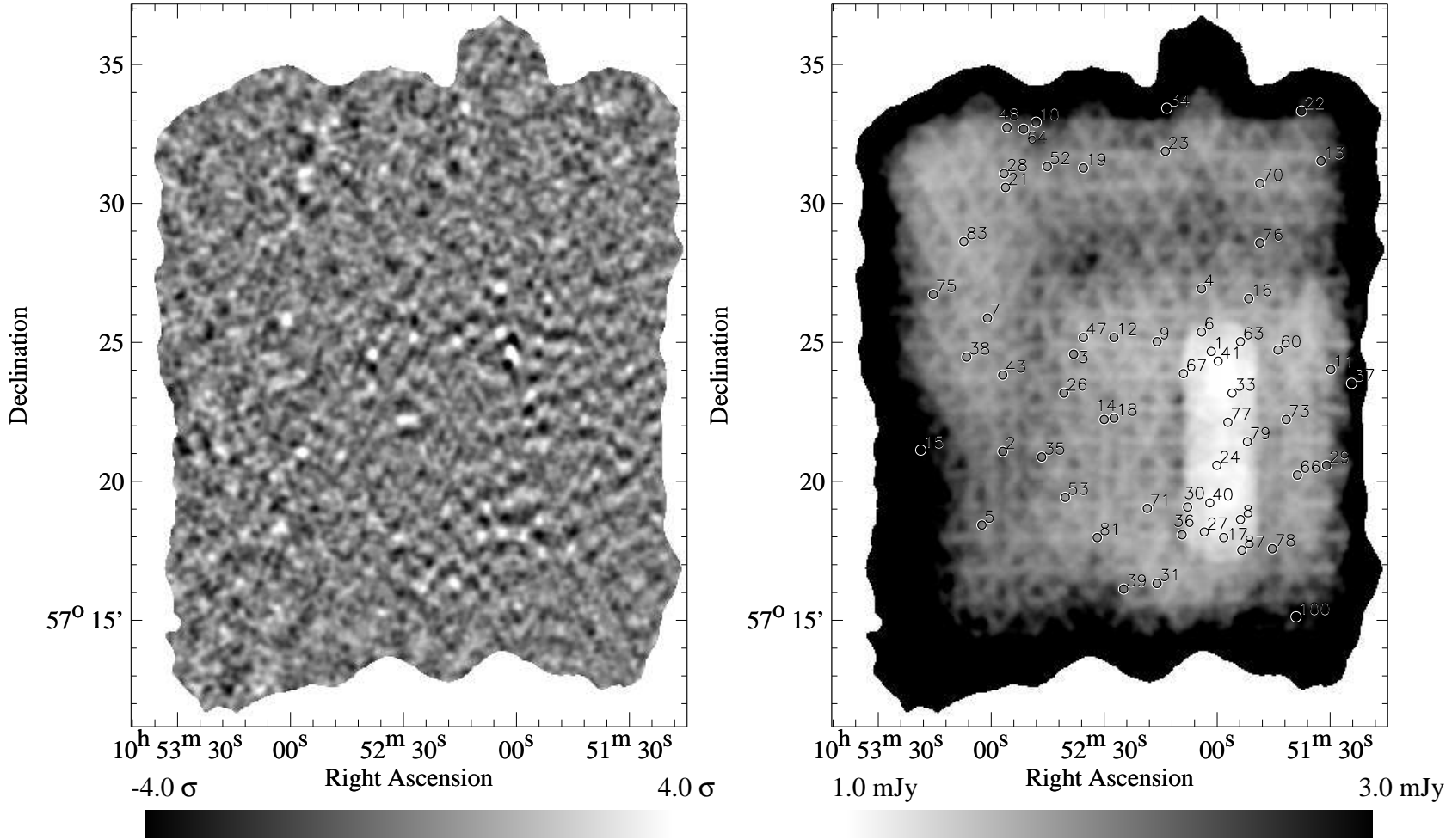,width=1.4\textwidth}
\label{fig:lock}
\caption{The $850\,\mathrm{\mu m}$ S/N map with the off-beams folded
in at the timestream stage (i.e.~each chop is treated separately in order to optimise S/N)(left) and corresponding noise 
map (right) of the $485\,\mathrm{arcmin}^2$ for the LH are shown (specifically for Reduction D).  This area only includes $3\,\mathrm{arcsec}$ pixels that
were sampled more than 10 times.  On the noise map, 15-arcsec diameter circles indicate the positions of the 60 sources in the LH SHADES catalogue.  Notice that sources are less often detected in noisier areas of the map; in general they are found in areas corresponding to better weather and on a fine scale sources are preferentially found near the corners of the overlapping triangular rows.  The SCUBA 8-mJy LH Survey covers approximately one quarter of the map in the lower-righthand corner. Within that region is a deep rectangular test strip ($\sigma \sim 1$mJy) that was observed before the rest of the SCUBA 8-mJy Survey.  A similar noise map for Reduction B is found in \citet{Mortier}.}
\end{figure}
\end{landscape}

\begin{landscape}
\begin{figure}\label{fig:sxdf}
\psfig{file=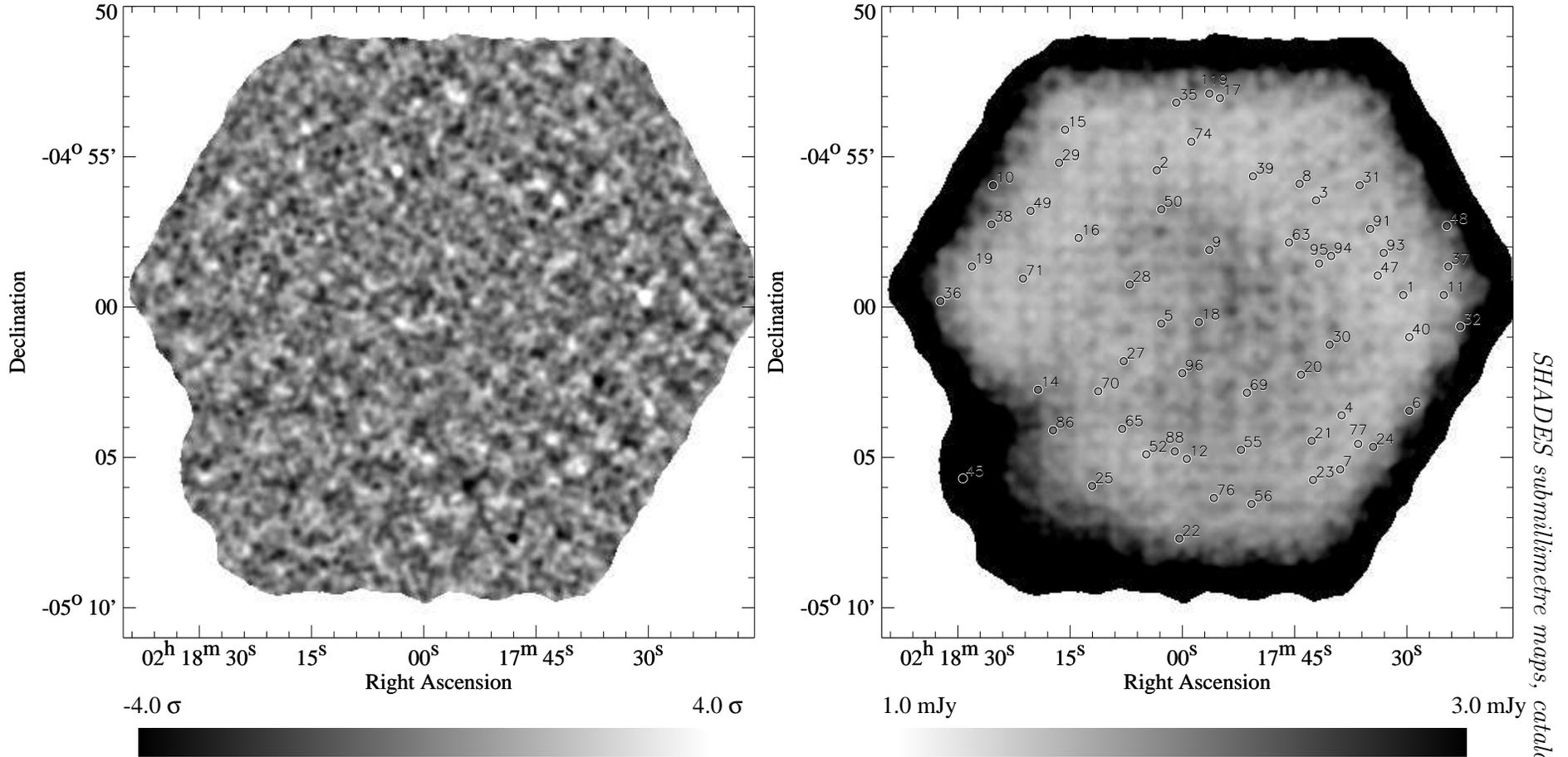,width=1.4\textwidth}
\caption{The $850\,\mathrm{\mu m}$ S/N map (with the off-beams folded in at the timestream level) (left) and corresponding noise map (right) of the $406\,\mathrm{arcmin}^2$ SXDF region are shown (specifically for Reduction D).  This area
 only includes $3\,\mathrm{arcsec}$ pixels that
were sampled more than 10 times.  On the noise map, 15-arcsec diameter circles indicate the positions of the 60 sources in the SXDF SHADES catalogue.  In general, sources are more often found on a `ring' (avoiding a noisier central region taken in predominantly poorer weather), while on a finer scale sources are preferentially found near the corners of the overlapping triangular pattern.  A similar noise map for Reduction B is found in \citet{Mortier}.}
\end{figure}
\end{landscape}

\begin{figure}
\psfig{file=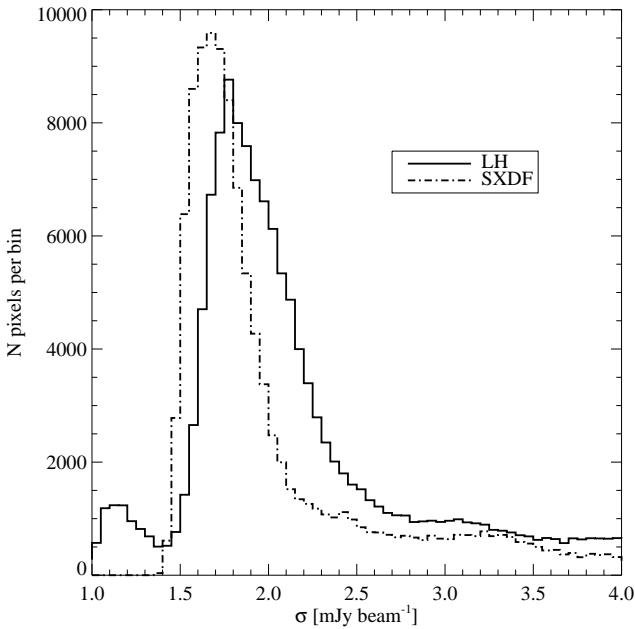,width=0.5\textwidth}
\caption{Histograms showing the noise distribution 
of the 14.7\arcsec FWHM beam-smoothed noise map (pixels with more than 10 hits from Reduction D) for the LH and SXDF fields in bins of $\Delta\,\sigma=0.05\,\mathrm{mJy}\,\mathrm{beam^{-1}}$.  The bump in the LH histogram between 1 and $1.5\,\mathrm{mJy}$ arises from the deep strip in the SCUBA 8-mJy Survey region.}
 \label{noise_hist_fig}
\end{figure}

\subsection{Maps}\label{clean}
The left panels of Figs.~1 and 2 show the point source S/N maps for the SHADES fields from Reduction D.  The positive and negative beams are clearly seen in these maps and appear as a result of the differential measurements taken at the different chop throws.  Noise maps are also presented (see right panels of Figs.~1 and 2) with the positions of the SHADES catalogue sources marked by circles.  Each sample of data that is added into a pixel is weighted by the variance of the bolometer timestream; the noise map here represents the square-root of the variance term.  The regular grid pattern in the noise maps corresponds to variation in the effective observing time as a function of location arising from the triangular pointing pattern in the survey and the decision to chop the telescope along sky coordinates (RA and Dec.) rather than allowing field rotation to smooth the observing time over the map.  

Several features in the maps are worth noting.  The LH map has a non-uniform chop strategy, since the earlier SCUBA 8-mJy Survey region \citep{Scott} was taken only with a single $30\,\mathrm{arcsec}$ chop (with nod) in Declination. This different chop strategy noticeably changes the character of the flux density map; in the SCUBA 8-mJy Survey region the noise is clearly spatially correlated in the vertical direction, and bright sources only have negative side-lobes above and below them. In the noise map, however, the triangular pattern is repeated across the entire field, since the SCUBA 8-mJy Survey was also built from a mosaic of overlapping jiggle-maps falling on the same grid adopted for SHADES. The deepest region of SHADES is a small rectangular region at the centre of the SCUBA 8-mJy Survey where additional data were available from a test strip observed before the commencement of the full SCUBA 8-mJy Survey.  Histograms of the noise maps (for Reduction D) are given in Fig.~\ref{noise_hist_fig}.  The bump in the LH histogram between 1 and $1.5\,\mathrm{mJy}$ arises from the deep strip in the SCUBA 8-mJy Survey region.

\section{The SHADES catalogue}\label{compareall}

The SHADES catalogue is compiled using the following steps.  First, a preliminary joint identification list is compiled by cross-identifying the four independent source lists.  Consensus SHADES positions and flux densities are then determined. Each source flux density is corrected for flux boosting (see Section~\ref{deboost}), and we reject sources based on their deboosted flux density distributions (see Section~\ref{membership}).  This trimming removes from the preliminary list virtually all of the sources which appear to be 2.5--$3.5\,\sigma$ in the maps.  The catalogue we construct here is a robust list, intended to be the starting point for follow-up observations leading to SED fitting and photometric redshift estimation for individual sources and for the population.

\subsection{Combining partially dependent data}\label{dependent}
The four data reductions were carried out independently, but since they use
the same data, their results are obviously not statistically independent.
It is relatively straightforward to determine if the results of the
different reductions are consistent and we will show that the
differences between reductions are small compared to the noise in any
one reduction.  Given that the reductions are consistent, it is
acceptable to choose the result of any single reduction as the final
answer, but combining the analyses is likely to lead to slightly
higher precision and reliability.

The statistics of combining the four reductions is not simple, because
the degree of statistical independence is not well characterised.
Differences between reductions would arise if systematic errors are
present, but might also come from random errors and slight differences
in weights of the input data.  Differences in the estimated
uncertainties would arise if one reduction is genuinely more
precise than the others, but also if one reduction over- or
under-estimates the uncertainties.  In combining our reductions we
have taken a variety of approaches, using the mean value, the median
or the most precise single reduction, and these choices will be
described below in the sections on flux densities, source positions, and
source number densities.  We quote as an uncertainty the
smallest value claimed by a single reduction. In doing so we are
taking the conservative position that while combining the reductions
does not decrease uncertainties nearly as much as combining fully
independent data would, it ought not to increase the uncertainty of our most 
precise estimate.

\subsection{Preliminary joint identification list}\label{prelim}

An extended source list for each reduction is made of all points having S/N $\geq 2.5$ in the maps.  The S/N threshold is kept deliberately low to avoid missing genuine sources at this stage.

A preliminary joint list is constructed by identifying sources for the four extended lists which are within $10\,\mathrm{arcsec}$ and which are seen with S/N $\geq 3$ in at least two reductions.  This preliminary list contains 94 source candidates in the SXDF and 87 in LH.

We compute a SHADES map flux density for each source using the following recipe.  We compute a raw flux density likelihood distribution by multiplying the individual Gaussians constructed from the flux and noise in each reduction and normalise the resulting distribution.  We take the SHADES map flux density to be the maximum likelihood.  We use the lowest quoted error as the $1\,\sigma$ SHADES map noise uncertainty.  Because the data are common to the four reductions, adding errors as inverse variances would seriously underestimate the net uncertainty, while simply averaging the uncertainties allows an imprecise single estimate to lower the combined uncertainty, which does not make sense.  Reductions B and D agree well on flux densities in both fields and also claim the smallest photometric uncertainty, so data from those two reductions dominate the weighted mean flux density for those sources which they extract.  We find that the deviations between any single reduction and the group flux density are smaller than the adopted measurement noise, as expected.  The details are in Section~\ref{compareflux}.  The ratio of SHADES map flux to minimum noise is listed as S/N in Table~\ref{tab:lock}. 

\begin{figure}
\psfig{file=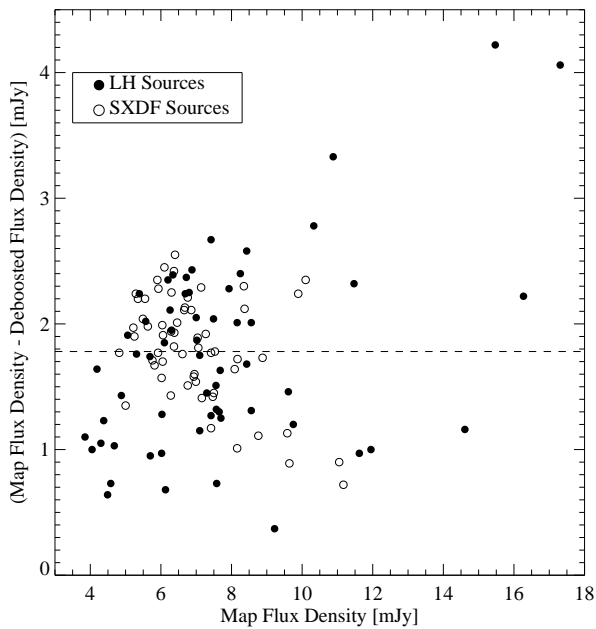,width=0.5\textwidth}
\caption{Map flux density minus deboosted flux density versus the map-detected flux density for the LH (filled symbols) and SXDF (open symbols) SHADES catalogue sources.  The line $y=0$ represents a 1:1 ratio between the flux densities.  Map-detected source flux densities are deboosted by an average of $1.8\,\mathrm{mJy}$ (the dashed line), although there are a range of deboosting values, because this correction depends on the noise as well as the signal.}
\label{fig:before_after}
\end{figure}

\subsection{Deboosted flux densities}\label{deboost}

We have employed the simple Bayesian recipe of \citet{Coppin} to correct the preliminary source list for effects of flux density boosting in submillimetre maps.  Details of how we deboost flux densities from the joint candidate list are given below.  After deboosting, we are left with a catalogue of 120 \textit{robust} SHADES sources, coincidentally 60 sources per field.

The submillimetre source count density in the flux range of the SHADES survey falls very rapidly compared to the width of the approximately Gaussian noise distribution of our maps. Therefore we expect to identify an excess of low flux density sources whose locations happen to coincide with positive noise and whose apparent flux densities  have therefore been increased above the survey's S/N limit.  This `flux boosting' is a well-known effect in low S/N submillimetre maps and must be quantified so that individual source flux densities can be corrected to represent the best estimate of the true underlying flux density of each source.  This boosting is sometimes referred to as Malmquist bias, although more properly that refers to seeing sources more luminous than average in a magnitude-limited survey because of a spread in luminosity, i.e.~there is still Malmquist bias even when there is no measurement error \citep{Hendry}.  Flux density boosting is also distinct from Eddington bias, which properly refers to the effect on the number counts rather than the individual source flux densities.

Each candidate source's flux density is deboosted following the \citet{Coppin} Bayesian recipe, resulting in a posterior flux density probability distribution which is altered from the Gaussian probability distribution inferred from the maps.  The probability distribution for an individual source's intrinsic flux density $S_\mathrm{i}$ given that it is found at the observed flux density $S_\mathrm{o}\,\pm\,\sigma_\mathrm{o}$, is 
$p(S_\mathrm{i}|S_\mathrm{o},\sigma_\mathrm{o})$, which is factored using Bayes' theorem:

\begin{equation}\label{bayes}
  p(S_\mathrm{i}|S_\mathrm{o},\sigma_\mathrm{o}) = \frac{p(S_\mathrm{i})p(S_\mathrm{o},\sigma_\mathrm{o}|S_\mathrm{i})}{p(S_\mathrm{o},\sigma_\mathrm{o})} .
\end{equation}

\noindent This expression states that the intrinsic flux density distribution
of the source is the likelihood of observing the data (right hand term
in the numerator) weighted by the prior distribution of flux densities
(left hand term in the numerator). A Gaussian noise distribution 
is used for the likelihood of the data (the photometric error distribution). 
We use an informative prior for pixel flux densities constructed from existing knowledge of the $850\,\mathrm{\mu m}$ extra-galactic source counts and the SHADES observing strategy.  This inherently gives more weight to lower flux density sources, since they are more numerous.  To construct the prior, we fit the broken power-law model given in \citet{Scott} to the counts of \citet{Borys2003} and constrained by the lensing counts at fainter fluxes, but have checked that mild deviations from this make no significant difference.  The model is of the form

\begin{equation}\label{doublepl}
\frac{dN}{dS} = \frac{N'}{S'}\left[ \left(\frac{S}{S'}\right)^\alpha + \left(\frac{S}{S'}\right)^\beta \right]^{-1}.
\end{equation}

\noindent  Artificial skies are generated consistent with Equation~\ref{doublepl}, Poisson statistics, random source locations and each sky is sampled with the SHADES chopping pattern and point spread function to obtain a pixel flux density distribution.

Because our maps are differential with zero mean, and because they contain noise, the expression in Equation~\ref{bayes} can have a negative tail.
The fraction of the posterior distribution $p(S_\mathrm{i}|S_\mathrm{o},\sigma_\mathrm{o})$
having $S_\mathrm{i} < 0$ 
is taken as the probability that a given
source is falsely detected.  

The idea of using Bayes' theorem to find a posterior estimate of the flux density using the source counts as a prior appears to have been first clearly written down by \citet{Jauncey}, as a way to correct survey-detected radio sources for flux density biasing (as first pointed out by \citealt{Eddington}).  Other papers which discuss similar ideas\footnote{In stellar astronomy the same effect on parallax measurements is known as Lutz-Kelker bias \citet{Lutz}.} include \citet{Murdoch}, \citet{Schmidt}, \citet{Hogg}, \citet{Wang}, and \citet{Teerikorpi}.

\begin{figure*}
\includegraphics[width=5.0in,angle=0]{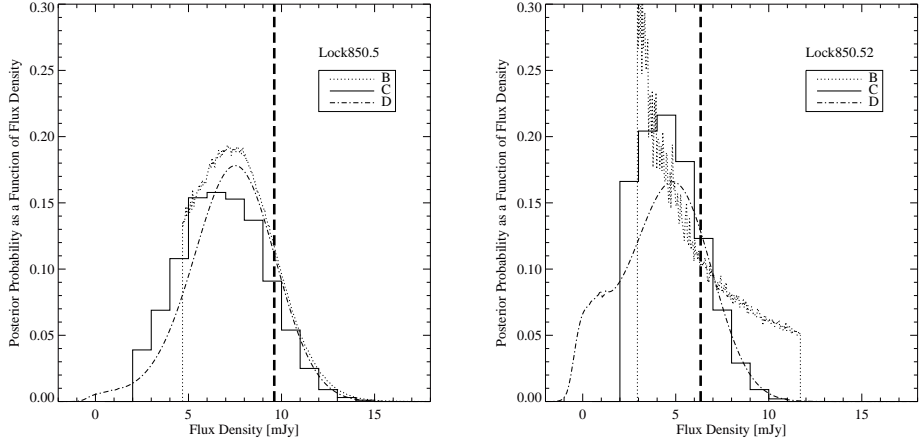}
\caption{Posterior flux density probability density functions calculated from Reduction D by the method described in Section~\ref{deboost} are shown as dot-dashed curves for a high S/N source, LOCK850.5, and a low S/N one, LOCK850.52.  The thick dashed vertical line indicates the average SHADES map flux density before deboosting.   Notice how asymmetric the low S/N posterior flux density distribution is.  For comparison, distributions calculated by alternate methods from  Reductions B and C are shown for the same sources.   Each group's map flux density is within $0.5\,\mathrm{mJy}$ of this average flux density before deboosting.  In these cases, the SHADES posterior flux density distribution (not shown) follows closely the shape of Reduction D's posterior flux density distribution.  All of the distributions shown have been normalised to have unit area.  The distributions from Reduction C are lower resolution due to the choice of bin size for the simulations used to determine the posterior distribution and are truncated below $2\,\mathrm{mJy}$ (see Section~\ref{redC}).   Posterior distributions from Reduction B are truncated outside the region between 0.5--2 times the map flux density.}
\label{fig:deboosted_compare}
\end{figure*}

On average, the deboosting reduces the source flux density by $\simeq 1.8\,\mathrm{mJy}$  (see Fig.~\ref{fig:before_after}), increases the width of the photometric error distribution by about 10 per cent, and renders the shape of the resulting distribution to be skewed and non-Gaussian.  The details of these effects depend both on the observed signal, $S_\mathrm{o}$, and the observed noise, $\sigma_\mathrm{o}$, and not just on the S/N.  The effects are larger for sources extracted from noisier regions of the maps.  We have plotted two examples of posterior distributions taken from the LH region in Fig.~\ref{fig:deboosted_compare}. The first is a bright source and the other is dim; one readily sees that the skew of the distribution is more pronounced in the dim case.  We note that previous submillimetre surveys such as those of \citet{Scott} and \citet{Eales} assessed the level of flux density boosting through simulations by comparing input source flux densities to extracted source flux densities for a $3\,\sigma$ catalogue cut and noise map RMS values of $\simeq2.5\,\mathrm{mJy}$ and 1\,mJy, respectively.  \citet{Scott} found a boosting factor of about 15 per cent at 8\,mJy and 10 per cent at 11\,mJy, while \citet{Eales} quote a median boost factor of 1.44 with a large scatter: these correction factors were detemined as a function of flux density only.  In SHADES, each source is individually corrected for flux density boosting by a factor determined as a function of the map-detected flux density \textit{as well as} the noise estimate.

The deboosting recipe has been successfully tested against follow-up photometry for individual sources in \citet{Coppin}.  Its performance in returning the input source distribution is tested in Section~\ref{deboosttest2}.

Since some of the LH data were taken originally for the SCUBA 8-mJy Survey \citep{Scott} using a single chop throw, the effect of using this observing scheme on the reported deboosted flux densities was investigated by recalculating the prior and flux density posterior probability distributions.  Reported flux densities were different for less than 10 per cent of the SHADES sources, and at most by a negligible $\pm 0.1\,\mathrm{mJy}$.  In addition, 3 extra sources near the rejection threshold made it into the final LH source catalogue.  For simplicity, we used the prior based on the multiple chop SHADES observing scheme to deboost all of the sources, since it is representative of the majority of the data.

The possible effects of clustering have not been included in creating the prior distribution used in deboosting fluxes. We have checked that  clustering at the levels anticipated for SHADES  sources has a negligible effect on the deboosted flux density distributions.  Using 50 realisations of the phenomenological galaxy formation model used in \citet{vanKampen}, with a clustering strength of $\theta_{o}\simeq10\,\mathrm{arcsec}$, we create a noiseless distribution of map pixel fluxes and use it to construct a new prior distribution and deboost sources with similar flux densities and S/N as the SHADES sources.  We compare the posterior flux density distributions to those calculated using a prior with the same input source count model, but with randomised positions, i.e. no clustering.   We find negligible differences between the distributions.  A larger effect on the shape of the posterior flux density probability distribution comes from using a much steeper source count model at the faint end of the number counts, which has a small but noticeable effect on the shape of the posterior flux density probability distribution at flux densities below $1\,\mathrm{mJy}$, making sources more likely to pass the catalogue cut.  We therefore claim to be making the most conservative catalogue cut, as we use less steep source counts at the faint end that we believe are a balance between faint lensing and bright blank-field counts.

\subsection{$850\,\mathrm{\mu m}$ catalogue membership}\label{membership}

The final cut in catalogue membership is the requirement that each accepted source has less than 5 per cent of its posterior probability distribution below $0\,\mathrm{mJy}$, or $P(S_\mathrm{i}\leq0\,\mathrm{mJy})<5$ per cent.  This threshold is a good balance between detecting sources while keeping the number of spurious detections to a minimum.

\begin{figure}
\psfig{file=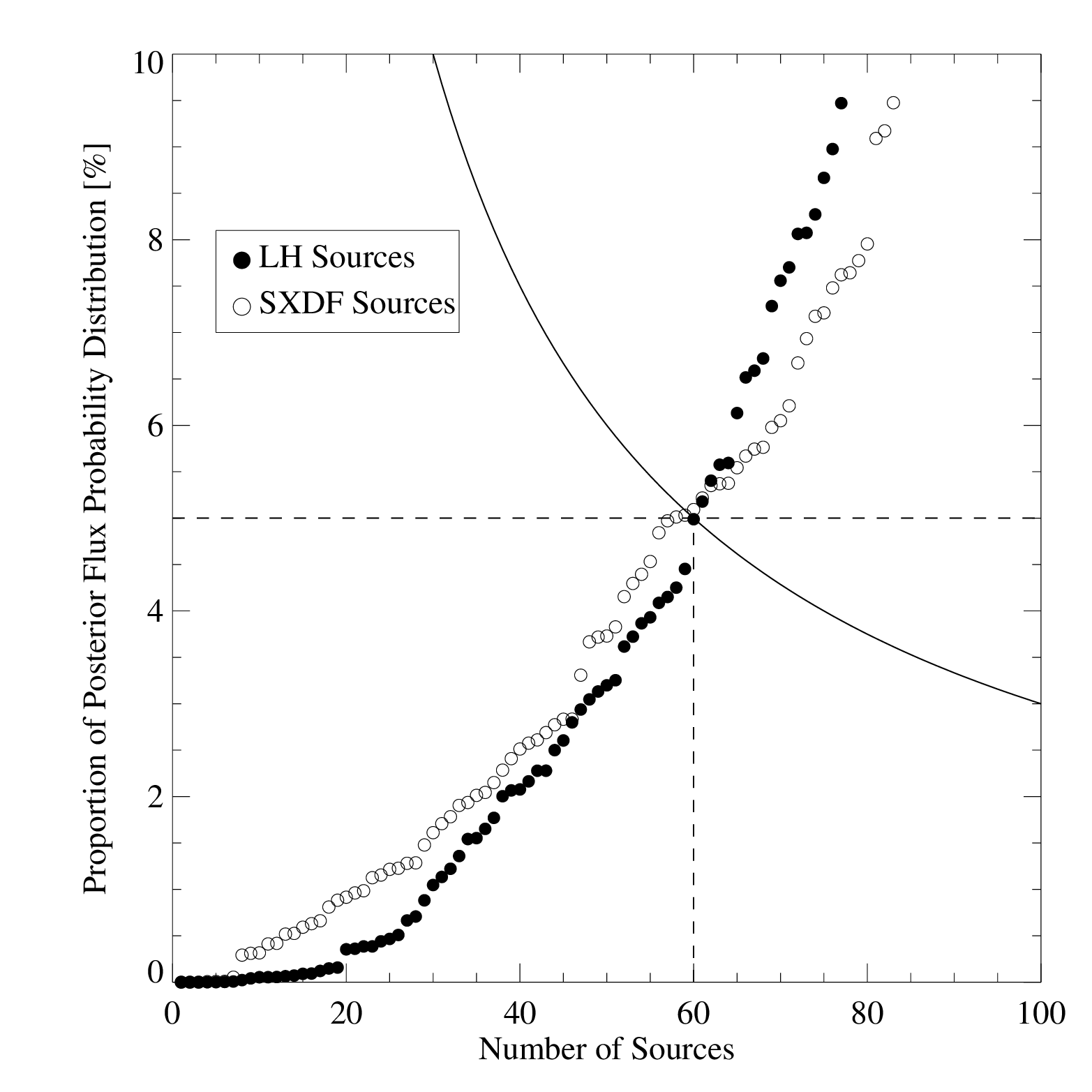,width=0.5\textwidth}
\caption{Percentage likelihood of zero flux density.  This plot shows the 
posterior probability that each source has a flux density $<0\,\mathrm{mJy}$.  
 This probability is plotted for each source {\it candidate} in the LH (filled symbols) and SXDF (open symbols) in 
ascending order.  Only those candidate sources for which 
the null probability is less than 5 per cent (corresponding to 
locations in the figure below the horizontal dashed line), 
are kept in the SHADES catalogue. In each of the LH and SXDF
 there are 60 such sources.  Notice the comparatively small 
number of sources in the SXDF with very low null probabilities compared with
the LH.  Notice also that if we had chosen a cut below 5 per cent there would have been more LH than SXDF sources, while the opposite would have been true if we had chosen a cut above 5 per cent.  Given the number of beams in each map, one expects typically five $3\,\sigma$ peaks at random, and perhaps three of these survive the deboosting process.  The solid line shows the percentage of the catalogue comprised of such sources as a function of the number of sources in the catalogue, $P=3/N$, which coincidentally crosses in the same place.}
\label{fig:FDR_lock}
\end{figure}

One could tune the catalogue membership using the thresholding technique called the False Discovery Rate (FDR; see \citealt{BH} and \citealt{Miller}) in order to control the average fraction of spurious sources in the catalogue.  In Fig.~\ref{fig:FDR_lock} we have plotted the posterior null probability for each source candidate (i.e.~the percentage of the posterior flux density probability distribution which is below $0\,\mathrm{mJy}$) ranked in ascending order.  This plot illustrates how our choice of probability cut-off for individual sources occurs comfortably before the regime where the FDR increases dramatically.  It is also coincidentally where the number of sources times the FDR approximately equals 1.  These plots show that as one pushes beyond $\simeq 60$, the number of spurious sources becomes $\gg 1$.

Given the number of beams in each map, 
one expects approximately five $3\,\sigma$ peaks at random.  If we 
relaxed the null probability cut we would increase the number of 
sources, but the chances of random noise peaks making it into the 
catalogue would then become important.  The {\it average\ } 
of the null probabilities is 1.5 (2.0) per cent in the LH (SXDF), 
and this can be interpreted as an overall FDR for the catalogue.
 
\begin{figure*}
\includegraphics[width=6.0in,angle=0]{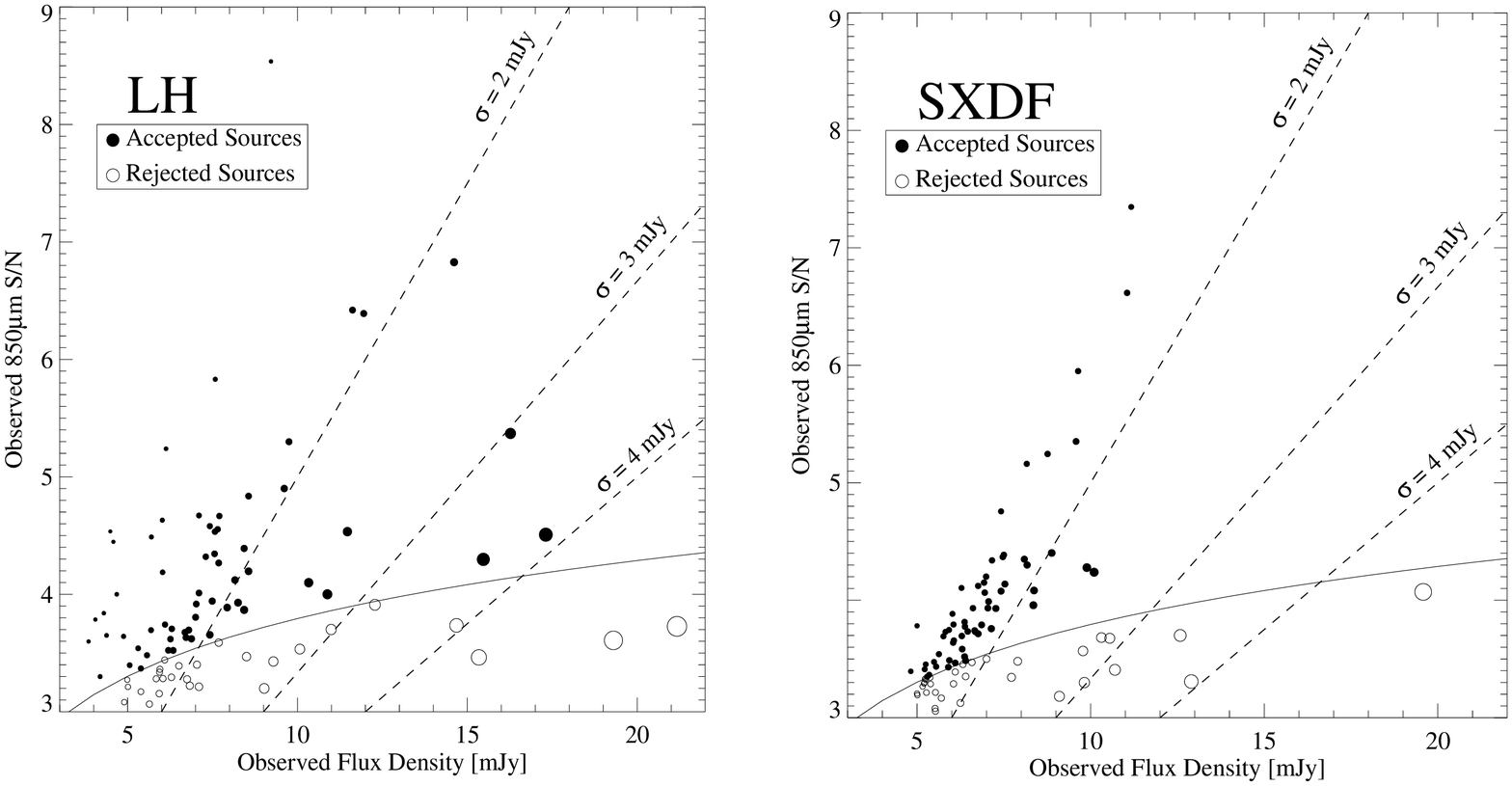}
\caption{Effective cuts in the LH (left) and SXDF (right) source catalogues are shown.  For all candidate sources, the SHADES observed S/N is plotted against observed flux density as circles (the size of a circle is proportional to the map noise level where a source was found).  The dashed lines show observed noise levels of 2, 3, and $4\,\mathrm{mJy}$.  Sources are retained in the catalogue if the total posterior probability that the flux density is zero is less than 5 per cent (filled circles).  Only the sources lying above the solid curve satisfy this criterion.  The effect of the rapidly falling source count spectrum is visible as the curvature of this cut-off, \textit{i.e.} the rise in the required S/N with increasing noise.  An apparent $3.5\,\sigma$ source in a noisy region of the map has a higher flux density than a $3.5\,\sigma$ source found in a quiet region; bright sources are rare, so this source is less likely to be genuine than a quieter dim source is.  Notice that there are dramatically fewer sources detected at flux densities above $10\,\mathrm{mJy}$ in the SXDF as compared to the LH field, even though the number of fainter sources is similar.  Also note that essentially none of the SXDF detected sources have noise in the 1--$1.5\,\mathrm{mJy}$ range, as compared to the LH, which includes the lower noise SCUBA 8-mJy Survey region.}
\label{fig:flux_noise}
\end{figure*}

\begin{figure}
\psfig{file=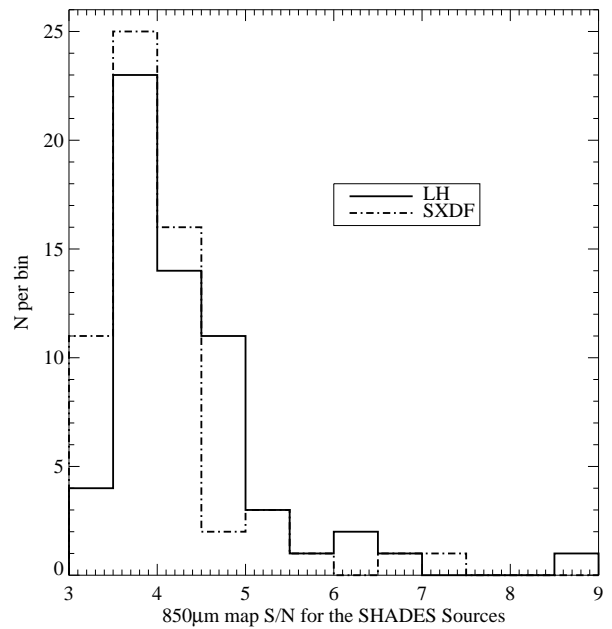,width=0.5\textwidth}
\caption{Distribution of the map-detected S/N ratios 
(SHADES catalogue values) for the 60 LH and 60 SXDF
sources in bins of $\Delta\,\mathrm{S/N}=0.5$.  
We note that all of the SHADES catalogue 
sources were initially detected above a S/N of 
3.2, whereas the preliminary list 
contained sources $\geq2.5$. }
\label{both_snr_histogram}
\end{figure}

Effective flux density and noise cuts in each catalogue are shown in Fig.~\ref{fig:flux_noise}.  \textit{No} sources with observed S/N $<3.2$ are kept in the final catalogues, with the majority of the detections lying above $3.5\,\sigma$ (see Fig.~\ref{both_snr_histogram} for the S/N distribution of the SHADES catalogue sources).  
  
The flux densities we quote in Table~\ref{tab:lock} are median flux density estimates and the quoted errors correspond to the central 68 per cent of the posterior flux density distribution.

We remark that there are substantially 
more \textit{high} S/N sources with null probabilities 
below 5 per cent (i.e~accepted in the SHADES catalogue) 
in the SHADES maps of the LH region than 
in the SXDF (see Fig.~\ref{both_snr_histogram}).  

\subsection{Final SHADES catalogue}

The SHADES catalogue is given in Table~\ref{tab:lock}.  Gaps in the source numbering sequence indicate sources that were rejected from the preliminary catalogue because they either failed to be detected by at least two groups with S/N $\geq 3$, or because $P(S_\mathrm{i}\leq0\,\mathrm{mJy})>5$ per cent.  Comments on particular sources are noted in Appendix~\ref{notes}.

For the first time in a submillimetre-selected survey, a careful estimate of the unbiased flux density of each source is provided (cf. Section~\ref{deboost}).  


Table~\ref{tab:lock} also contains $3\,\sigma$ upper limits for the $450\,\mathrm{\mu m}$ flux densities of each SHADES source (see Appendix~\ref{survival} for details).  

\scriptsize
\begin{center}
\begin{onecolumn}
\begin{longtable}{llccccccl}
\hline
\multicolumn{1}{l}{Name (IAU)} & \multicolumn{1}{l}{Nickname} & \multicolumn{1}{c}{RA} & \multicolumn{1}{c}{Dec.} & \multicolumn{1}{c}{$S_{850}$} & \multicolumn{1}{c}{map S/N} & \multicolumn{1}{c}{map $\sigma_{850}$} & \multicolumn{1}{c}{$S_{450}$} & \multicolumn{1}{l}{Other IDs}\\
\multicolumn{1}{c}{} & \multicolumn{1}{c}{} & \multicolumn{1}{c}{(J2000)} & \multicolumn{1}{c}{(J2000)} & \multicolumn{1}{c}{(mJy)} & \multicolumn{1}{c}{} & \multicolumn{1}{c}{(mJy)} & \multicolumn{1}{c}{(mJy)} & \multicolumn{1}{c}{/Notes} \\\hline
\endhead

\multicolumn{9}{l}{{Continued on Next Page\ldots}}\\
\endfoot

\\[-1.8ex]\hline\hline
\endlastfoot

SHADES J105201+572443 & LOCK850.1 & \(10^{\mathrm{h}}52^{\mathrm{m}}01{\mathrm{\fs}}42\) & \(57^{\circ}24'43{\farcs}0\) & 8.85 ($\pm^{1.0}_{1.0}$) & 8.5 & 1.1 & $<47\,\diamond$ & LE850.1,LE1100.14,$\heartsuit\chi_\mathrm{s}$ \\
SHADES J105257+572105 & LOCK850.2 & \(10^{\mathrm{h}}52^{\mathrm{m}}57{\mathrm{\fs}}32\) & \(57^{\circ}21'05{\farcs}8\) & 13.45 ($\pm^{2.1}_{2.1}$) & 6.8  & 2.1 & $<123$ & LE1100.1,$\heartsuit\aleph$\\
SHADES J105238+572436 & LOCK850.3 & \(10^{\mathrm{h}}52^{\mathrm{m}}38{\mathrm{\fs}}25\) & \(57^{\circ}24'36{\farcs}5\) & 10.95 ($\pm^{1.8}_{1.9}$) & 6.4  & 1.9 & $<34$ & LE850.2,LE1100.8,$\heartsuit$ \\
SHADES J105204+572658 & LOCK850.4 & \(10^{\mathrm{h}}52^{\mathrm{m}}04{\mathrm{\fs}}17\) & \(57^{\circ}26'58{\farcs}9\) & 10.65 ($\pm^{1.7}_{1.8}$) & 6.4 & 1.8 & $<134\,\diamond$ & LE850.14,$\heartsuit$ \\
SHADES J105302+571827 & LOCK850.5 & \(10^{\mathrm{h}}53^{\mathrm{m}}02{\mathrm{\fs}}62\) & \(57^{\circ}18'27{\farcs}0\) & 8.15 ($\pm^{2.0}_{2.1}$) & 4.9 & 2.0 & $<107$ & $\heartsuit\aleph$\\
SHADES J105204+572526 & LOCK850.6 & \(10^{\mathrm{h}}52^{\mathrm{m}}04{\mathrm{\fs}}13\) & \(57^{\circ}25'26{\farcs}3\) & 6.85 ($\pm^{1.3}_{1.3}$) & 5.8 & 1.3 & $<77$ & LE850.4,$\heartsuit$ \\
SHADES J105301+572554 & LOCK850.7 & \(10^{\mathrm{h}}53^{\mathrm{m}}01{\mathrm{\fs}}40\) & \(57^{\circ}25'54{\farcs}2\) & 8.55 ($\pm^{1.8}_{1.9}$) & 5.3 & 1.8 & $<85$ & $\heartsuit$\\
SHADES J105153+571839 & LOCK850.8 & \(10^{\mathrm{h}}51^{\mathrm{m}}53{\mathrm{\fs}}86\) & \(57^{\circ}18'39{\farcs}8\) & 5.45 ($\pm^{1.1}_{1.2}$) & 5.2 & 1.2 & $<31$ & LE850.27,$\heartsuit\chi_\mathrm{t}$ \\
SHADES J105216+572504 & LOCK850.9 & \(10^{\mathrm{h}}52^{\mathrm{m}}16{\mathrm{\fs}}09\) & \(57^{\circ}25'04{\farcs}1\) & 5.95 ($\pm^{1.6}_{1.6}$) & 4.7 & 1.5 & $<68$ & LE850.29,$\heartsuit$ \\
SHADES J105248+573258 & LOCK850.10 & \(10^{\mathrm{h}}52^{\mathrm{m}}48{\mathrm{\fs}}61\) & \(57^{\circ}32'58{\farcs}6\) & 9.15 ($\pm^{2.7}_{2.9}$) & 4.5 & 2.5 & $<365$ & $\dagger$\\
SHADES J105129+572405 & LOCK850.11 & \(10^{\mathrm{h}}51^{\mathrm{m}}29{\mathrm{\fs}}53\) & \(57^{\circ}24'05{\farcs}2\) & 6.25 ($\pm^{1.7}_{1.8}$) & 4.5 & 1.7 & $<62$ & $\heartsuit\aleph$\\
SHADES J105227+572513 & LOCK850.12 & \(10^{\mathrm{h}}52^{\mathrm{m}}27{\mathrm{\fs}}61\) & \(57^{\circ}25'13{\farcs}1\) & 6.15 ($\pm^{1.7}_{1.7}$) & 4.6 & 1.6 & $<35$ & LE850.16,LE1100.16,$\heartsuit\chi_\mathrm{t}$ \\
SHADES J105132+573134 & LOCK850.13 & \(10^{\mathrm{h}}51^{\mathrm{m}}32{\mathrm{\fs}}33\) & \(57^{\circ}31'34{\farcs}8\) & 5.65 ($\pm^{2.3}_{2.9}$) & 3.9 & 2.0 & $<109$ & $\dagger\ast$\\
SHADES J105230+572215 & LOCK850.14 & \(10^{\mathrm{h}}52^{\mathrm{m}}30{\mathrm{\fs}}11\) & \(57^{\circ}22'15{\farcs}6\) & 7.25 ($\pm^{1.8}_{1.9}$) & 4.8 & 1.8 & $<96$ & LE850.6,LE1100.5,$\heartsuit$ \\
SHADES J105319+572110 & LOCK850.15 & \(10^{\mathrm{h}}53^{\mathrm{m}}19{\mathrm{\fs}}20\) & \(57^{\circ}21'10{\farcs}6\) & 13.25 ($\pm^{4.3}_{5.0}$) & 4.5 & 3.8 & $<149$ & $\heartsuit\chi_\mathrm{t}$\\
SHADES J105151+572637 & LOCK850.16 & \(10^{\mathrm{h}}51^{\mathrm{m}}51{\mathrm{\fs}}45\) & \(57^{\circ}26'37{\farcs}0\) & 5.85 ($\pm^{1.8}_{1.9}$) & 4.3 & 1.7 & $<67\,\diamond$ & LE850.7,$\heartsuit$ \\
SHADES J105158+571800 & LOCK850.17 & \(10^{\mathrm{h}}51^{\mathrm{m}}58{\mathrm{\fs}}25\) & \(57^{\circ}18'00{\farcs}8\) & 4.75 ($\pm^{1.3}_{1.3}$) & 4.5 & 1.3 & $<55$ & LE850.3,$\heartsuit$ \\
SHADES J105227+572217 & LOCK850.18 & \(10^{\mathrm{h}}52^{\mathrm{m}}27{\mathrm{\fs}}69\) & \(57^{\circ}22'17{\farcs}8\) & 6.05 ($\pm^{1.9}_{2.1}$) & 4.3 & 1.8 & $<84$ & SDS.20,$\dagger$ \\
SHADES J105235+573119 & LOCK850.19 & \(10^{\mathrm{h}}52^{\mathrm{m}}35{\mathrm{\fs}}71\) & \(57^{\circ}31'19{\farcs}1\) & 5.15 ($\pm^{2.0}_{2.4}$) & 3.9 & 1.8 & $<87$ & $\heartsuit$\\
SHADES J105256+573038 & LOCK850.21 & \(10^{\mathrm{h}}52^{\mathrm{m}}56{\mathrm{\fs}}86\) & \(57^{\circ}30'38{\farcs}1\) & 4.15 ($\pm^{2.0}_{2.5}$) & 3.6 & 1.7 & $<70$ & $\heartsuit\aleph$\\
SHADES J105137+573323 & LOCK850.22 & \(10^{\mathrm{h}}51^{\mathrm{m}}37{\mathrm{\fs}}55\) & \(57^{\circ}33'23{\farcs}3\) & 7.55 ($\pm^{3.2}_{4.2}$) & 4.0 & 2.7 & $<76$ & $\heartsuit\aleph$\\
SHADES J105213+573154 & LOCK850.23 & \(10^{\mathrm{h}}52^{\mathrm{m}}13{\mathrm{\fs}}74\) & \(57^{\circ}31'54{\farcs}1\) & 4.35 ($\pm^{1.9}_{2.4}$) & 3.7 & 1.7 & $<60$ & $\heartsuit$\\
SHADES J105200+572038 & LOCK850.24 & \(10^{\mathrm{h}}52^{\mathrm{m}}00{\mathrm{\fs}}23\) & \(57^{\circ}20'38{\farcs}1\) & 2.75 ($\pm^{1.2}_{1.2}$) & 3.6 & 1.1 & $<32$ & LE850.32,$\heartsuit$ \\
SHADES J105240+572312 & LOCK850.26 & \(10^{\mathrm{h}}52^{\mathrm{m}}40{\mathrm{\fs}}95\) & \(57^{\circ}23'12{\farcs}0\) & 5.85 ($\pm^{2.4}_{2.9}$) & 3.9 & 2.1 & $<48$ & $\ddagger$ \\
SHADES J105203+571813 & LOCK850.27 & \(10^{\mathrm{h}}52^{\mathrm{m}}03{\mathrm{\fs}}57\) & \(57^{\circ}18'13{\farcs}5\) & 5.05 ($\pm^{1.3}_{1.3}$) & 4.6 & 1.3 & $<32$ & LE1100.4,$\dagger\ast$\\
SHADES J105257+573107 & LOCK850.28 & \(10^{\mathrm{h}}52^{\mathrm{m}}57{\mathrm{\fs}}00\) & \(57^{\circ}31'07{\farcs}1\) & 6.45 ($\pm^{1.7}_{1.8}$) & 4.7 & 1.7 & $<56$ & $\dagger\ast$ \\
SHADES J105130+572036 & LOCK850.29 & \(10^{\mathrm{h}}52^{\mathrm{m}}30{\mathrm{\fs}}92\) & \(57^{\circ}20'36{\farcs}0\) & 6.75 ($\pm^{2.0}_{2.2}$) & 4.4 & 1.9 & $<64$ & LE850.11,$\heartsuit$ \\
SHADES J105207+571906 & LOCK850.30 & \(10^{\mathrm{h}}52^{\mathrm{m}}07{\mathrm{\fs}}79\) & \(57^{\circ}19'06{\farcs}6\) & 4.75 ($\pm^{1.5}_{1.6}$) & 4.2 & 1.4 & $<86$ & LE850.12,$\heartsuit$ \\
SHADES J105216+571621 & LOCK850.31 & \(10^{\mathrm{h}}52^{\mathrm{m}}16{\mathrm{\fs}}06\) & \(57^{\circ}16'21{\farcs}1\) & 6.05 ($\pm^{1.8}_{2.0}$) & 4.3 & 1.7 & $<80$ & $\heartsuit$\\
SHADES J105155+572311 & LOCK850.33 & \(10^{\mathrm{h}}51^{\mathrm{m}}55{\mathrm{\fs}}98\) & \(57^{\circ}23'11{\farcs}8\) & 3.85 ($\pm^{1.0}_{1.1}$) & 4.4 & 1.0 & $<49\,\diamond$ & LE850.18,$\dagger\ast\aleph$ \\
SHADES J105213+573328 & LOCK850.34 & \(10^{\mathrm{h}}52^{\mathrm{m}}13{\mathrm{\fs}}50\) & \(57^{\circ}33'28{\farcs}1\) & 14.05 ($\pm^{3.1}_{3.2}$) & 5.4 & 3.0 & $<100$ & $\dagger\ast$\\
SHADES J105246+572056 & LOCK850.35 & \(10^{\mathrm{h}}52^{\mathrm{m}}46{\mathrm{\fs}}92\) & \(57^{\circ}20'56{\farcs}3\) & 6.15 ($\pm^{2.2}_{2.4}$) & 4.1 & 2.0 & $<91$ & $\dagger\ast\aleph$\\
SHADES J105209+571806 & LOCK850.36 & \(10^{\mathrm{h}}52^{\mathrm{m}}09{\mathrm{\fs}}34\) & \(57^{\circ}18'06{\farcs}8\) & 6.35 ($\pm^{1.7}_{1.8}$) & 4.6 & 1.7 & $<67$ & $\dagger\ast$\\
SHADES J105124+572334 & LOCK850.37 & \(10^{\mathrm{h}}51^{\mathrm{m}}24{\mathrm{\fs}}13\) & \(57^{\circ}23'34{\farcs}9\) & 7.55 ($\pm^{2.9}_{3.5}$) & 4.1 & 2.5 & $<20$ & $\heartsuit$\\
SHADES J105307+572431 & LOCK850.38 & \(10^{\mathrm{h}}53^{\mathrm{m}}07{\mathrm{\fs}}10\) & \(57^{\circ}24'31{\farcs}4\) & 4.35 ($\pm^{2.2}_{2.7}$) & 3.6 & 1.9 & $<93$ & $\dagger\ast$ \\
SHADES J105224+571609 & LOCK850.39 & \(10^{\mathrm{h}}52^{\mathrm{m}}24{\mathrm{\fs}}85\) & \(57^{\circ}16'09{\farcs}8\) & 6.55 ($\pm^{2.2}_{2.7}$) & 8.6 & 2.0 & $<9$ & $\dagger\aleph$ \\
SHADES J105202+571915 & LOCK850.40 & \(10^{\mathrm{h}}52^{\mathrm{m}}02{\mathrm{\fs}}01\) & \(57^{\circ}19'15{\farcs}8\) & 3.05 ($\pm^{1.1}_{1.2}$) & 3.8 & 1.1 & $<40$ & LE850.21,$\heartsuit\chi_\mathrm{t}$ \\
SHADES J105159+572423 & LOCK850.41 & \(10^{\mathrm{h}}51^{\mathrm{m}}59{\mathrm{\fs}}86\) & \(57^{\circ}24'23{\farcs}6\) & 3.85 ($\pm^{0.9}_{1.0}$) & 4.5 & 1.0 & $<16$ & LE850.8,LE1100.17,$\dagger\chi_\mathrm{s}$ \\
SHADES J105257+572351 & LOCK850.43 & \(10^{\mathrm{h}}52^{\mathrm{m}}57{\mathrm{\fs}}17\) & \(57^{\circ}23'51{\farcs}8\) & 4.95 ($\pm^{2.1}_{2.6}$) & 3.8 & 1.8 & $<80$ & $\dagger\aleph$\\
SHADES J105235+572514 & LOCK850.47 & \(10^{\mathrm{h}}52^{\mathrm{m}}35{\mathrm{\fs}}63\) & \(57^{\circ}25'14{\farcs}0\) & 3.55 ($\pm^{1.7}_{2.1}$) & 3.5 & 1.5 & $<21$ & SDS.16,$\dagger\ast\aleph$ \\
SHADES J105256+573245 & LOCK850.48 & \(10^{\mathrm{h}}52^{\mathrm{m}}56{\mathrm{\fs}}24\) & \(57^{\circ}32'45{\farcs}8\) & 5.45 ($\pm^{2.1}_{2.5}$) & 3.9 & 1.9 & $<79$ & $\dagger\ast\aleph$\\
SHADES J105245+573121 & LOCK850.52 & \(10^{\mathrm{h}}52^{\mathrm{m}}45{\mathrm{\fs}}53\) & \(57^{\circ}31'21{\farcs}9\) & 3.95 ($\pm^{2.2}_{2.7}$) & 3.5 & 1.8 & $<106$ & $\dagger\ast$\\
SHADES J105240+571928 & LOCK850.53 & \(10^{\mathrm{h}}52^{\mathrm{m}}40{\mathrm{\fs}}49\) & \(57^{\circ}19'28{\farcs}4\) & 4.45 ($\pm^{2.3}_{2.9}$) & 3.6 & 1.9 & $<90$ & $\ddagger$\\
SHADES J105143+572446 & LOCK850.60 & \(10^{\mathrm{h}}51^{\mathrm{m}}43{\mathrm{\fs}}58\) & \(57^{\circ}24'46{\farcs}0\) & 3.15 ($\pm^{1.7}_{2.0}$) & 3.4 & 1.5 & $<44$ & LE850.10,$\ddagger\ast\aleph$ \\
SHADES J105153+572505 & LOCK850.63 & \(10^{\mathrm{h}}51^{\mathrm{m}}53{\mathrm{\fs}}91\) & \(57^{\circ}25'05{\farcs}1\) & 3.65 ($\pm^{1.2}_{1.3}$) & 4.0 & 1.2 & $<50$ & $\dagger$\\
SHADES J105251+573242 & LOCK850.64 & \(10^{\mathrm{h}}52^{\mathrm{m}}51{\mathrm{\fs}}81\) & \(57^{\circ}32'42{\farcs}2\) & 5.85 ($\pm^{2.5}_{3.2}$) & 3.9 & 2.2 & $<95$ & $\dagger$\\
SHADES J105138+572017 & LOCK850.66 & \(10^{\mathrm{h}}51^{\mathrm{m}}38{\mathrm{\fs}}69\) & \(57^{\circ}20'17{\farcs}2\) & 4.25 ($\pm^{1.9}_{2.2}$) & 3.7 & 1.6 & $<40$ & $\dagger\aleph$\\
SHADES J105209+572355 & LOCK850.67 & \(10^{\mathrm{h}}52^{\mathrm{m}}09{\mathrm{\fs}}00\) & \(57^{\circ}23'55{\farcs}1\) & 2.55 ($\pm^{1.5}_{1.5}$) & 3.3 & 1.3 & $<63$ & $\dagger\aleph$\\
SHADES J105148+573046 & LOCK850.70 & \(10^{\mathrm{h}}51^{\mathrm{m}}48{\mathrm{\fs}}52\) & \(57^{\circ}30'46{\farcs}7\) & 3.85 ($\pm^{2.2}_{2.5}$) & 3.5 & 1.8 & $<106$ & $\ddagger\ast$\\
SHADES J105218+571903 & LOCK850.71 & \(10^{\mathrm{h}}52^{\mathrm{m}}18{\mathrm{\fs}}62\) & \(57^{\circ}19'03{\farcs}8\) & 3.95 ($\pm^{1.8}_{2.0}$) & 3.7 & 1.5 & $<99$ & $\dagger\ast$\\
SHADES J105141+572217 & LOCK850.73 & \(10^{\mathrm{h}}51^{\mathrm{m}}41{\mathrm{\fs}}66\) & \(57^{\circ}22'17{\farcs}6\) & 3.55 ($\pm^{1.9}_{2.3}$) & 3.5 & 1.6 & $<49$ & $\dagger\ast$\\
SHADES J105315+572645 & LOCK850.75 & \(10^{\mathrm{h}}53^{\mathrm{m}}15{\mathrm{\fs}}93\) & \(57^{\circ}26'45{\farcs}5\) & 4.45 ($\pm^{2.2}_{2.6}$) & 3.7 & 1.8 & $<50$ & $\dagger\ast$\\
SHADES J105148+572838 & LOCK850.76 & \(10^{\mathrm{h}}51^{\mathrm{m}}48{\mathrm{\fs}}52\) & \(57^{\circ}28'38{\farcs}7\) & 4.75 ($\pm^{2.5}_{3.1}$) & 3.7 & 2.0 & $<90$ & LE1100.15,$\heartsuit\aleph$ \\
SHADES J105157+572210 & LOCK850.77 & \(10^{\mathrm{h}}51^{\mathrm{m}}57{\mathrm{\fs}}00\) & \(57^{\circ}22'10{\farcs}1\) & 3.25 ($\pm^{1.2}_{1.3}$) & 3.8 & 1.1 & $<39$ & $\dagger\aleph$\\
SHADES J105145+571738 & LOCK850.78 & \(10^{\mathrm{h}}51^{\mathrm{m}}45{\mathrm{\fs}}33\) & \(57^{\circ}17'38{\farcs}7\) & 4.55 ($\pm^{2.2}_{2.7}$) & 3.7 & 1.8 & $<56$ & $\dagger$\\
SHADES J105152+572127 & LOCK850.79 & \(10^{\mathrm{h}}51^{\mathrm{m}}52{\mathrm{\fs}}10\) & \(57^{\circ}21'27{\farcs}4\) & 3.15 ($\pm^{1.3}_{1.5}$) & 3.7 & 1.2 & $<41$ & $\dagger\aleph$\\
SHADES J105231+571800 & LOCK850.81 & \(10^{\mathrm{h}}52^{\mathrm{m}}31{\mathrm{\fs}}99\) & \(57^{\circ}18'00{\farcs}4\) & 5.35 ($\pm^{1.9}_{2.3}$) & 4.0 & 1.8 & $<92$ & $\dagger$\\
SHADES J105307+572839 & LOCK850.83 & \(10^{\mathrm{h}}53^{\mathrm{m}}07{\mathrm{\fs}}94\) & \(57^{\circ}28'39{\farcs}1\) & 3.15 ($\pm^{2.0}_{2.1}$) & 3.4 & 1.6 & $<69$ & $\ddagger\ast$\\
SHADES J105153+571733 & LOCK850.87 & \(10^{\mathrm{h}}51^{\mathrm{m}}53{\mathrm{\fs}}30\) & \(57^{\circ}17'33{\farcs}4\) & 3.45 ($\pm^{1.5}_{1.7}$) & 3.6 & 1.3 & $<54$ & $\dagger$\\
SHADES J105139+571509 & LOCK850.100 & \(10^{\mathrm{h}}51^{\mathrm{m}}39{\mathrm{\fs}}06\) & \(57^{\circ}15'09{\farcs}8\) & 11.25 ($\pm^{4.2}_{5.3}$) & 4.3 & 3.6 & $<75$ & $\ddagger\ast$\\
\hline\hline
SHADES J021730-045937 & SXDF850.1 & \(02^{\mathrm{h}}17^{\mathrm{m}}30{\mathrm{\fs}}53\) & \(-04^{\circ}59'37{\farcs}0\) & 10.45 ($\pm^{1.5}_{1.4}$) & 7.3 & 1.5 & $<65$ & $\heartsuit$\\
SHADES J021803-045527 & SXDF850.2 & \(02^{\mathrm{h}}18^{\mathrm{m}}03{\mathrm{\fs}}51\) & \(-04^{\circ}55'27{\farcs}2\) & 10.15 ($\pm^{1.6}_{1.6}$) & 6.6 & 1.7 & $<98$ & $\heartsuit$\\
SHADES J021742-045628 & SXDF850.3 & \(02^{\mathrm{h}}17^{\mathrm{m}}42{\mathrm{\fs}}14\) & \(-04^{\circ}56'28{\farcs}2\) & 8.75 ($\pm^{1.5}_{1.6}$) & 6.0 & 1.6 & $<81$ & $\heartsuit$\\
SHADES J021738-050337 & SXDF850.4 & \(02^{\mathrm{h}}17^{\mathrm{m}}38{\mathrm{\fs}}62\) & \(-05^{\circ}03'37{\farcs}5\) & 4.45 ($\pm^{1.7}_{2.0}$) & 3.9 & 1.6 & $<73$ & $\heartsuit$\\
SHADES J021802-050032 & SXDF850.5 & \(02^{\mathrm{h}}18^{\mathrm{m}}02{\mathrm{\fs}}88\) & \(-05^{\circ}00'32{\farcs}8\) & 8.45 ($\pm^{1.7}_{1.9}$) & 5.4 & 1.8 & $<44\,\diamond$ & $\heartsuit\aleph$\\
SHADES J021729-050326 & SXDF850.6 & \(02^{\mathrm{h}}17^{\mathrm{m}}29{\mathrm{\fs}}77\) & \(-05^{\circ}03'26{\farcs}8\) & 8.15 ($\pm^{2.2}_{2.2}$) & 4.7 & 2.1 & $<81$ & $\dagger\chi_\mathrm{s}$\\
SHADES J021738-050523 & SXDF850.7 & \(02^{\mathrm{h}}17^{\mathrm{m}}38{\mathrm{\fs}}92\) & \(-05^{\circ}05'23{\farcs}7\) & 7.15 ($\pm^{1.5}_{1.6}$) & 5.2 & 1.6 & $<61$ & $\heartsuit\chi_\mathrm{s}$\\
SHADES J021744-045554 & SXDF850.8 & \(02^{\mathrm{h}}17^{\mathrm{m}}44{\mathrm{\fs}}43\) & \(-04^{\circ}55'54{\farcs}7\) & 6.05 ($\pm^{1.8}_{1.9}$) & 4.4 & 1.7 & $<45$ & $\heartsuit$\\
SHADES J021756-045806 & SXDF850.9 & \(02^{\mathrm{h}}17^{\mathrm{m}}56{\mathrm{\fs}}42\) & \(-04^{\circ}58'06{\farcs}7\) & 6.45 ($\pm^{2.0}_{2.1}$) & 4.3 & 1.9 & $<43$ & $\dagger\ast\aleph$\\
SHADES J021825-045557 & SXDF850.10 & \(02^{\mathrm{h}}18^{\mathrm{m}}25{\mathrm{\fs}}25\) & \(-04^{\circ}55'57{\farcs}2\) & 7.75 ($\pm^{2.6}_{3.1}$) & 4.2 & 2.4 & $<134$ & $\heartsuit\chi_\mathrm{t}$\\
SHADES J021725-045937 & SXDF850.11 & \(02^{\mathrm{h}}17^{\mathrm{m}}25{\mathrm{\fs}}12\) & \(-04^{\circ}59'37{\farcs}4\) & 4.55 ($\pm^{1.9}_{2.2}$) & 3.8 & 1.7 & $<79$ & $\dagger\ast$\\
SHADES J021759-050503 & SXDF850.12 & \(02^{\mathrm{h}}17^{\mathrm{m}}59{\mathrm{\fs}}37\) & \(-05^{\circ}05'03{\farcs}7\) & 5.75 ($\pm^{1.7}_{1.8}$) & 4.3 & 1.7 & $<115$ & $\heartsuit$\\
SHADES J021819-050244 & SXDF850.14 & \(02^{\mathrm{h}}18^{\mathrm{m}}19{\mathrm{\fs}}26\) & \(-05^{\circ}02'44{\farcs}2\) & 4.85 ($\pm^{1.9}_{2.1}$) & 3.9 & 1.7 & $<121$ & $\dagger\ast$\\
SHADES J021815-045405 & SXDF850.15 & \(02^{\mathrm{h}}18^{\mathrm{m}}15{\mathrm{\fs}}70\) & \(-04^{\circ}54'05{\farcs}2\) & 6.25 ($\pm^{1.6}_{1.6}$) & 4.8 & 1.6 & $<42$ & $\heartsuit$\\
SHADES J021813-045741 & SXDF850.16 & \(02^{\mathrm{h}}18^{\mathrm{m}}13{\mathrm{\fs}}89\) & \(-04^{\circ}57'41{\farcs}7\) & 4.85 ($\pm^{1.7}_{1.8}$) & 4.1 & 1.5 & $<70$ & $\heartsuit$\\
SHADES J021754-045302 & SXDF850.17 & \(02^{\mathrm{h}}17^{\mathrm{m}}54{\mathrm{\fs}}98\) & \(-04^{\circ}53'02{\farcs}8\) & 7.65 ($\pm^{1.7}_{1.7}$) & 5.2 & 1.7 & $<71$ & $\dagger\ast$\\
SHADES J021757-050029 & SXDF850.18 & \(02^{\mathrm{h}}17^{\mathrm{m}}57{\mathrm{\fs}}79\) & \(-05^{\circ}00'29{\farcs}8\) & 6.45 ($\pm^{2.0}_{2.2}$) & 4.3 & 1.9 & $<54$ & $\heartsuit\aleph$\\
SHADES J021828-045839 & SXDF850.19 & \(02^{\mathrm{h}}18^{\mathrm{m}}28{\mathrm{\fs}}15\) & \(-04^{\circ}58'39{\farcs}2\) & 4.35 ($\pm^{1.8}_{2.1}$) & 3.8 & 1.6 & $<54$ & $\heartsuit\aleph\chi_\mathrm{s}$\\
SHADES J021744-050216 & SXDF850.20 & \(02^{\mathrm{h}}17^{\mathrm{m}}44{\mathrm{\fs}}18\) & \(-05^{\circ}02'16{\farcs}0\) & 4.45 ($\pm^{2.0}_{2.2}$) & 3.8 & 1.7 & $<82$ & $\heartsuit$\\
SHADES J021742-050427 & SXDF850.21 & \(02^{\mathrm{h}}17^{\mathrm{m}}42{\mathrm{\fs}}80\) & \(-05^{\circ}04'27{\farcs}7\) & 5.25 ($\pm^{2.0}_{2.2}$) & 4.0 & 1.8 & $<51$ & $\heartsuit\chi_\mathrm{s}$\\
SHADES J021800-050741 & SXDF850.22 & \(02^{\mathrm{h}}18^{\mathrm{m}}00{\mathrm{\fs}}38\) & \(-05^{\circ}07'41{\farcs}5\) & 6.25 ($\pm^{2.3}_{2.6}$) & 4.1 & 2.1 & $<172$ & $\heartsuit\aleph$\\
SHADES J021742-050545 & SXDF850.23 & \(02^{\mathrm{h}}17^{\mathrm{m}}42{\mathrm{\fs}}53\) & \(-05^{\circ}05'45{\farcs}5\) & 5.25 ($\pm^{1.7}_{2.0}$) & 4.1 & 1.6 & $<59$ & $\heartsuit$\\
SHADES J021734-050437 & SXDF850.24 & \(02^{\mathrm{h}}17^{\mathrm{m}}34{\mathrm{\fs}}58\) & \(-05^{\circ}04'37{\farcs}7\) & 5.15 ($\pm^{2.0}_{2.3}$) & 3.9 & 1.8 & $<69$ & $\heartsuit$\\
SHADES J021812-050555 & SXDF850.25 & \(02^{\mathrm{h}}18^{\mathrm{m}}12{\mathrm{\fs}}12\) & \(-05^{\circ}05'55{\farcs}7\) & 4.05 ($\pm^{2.1}_{2.5}$) & 3.6 & 1.8 & $<59$ & $\dagger\ast$\\
SHADES J021807-050148 & SXDF850.27 & \(02^{\mathrm{h}}18^{\mathrm{m}}07{\mathrm{\fs}}86\) & \(-05^{\circ}01'48{\farcs}5\) & 5.65 ($\pm^{2.0}_{2.3}$) & 4.1 & 1.8 & $<34$ & $\heartsuit\aleph\chi_\mathrm{s}$\\
SHADES J021807-045915 & SXDF850.28 & \(02^{\mathrm{h}}18^{\mathrm{m}}07{\mathrm{\fs}}04\) & \(-04^{\circ}59'15{\farcs}5\) & 4.85 ($\pm^{2.2}_{2.7}$) & 3.8 & 1.9 & $<75$ & $\dagger\aleph\chi_\mathrm{t}$\\
SHADES J021816-045511 & SXDF850.29 & \(02^{\mathrm{h}}18^{\mathrm{m}}16{\mathrm{\fs}}47\) & \(-04^{\circ}55'11{\farcs}8\) & 5.35 ($\pm^{1.8}_{1.9}$) & 4.1 & 1.7 & $<135$ & $\dagger\ast$\\
SHADES J021740-050116 & SXDF850.30 & \(02^{\mathrm{h}}17^{\mathrm{m}}40{\mathrm{\fs}}31\) & \(-05^{\circ}01'16{\farcs}2\) & 5.75 ($\pm^{2.0}_{2.2}$) & 4.1 & 1.8 & $<85$ & $\heartsuit$\\
SHADES J021736-045557 & SXDF850.31 & \(02^{\mathrm{h}}17^{\mathrm{m}}36{\mathrm{\fs}}30\) & \(-04^{\circ}55'57{\farcs}5\) & 6.05 ($\pm^{1.7}_{2.0}$) & 4.4 & 1.7 & $<30$ & $\heartsuit$\\
SHADES J021722-050038 & SXDF850.32 & \(02^{\mathrm{h}}17^{\mathrm{m}}22{\mathrm{\fs}}89\) & \(-05^{\circ}00'38{\farcs}1\) & 6.05 ($\pm^{2.4}_{3.0}$) & 4.0 & 2.1 & $<101$ & $\dagger\ast$\\
SHADES J021800-045311 & SXDF850.35 & \(02^{\mathrm{h}}18^{\mathrm{m}}00{\mathrm{\fs}}89\) & \(-04^{\circ}53'11{\farcs}2\) & 5.35 ($\pm^{1.8}_{2.1}$) & 4.1 & 1.7 & $<62$ & $\heartsuit$\\
SHADES J021832-045947 & SXDF850.36 & \(02^{\mathrm{h}}18^{\mathrm{m}}32{\mathrm{\fs}}27\) & \(-04^{\circ}59'47{\farcs}2\) & 5.45 ($\pm^{1.8}_{1.9}$) & 4.2 & 1.7 & $<76$ & $\dagger\ast$\\
SHADES J021724-045839 & SXDF850.37 & \(02^{\mathrm{h}}17^{\mathrm{m}}24{\mathrm{\fs}}45\) & \(-04^{\circ}58'39{\farcs}9\) & 4.55 ($\pm^{2.2}_{2.6}$) & 3.7 & 1.8 & $<63$ & $\heartsuit\aleph$\\
SHADES J021825-045714 & SXDF850.38 & \(02^{\mathrm{h}}18^{\mathrm{m}}25{\mathrm{\fs}}43\) & \(-04^{\circ}57'14{\farcs}7\) & 3.85 ($\pm^{2.3}_{2.7}$) & 3.5 & 1.8 & $<76$ & $\heartsuit$\\
SHADES J021750-045540 & SXDF850.39 & \(02^{\mathrm{h}}17^{\mathrm{m}}50{\mathrm{\fs}}60\) & \(-04^{\circ}55'40{\farcs}2\) & 4.05 ($\pm^{1.7}_{2.1}$) & 3.7 & 1.6 & $<61$ & $\dagger\ast$\\
SHADES J021729-050059 & SXDF850.40 & \(02^{\mathrm{h}}17^{\mathrm{m}}29{\mathrm{\fs}}67\) & \(-05^{\circ}00'59{\farcs}2\) & 3.65 ($\pm^{1.5}_{1.6}$) & 3.8 & 1.3 & $<40$ & $\heartsuit$\\
SHADES J021829-050540 & SXDF850.45 & \(02^{\mathrm{h}}18^{\mathrm{m}}29{\mathrm{\fs}}33\) & \(-05^{\circ}05'40{\farcs}7\) & 21.95 ($\pm^{6.2}_{6.8}$) & 4.9 & 5.6 & $<186$ & $\dagger\ast\chi_\mathrm{t}$\\
SHADES J021733-045857 & SXDF850.47 & \(02^{\mathrm{h}}17^{\mathrm{m}}33{\mathrm{\fs}}89\) & \(-04^{\circ}58'57{\farcs}7\) & 3.05 ($\pm^{1.6}_{1.9}$) & 3.4 & 1.4 & $<54$ & $\heartsuit\chi_\mathrm{t}$\\
SHADES J021724-045717 & SXDF850.48 & \(02^{\mathrm{h}}17^{\mathrm{m}}24{\mathrm{\fs}}62\) & \(-04^{\circ}57'17{\farcs}7\) & 7.65 ($\pm^{2.5}_{2.9}$) & 4.3 & 2.3 & $<125$ & $\dagger\ast$\\
SHADES J021820-045648 & SXDF850.49 & \(02^{\mathrm{h}}18^{\mathrm{m}}20{\mathrm{\fs}}26\) & \(-04^{\circ}56'48{\farcs}5\) & 3.35 ($\pm^{2.0}_{2.2}$) & 3.4 & 1.6 & $<75$ & $\ddagger\ast\ast$\\
SHADES J021802-045645 & SXDF850.50 & \(02^{\mathrm{h}}18^{\mathrm{m}}02{\mathrm{\fs}}86\) & \(-04^{\circ}56'45{\farcs}5\) & 5.35 ($\pm^{2.0}_{2.5}$) & 3.9 & 1.9 & $<74$ & $\heartsuit\aleph$\\
SHADES J021804-050453 & SXDF850.52 & \(02^{\mathrm{h}}18^{\mathrm{m}}04{\mathrm{\fs}}90\) & \(-05^{\circ}04'53{\farcs}7\) & 3.25 ($\pm^{1.8}_{2.1}$) & 3.4 & 1.5 & $<84$ & $\ddagger\ast\ast$\\
SHADES J021752-050446 & SXDF850.55 & \(02^{\mathrm{h}}17^{\mathrm{m}}52{\mathrm{\fs}}19\) & \(-05^{\circ}04'46{\farcs}5\) & 3.95 ($\pm^{2.2}_{2.7}$) & 3.5 & 1.8 & $<80$ & $\ddagger\ast$\\
SHADES J021750-050631 & SXDF850.56 & \(02^{\mathrm{h}}17^{\mathrm{m}}50{\mathrm{\fs}}68\) & \(-05^{\circ}06'31{\farcs}8\) & 3.65 ($\pm^{2.2}_{2.5}$) & 3.5 & 1.8 & $<154$ & $\dagger\aleph$\\
SHADES J021745-045750 & SXDF850.63 & \(02^{\mathrm{h}}17^{\mathrm{m}}45{\mathrm{\fs}}80\) & \(-04^{\circ}57'50{\farcs}5\) & 4.15 ($\pm^{1.7}_{2.1}$) & 3.7 & 1.6 & $<29$ & $\heartsuit\aleph$\\
SHADES J021807-050403 & SXDF850.65 & \(02^{\mathrm{h}}18^{\mathrm{m}}07{\mathrm{\fs}}94\) & \(-05^{\circ}04'03{\farcs}2\) & 4.35 ($\pm^{1.9}_{2.3}$) & 3.7 & 1.7 & $<56$ & $\dagger\ast$\\
SHADES J021751-050250 & SXDF850.69 & \(02^{\mathrm{h}}17^{\mathrm{m}}51{\mathrm{\fs}}40\) & \(-05^{\circ}02'50{\farcs}8\) & 3.65 ($\pm^{2.1}_{2.4}$) & 3.5 & 1.7 & $<77$ & $\ddagger\ast\aleph$\\
SHADES J021811-050247 & SXDF850.70 & \(02^{\mathrm{h}}18^{\mathrm{m}}11{\mathrm{\fs}}20\) & \(-05^{\circ}02'47{\farcs}2\) & 4.05 ($\pm^{1.9}_{2.3}$) & 3.6 & 1.7 & $<60$ & $\dagger$\\
SHADES J021821-045903 & SXDF850.71 & \(02^{\mathrm{h}}18^{\mathrm{m}}21{\mathrm{\fs}}24\) & \(-04^{\circ}59'03{\farcs}2\) & 4.15 ($\pm^{1.9}_{2.4}$) & 3.7 & 1.7 & $<54$ & $\heartsuit$\\
SHADES J021758-045428 & SXDF850.74 & \(02^{\mathrm{h}}17^{\mathrm{m}}58{\mathrm{\fs}}73\) & \(-04^{\circ}54'28{\farcs}8\) & 3.35 ($\pm^{1.8}_{2.1}$) & 3.5 & 1.5 & $<61$ & $\ddagger\ast$\\
SHADES J021755-050621 & SXDF850.76 & \(02^{\mathrm{h}}17^{\mathrm{m}}55{\mathrm{\fs}}78\) & \(-05^{\circ}06'21{\farcs}8\) & 4.45 ($\pm^{2.0}_{2.4}$) & 3.7 & 1.7 & $<124$ & $\dagger\aleph$\\
SHADES J021736-050432 & SXDF850.77 & \(02^{\mathrm{h}}17^{\mathrm{m}}36{\mathrm{\fs}}43\) & \(-05^{\circ}04'32{\farcs}2\) & 3.05 ($\pm^{2.0}_{2.1}$) & 3.3 & 1.6 & $<50$ & $\dagger$\\
SHADES J021817-050404 & SXDF850.86 & \(02^{\mathrm{h}}18^{\mathrm{m}}17{\mathrm{\fs}}18\) & \(-05^{\circ}04'04{\farcs}7\) & 3.65 ($\pm^{1.9}_{2.2}$) & 3.5 & 1.6 & $<45$ & $\ddagger\chi_\mathrm{t}$\\
SHADES J021800-050448 & SXDF850.88 & \(02^{\mathrm{h}}18^{\mathrm{m}}00{\mathrm{\fs}}99\) & \(-05^{\circ}04'48{\farcs}5\) & 4.55 ($\pm^{2.1}_{2.5}$) & 3.7 & 1.8 & $<99$ & $\dagger$\\
SHADES J021734-045723 & SXDF850.91 & \(02^{\mathrm{h}}17^{\mathrm{m}}34{\mathrm{\fs}}81\) & \(-04^{\circ}57'23{\farcs}9\) & 3.55 ($\pm^{2.1}_{2.5}$) & 3.4 & 1.7 & $<79$ & $\ddagger$\\
SHADES J021733-045813 & SXDF850.93 & \(02^{\mathrm{h}}17^{\mathrm{m}}33{\mathrm{\fs}}08\) & \(-04^{\circ}58'13{\farcs}5\) & 3.15 ($\pm^{2.0}_{2.1}$) & 3.4 & 1.6 & $<70$ & $\ddagger\ast\aleph$\\
SHADES J021740-045817 & SXDF850.94 & \(02^{\mathrm{h}}17^{\mathrm{m}}40{\mathrm{\fs}}08\) & \(-04^{\circ}58'17{\farcs}7\) & 4.15 ($\pm^{1.8}_{2.1}$) & 3.7 & 1.6 & $<49$ & $\ddagger$\\
SHADES J021741-045833 & SXDF850.95 & \(02^{\mathrm{h}}17^{\mathrm{m}}41{\mathrm{\fs}}72\) & \(-04^{\circ}58'33{\farcs}7\) & 3.45 ($\pm^{1.9}_{2.2}$) & 3.5 & 1.6 & $<92$ & $\ddagger\aleph$\\
SHADES J021800-050212 & SXDF850.96 & \(02^{\mathrm{h}}18^{\mathrm{m}}00{\mathrm{\fs}}00\) & \(-05^{\circ}02'12{\farcs}8\) & 4.75 ($\pm^{2.1}_{2.5}$) & 3.8 & 1.8 & $<58$ & $\ddagger\aleph$\\
SHADES J021756-045255 & SXDF850.119 & \(02^{\mathrm{h}}17^{\mathrm{m}}56{\mathrm{\fs}}35\) & \(-04^{\circ}52'55{\farcs}2\) & 4.55 ($\pm^{2.1}_{2.5}$) & 3.7 & 1.8 & $<70$ & $\ddagger$\\
\caption{The $850\,\mathrm{\mu m}$ SHADES catalogue for the LH and SXDF regions.  Uncertainties in physical positions are $3.2\,\mathrm{arcsec}$ in RA and $3.2\,\mathrm{arcsec}$ in Dec. (the same uncertainties have been adopted for all SHADES sources in both fields).  Estimates of the true unbiased median flux density of each source is given, with accompanying error bars representing the 68 per cent confidence bounds of the (non-Gaussian) deboosted flux density distribution (cf.~Section~\ref{deboost} for the Bayesian estimate of deboosted flux density).  We also provide the combined map S/N and noise estimates (i.e.~values prior to deboosting).  We do not claim any $450\,\mathrm{\mu m}$ detections of SHADES sources; $3\,\sigma$ upper limits limits are given in the penultimate column.  Corresponding SCUBA 8-mJy Survey IDs (\citealt{Scott}) and new IDs from the reduction of \citet{SDS} are listed in the final column for reference, along with an indication of how many groups detected each source.  See Appendix~\ref{notes} for detailed notes on some of the sources.\\ \\ \textbf{Legend}:\\ $\heartsuit$ This source was identified by all four groups with a S/N $>3$.\\ $\dagger$ This source was identified by three groups with a S/N $>3$.\\ $\ddagger$ This source was identified by two groups with a S/N $>3$.\\ $\ast$ This source was identified by one additional group with $2.5<\mathrm{S/N}<3$.\\ $\ast\ast$ This source was identified by two additional groups with $2.5<\mathrm{S/N}<3$.\\ $\aleph$ This source is mildly affected by the noise spike (cf.~Appendix~\ref{noisespike}).\\ $\chi_\mathrm{s}$, $\chi_\mathrm{t}$ This source has a relatively poor spatial or temporal $\chi^{2}$ fit.\\ $\diamond$ This source has a hint of flux at $450\,\mathrm{\mu m}$ measured at the SHADES catalogue positions at a significance level $>3\,\sigma$ (see Fig.~\ref{fig:450hist}).}\label{tab:lock}
\end{longtable}
\end{onecolumn}
\end{center}
\begin{twocolumn}
\normalsize

\subsubsection{Comparison with the SCUBA 8-mJy Survey}
The new SHADES catalogue was cross-matched with the SCUBA 8-mJy Survey ${>}\,3.0\,\sigma$ source catalogue \citep{Scott}, which used a subset of our data.  We failed to re-detect 2/12 of the ${>}\,4\,\sigma$ sources (LE850.5 and LE850.9), 5/9 sources with published S/N between 3.5 and 4.0 (LE850.13, LE850.15, LE850.17, LE850.19, and LE850.20), and (not surprisingly) 12/15 sources with S/N between 3.0 and 3.5 (LE850.22--26, LE850.28, LE850.30, LE850.31, and LE850.33--36).  These findings are similar to those given in \citet{Ivison}, \citet{Mortier} and \citet{Ivison05}.  \citet{Ivison} rejected LE850.9,  LE850.10,  LE850.15, and LE850.20 due to the lack of associated radio counterparts, combined with the fact that they were found in noisy regions of the map ($\sigma\,{>}\,3\,\mathrm{mJy}$).  These sources are also rejected in our analysis, except in the case of LE850.10, since this source is re-detected in the SHADES data (LOCK850.60), albeit with a lower S/N than that found in the SCUBA 8-mJy Survey ($\simeq3.4\,\sigma$ as compared with the previous $4.2\,\sigma$ detection).  We find that there is less than a 4 per cent chance of LE850.10 having a true flux density of $\leq 0\,\mathrm{mJy}$ and therefore it survives the deboosting cut.  This source also has a tentative radio identification (Ivison et al. in preparation).  See Table~\ref{tab:lock} for the corresponding new SHADES measurements of the SCUBA 8-mJy Survey sources.  

The SHADES catalogue was also cross-matched with the re-reduction of the SCUBA 8-mJy Survey ${>}\,3.0\,\sigma$ source catalogue \citep{SDS}, which includes some additional data and improvements made to the reduction methods.  In summary, \citet{SDS} re-detected all of the 36 original SCUBA 8-mJy Survey ${>}\,3.0\,\sigma$ sources (though 4 sources originally detected above $3.0\,\sigma$ dropped down to below $3.0\,\sigma$:  LE850.25, LE850.29, LE850.30, and LE850.31), and found 8 new $3.0<\mathrm{S/N}<3.7$ sources.  SHADES re-detects sources SDS.16 ($3.7\,\sigma$) and SDS.20 ($3.4\,\sigma$), but fails to re-detect SDS.25, SDS.32, SDS.33, SDS.36, SDS.38, and SDS.40 (new $<3.4\,\sigma$ sources).  See Table~\ref{tab:lock} for the corresponding SHADES measurements of the sources found in the \citet{SDS} re-reduction of the SCUBA 8-mJy Survey data.  

\subsection{Flux density comparison}\label{compareflux}

We checked for systematic effects between the reduction group flux densities in order to quantify their effect on the adopted SHADES map flux density.  We limit the discussion that follows to a subset of sources that \textit{all} groups find at $\geq2.5\,\sigma$:  54 in the LH; and 58 in the SXDF.  Unlike in the astrometry comparison (see later Section~\ref{compareposition}), we have no `true' flux density to compare the mean flux densities against, so only an inter-group comparison can be performed.

Flux density comparison scatter plots were produced, such as Fig.~\ref{fig:flux_flux_fig} for each pair of groups.  We noticed immediately that one reduction showed a systematically lower flux density than the other groups that appeared to depend on the pointing strategy of the observations.  After finding and correcting this error, the scatter in source flux densities between all groups appeared small on average, with no systematic offset apparent in any one group compared to another (except at the high flux density end, where the photometric errors are also very large).  We feel it is important to mention explicitly those very few occasions when the results of our comparisons are used to guide the details of any reduction, so that a critical reader will understand how close to a blind analysis the SHADES reductions are.

One might expect the choice of sky opacity correction factors or FCFs (i.e.~using monthly FCFs versus nightly measurements) to come into play at about the 2 per cent level for such low S/N data taken in more or less uniform weather conditions.  The RMS scatter between flux densities reported for LH sources by Reductions B and D is $0.9\,\mathrm{mJy}$ (see Fig.~\ref{fig:flux_flux_fig}) and is similar between the other groups.  Similar results were found for the SXDF.  We find that the photometry errors we might have introduced through differences in judgment are 1/3 as large as the total uncertainty in flux density, which is noise-based.  We therefore claim that the gain differences are small (i.e.~less than 5 per cent) and therefore unimportant for these low S/N data (they become important for $\sim 10\,\sigma$ sources, of which we do not find any in this survey).  This demonstrates that we are extracting flux densities well.  Using monthly averaged FCFs (Reduction D) versus using nightly measurements (Reductions A, B and C)  appears to give indistinguishable answers, and therefore the systematic error introduced from the former technique and the instantaneous measurement uncertainty in the latter technique are {\em both} insignificant with respect to the photometric errors in our survey.

\begin{figure*}
\includegraphics[width=5.0in,angle=0]{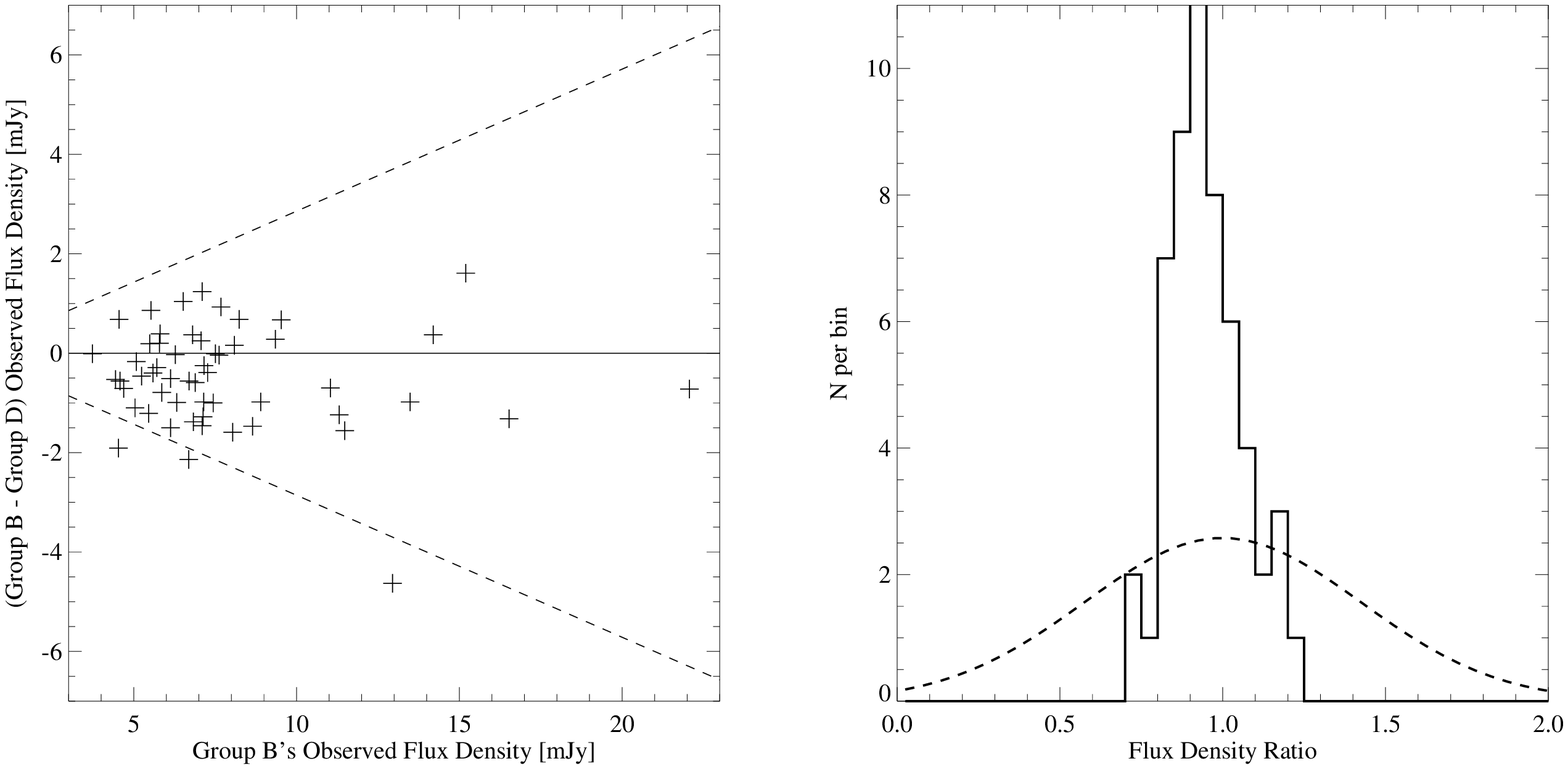}
\caption{The difference of the $850\,\mathrm{\mu m}$ flux density
as measured by Reductions B and D is plotted against the flux density measured 
by Reduction B for the LH sources that were found in all 
four reductions (left panel), and a histogram of flux density ratios of 
Reductions B and D (i.e.~B/D) for the LH (right panel).   
In the left panel the horizontal solid line indicates 
a ratio of unity, while 
the dashed lines show the $1\,\sigma$ scatter 
expected if each of the two 
reductions had {\it independent} photometry errors corresponding 
to a S/N of 3.5.  The scatter between these two 
reductions is substantially smaller than photometry errors, indicating 
that systematic errors associated with data reduction choices do 
not substantially influence our final flux density measurement values and 
uncertainties.  Reductions B and D are selected for comparison here 
because their calibration procedures 
differ most among the four reductions; any other pair of reductions is likely to demonstrate at least this level of agreement.   In the right panel 
the smooth (dashed) curve is the flux density ratio distribution 
expected for two \textit{independent} measurements with the 
approximate limiting SHADES S/N ($3.5\sigma$).  The narrowness of 
the histogram compared to the relative width of the Gaussian demonstrates 
that systematic errors in calibration are small compared to 
photometric uncertainties for all sources in the SHADES 
catalogue.  The two sets of flux density values are consistent 
with each other, while the best fit mean flux density difference 
is 4 per cent.  Since the uncertainty of the 
{\it mean of all 60 sources} in the SHADES LH catalogue 
is also 4 per cent, systematic flux density errors associated 
with differences in data reduction strategy are completely unimportant 
for individual sources. }
\label{fig:flux_flux_fig}
\end{figure*}

\subsection{Astrometry comparison}\label{compareposition}

Intercomparing the positions found by the four independent reductions allows us to check for systematic errors and estimate the uncertainty of our astrometry.  When comparing flux densities, differences between reductions can be measured but the `true' $850\,\mathrm{\mu m}$ flux density is not known, so systematic errors may be difficult to isolate.  The situation is much simpler with positions.  When sources have clearly identified radio counterparts the precision of the positions determined from the radio data is much higher than that available from the $850\,\mathrm{\mu m}$ data.  Effectively, one knows the `true underlying position' for these sources and sensitive tests for systematic errors are possible.  In this section we analyse the positions of a subset of the sources which have clear and compact radio counterparts determined by Ivison et al. (in preparation) and which have been identified at ${>}\,2.5\,\sigma$ in all four reductions.   These criteria, designed to facilitate clear comparison of the four reductions,  yield 17 and 24 sources in the LH and SXDF, respectively.   The full analysis of the alignment of SHADES sources with the corresponding radio data is made in Ivison et al. (in preparation).

Upon initial comparison it was clear that the positions of sources in the LH determined by one reduction  were displaced to positive RA by just over $2\,\mathrm{arcsec}$ compared to positions determined by any of the other reductions or compared to the radio positions.  This systematic effect was traced to errors in the use of the {\tt hastrom}  routine in {\tt idl-astrolib} and has been corrected.  Other than the correction of this small error, nothing has been adjusted to bring the reductions into agreement with each other or with the radio-determined positions.

The positions determined by one reduction (arbitrarily, Reduction D) in comparison to the mean $850\,\mathrm{\mu m}$ position for all the sources in the SXDF are plotted in the left hand panel of Fig.~\ref{fig:shades_posn_fig}.  The sources with compact radio IDs from Ivison et al. (in preparation) are shown as plusses, while other sources in the SXDF are shown as diamonds.  Both sub-samples have similar means and distributions, so we conclude that the analysis of positions for the restricted sub-sample provides a description of astrometry errors which is valid for the full list.   Table~\ref{tab:shades_posn} lists the RMS displacements of source positions determined by each reduction from the unweighted mean position determined by the four reductions, separately for the two SHADES fields.  Pixelisation inevitably adds $d/\sqrt{12}$  where $d$ is the pixel size, in quadrature to the RMS astrometry errors\footnote{The location of a detected source is uniform between $-1/2$ and 1/2 of the centre of a pixel of size 1.  The variance of the error made by quoting the pixel centre is the mean-square error $=\int^{1/2}_{-1/2}x^{2}dx=1/12$.}.  This is $0.9\,\mathrm{arcsec}$ for $3\,\mathrm{arcsec}$ pixels and has not been subtracted from the data in Table~\ref{tab:shades_posn}. Even so, Reductions C and D, with $3\,\mathrm{arcsec}$ pixels, have RMS displacements from the $850\,\mathrm{\mu m}$ mean which are, if anything, lower than the displacements of the reductions using smaller pixels.  It is perhaps not surprising that using small pixels (smaller than $\simeq\frac{\mathrm{FWHM}}{5}$ at least) in reconstruction of the data does not appear to add any astrometric precision (cf.~\citealt{condon_precise}).

The right hand panel of Fig.~\ref{fig:shades_posn_fig} shows the deviation of the mean offset of the position determined by the four SHADES reductions relative to the presumably correct radio-determined positions from Ivison et al. (in preparation).  Notice that the scatter is much larger than the scatter in the left hand panel.  We conclude that the four reductions are accurately extracting the location of the peak in the submillimetre flux density from the data, and that this peak is displaced from the true source location due to noise in the $850\,\mathrm{\mu m}$ data.  In a careful comparison, Ivison et al. (in preparation) confirm that the offsets scale as expected with the submillimetre beam size and S/N.  We quote a positional uncertainty in RA and Dec. offsets of $3.2\,\mathrm{arcsec}$ and find no evidence for an overall mean astrometric error.  It is interesting to note that the unweighted mean position from the four reductions is a better predictor of radio position than is obtained from any single reduction.  Taken together the panels in Figure~\ref{fig:shades_posn_fig} indicate that the reductions do a good job of determining the positions implied by the submillimetre data, but that those positions have a few arcsec scatter with respect to the true underlying source positions.

\begin{table*}
\caption{Astrometry Precision: The submillimetre positions of a subset of SHADES sources are compared to each other and to the radio positions found  in Ivison et al. (in preparation). 
 The columns marked `Std. dev. w.r.t. $850\,\mathrm{\mu m}$' 
show the standard deviations in arcseconds of each
reduction with respect to the unweighted mean of all four reductions.
The column marked `Std. dev. w.r.t. Radio' shows the standard deviations in arcseconds with respect to positions determined by Ivison et al. (in preparation).  Columns marked `Net' list the  quadrature sums of the RA and Dec values across both fields.  The positions of the radio sources  are known to a higher accuracy than is possible using the SCUBA data alone, assuming that all radio counterpart  IDs are correct.  The
pattern of variances in the table indicates that the peak in the submillimetre 
data is displaced from the true source position due to noise, but that all
reductions find consistently the same displaced peak location because the submillimetre 
noise is common to all four reductions.  The unweighted mean position of the four reductions is a better predictor of radio position than any single reduction.}
\vskip .2in
\begin{tabular}{lcccccc}
\hline 
\multicolumn{1}{|c|}{} & \multicolumn{5}{c|}{Std. dev. w.r.t. $850\,\mathrm{\mu m}$} & \multicolumn{1}{|c|}{Std. dev. w.r.t. Radio} \\ 

\hline

\multicolumn{1}{|c|}{SHADES }  & \multicolumn{2}{c|}{Lockman} &
\multicolumn{2}{c|}{SXDF} & \multicolumn{1}{c|}{Net 850} & \multicolumn{1}{c|}{Net Radio} \\ 

\multicolumn{1}{|c|}{Reduction}  & \multicolumn{1}{c}{RA} &
\multicolumn{1}{c|}{Dec.} & \multicolumn{1}{c}{RA} &
\multicolumn{1}{c|}{Dec.} & \multicolumn{1}{c|}{RMS} & \multicolumn{1}{c|}{RMS} \\ 
\hline\hline
\multicolumn{1}{|c|} A & 2.33 & 1.94 & 1.80 & \multicolumn{1}{c|}{2.91} &\multicolumn{1}{c|} {2.29} & \multicolumn{1}{c|} {3.46}\\
\multicolumn{1}{|c|} B & 1.13 & 1.20 & 1.40 & \multicolumn{1}{c|}{2.23} &\multicolumn{1}{c|}{1.55} & \multicolumn{1}{c|} {3.06} \\
\multicolumn{1}{|c|} C & 1.26 & 1.14 & 1.49 & \multicolumn{1}{c|}{1.71} & \multicolumn{1}{c|}{1.42} & \multicolumn{1}{c|} {2.77} \\
\multicolumn{1}{|c|} D & 0.93 & 1.43 & 1.38 & \multicolumn{1}{c|}{1.36} & \multicolumn{1}{c|}{1.29} & \multicolumn{1}{c|} {3.01} \\
\hline

\label{tab:shades_posn}
\end{tabular}
\end{table*}

\begin{figure*}
\psfig{file=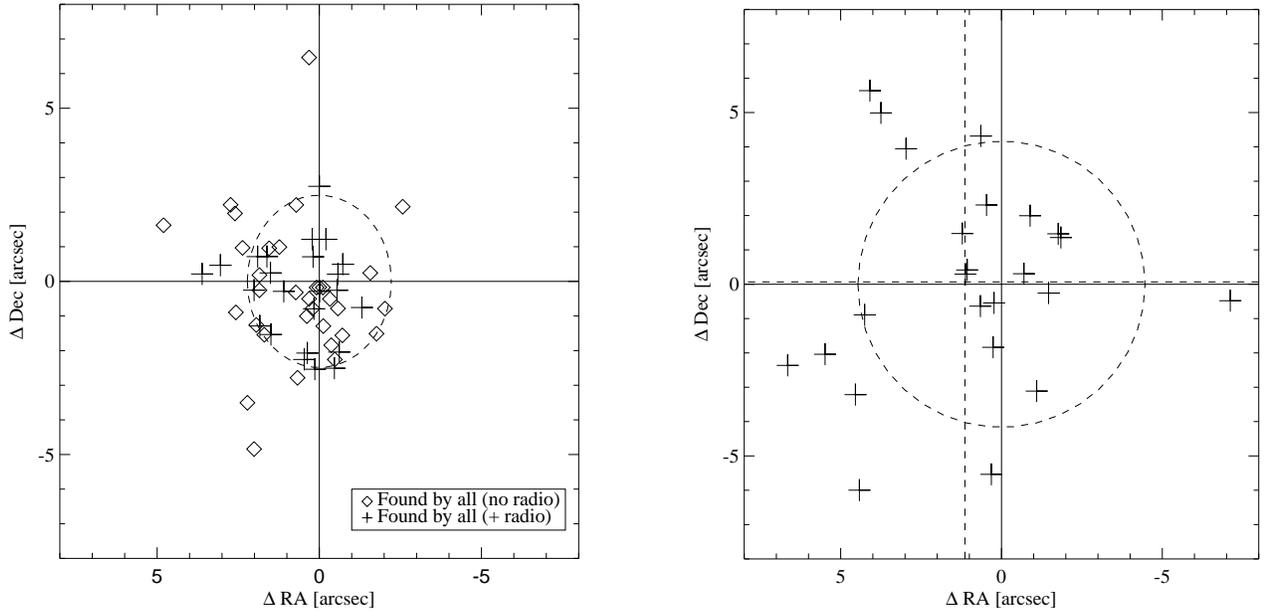,width=1.\textwidth}
\caption{The left panel shows the offset of the source positions inferred by
Reduction D from the {\it mean} of the positions found by all four
 reductions for all sources in the SXDF which are found in
every reduction.  The plusses show those points which are also included 
in the comparison with radio positions (see right panel and Table~\ref{tab:shades_posn}). The diamonds are the remaining sources and they do not appear to be distributed very differently than those we have compared to the radio counterparts.  The dashed ellipse indicates $\sqrt{2}$ times the one-dimensional variances in RA and Dec.; it contains $\simeq 2/3$ of the sources, as expected.  The right panel shows the offset in position of the \textit{mean} of
the four SHADES reductions relative to the radio data of Ivison et al. (in preparation) for a subset of sources which have clear well defined radio positions.  The radio data
are collected at substantially higher angular resolution and S/N, so the
scatter here is presumed to be dominated by scatter in the
submillimetre data.  As in the left panel, the semi-major and semi-minor axes of the dashed ellipse are $\sqrt{2}$ times the positional variances in RA and Dec.  This error ellipse should contain $\simeq 2/3$ of the sources.  The vertical and horizontal dashed 
lines show the mean displacements of SHADES locations from the radio;
these differences are \textit{not} statistically significant.  
Taken together these panels indicate that the reductions do a good job of determining the positions implied by the submillimetre data, but that those positions have a few arcsec scatter with respect to the true underlying source positions.  See Ivison et al. (in preparation) for a more detailed comparison.
}
\label{fig:shades_posn_fig}
\end{figure*}

\subsection{Source deblending}\label{deblend}

The source extraction methods of Reduction C (D) are insensitive to finding sources closer than $10\,(18)\,\mathrm{arcsec}$.  There is thus a potential problem with source blending, since the detection of very near neighbours would be evidence for clustering of SMGs.

Motivated by the discovery of two additional sources near two bright SMGs detected in the GOODS-N SCUBA map by \citet{Pope}, double Gaussians with variable positions and amplitudes were fitted to the 5 brightest sources in each field found by Reduction D.  However, no convincing evidence was found for favouring two sources over one.  The $\chi^{2}$ was typically lower when fitting double Gaussians, but the second source was never bright enough to be classified as a detection under our criteria.  We note here (and in Appendix~\ref{notes}) that SXDF850.1 appears quite extended in the NS direction, although the $\chi^{2}$ is not lower for two sources than for one.

\section{Results:  Differential Source Counts}\label{complete}

An important quantity which can be derived from the data is an estimate of the number of sources as a function of flux density.  SHADES is a single, uniform survey which has approximately doubled the total area of \textit{all} SCUBA-observed blank-field surveys.  While the dynamic range in flux densities is not as broad as is available from a compilation of data which includes the deepest surveys (particularly those associated with foreground gravitational lenses),  the results here are the most robust obtained so far in the range of flux densities where SHADES is sensitive.  In this section we provide reliable estimates of the \textit{differential} $850\,\mathrm{\mu m}$ number counts for the first time.  Differential counts offer an advantage compared to integral source counts, since each estimate of the number of sources in a flux density bin does not depend on the counts at brighter flux densities and thus will be much less correlated.  This makes fitting the results to models of source counts more straightforward.  Nevertheless, in the following section, \ref{numcts}, we integrate the differential counts to estimate the cumulative source distribution for comparison to previous data.

Although all of the sources in the SHADES catalogue are at least $3\,\sigma$ in the maps, and therefore have raw measured fluxes which are typically above 6 mJy, the well known steep submillimetre source flux density distribution implies that a source which is detected at 6--8\,mJy is as likely to be a 3--4\,mJy source accidentally observed with a positive noise fluctuation as it is to be a genuine 7\,mJy source (i.e.~Malmquist-like bias; cf. Section~\ref{deboost}).   The number of sources we have observed depends on the number density of sources down to quite faint flux densities, well below our nominal 6\,mJy limit.  Fits of differential source counts to our data will therefore provide constraints on the number of sources per square degree starting at about 3\,mJy.

To obtain estimates of differential source counts one must estimate completeness, flux density bias, survey area, spurious detection rates, and Poisson counting errors.   We have developed two analysis approaches to perform  these tasks.  One of these is a `direct estimate', which works with a list of sources and their deboosted fluxes  and sums the associated probability densities to obtain the parent source density spectrum.  The second method, `parametric fitting', self-consistently estimates the prior source density spectrum, the FDR and the source deboosting.  Direct counts use a fixed informative prior to perform the deboosting. This is a maximal use of existing information from past SCUBA surveys. The parametric approach leaves the prior as a free parameter. This second technique is probably more conservative, and the amount of noise in the answer strongly depends on how the prior is parameterised.

Although the SHADES catalogue is extremely robust in terms of a low expected false detection rate, it has a complicated selection function and is not necessarily optimal for measuring the source counts.  Because it is based on four different analyses the selection criteria are hard to Monte Carlo, so we do not use the SHADES catalogue to derive constraints on $850\,\mathrm{\mu m}$ source counts.  In particular, one would like to use statistical information from sources which may not individually be detected with high significance.  Therefore, source count spectra have been determined independently from the provisional source lists arising from each reduction.  Variations of the direct estimation approach have been applied to data from Reductions B and D, while parametric fitting has been applied to Reduction C.


The fits we present here are all derived from catalogues of sources detected above at least $2.5\,\sigma$.  Thus, although the number count estimates are statistical they are fundamentally different from so-called $P(D)$ analyses in which the distribution of pixel flux densities is fit directly to a source count model.  If we reduced our catalogue to lower and lower thresholds we would effectively recover the $P(D)$ results.  However, this is left to a future paper.

Table~\ref{tab:bias} highlights the key steps and differences in each
reduction's number count estimation.

\begin{table*}

\caption{Methods of accounting for bias.  Here subscript `i' refers to the input or unbiased flux density, subscript `o' refers to observed quantities in thresholded maps, and subscript `d' refers to a true detection.}

\begin{center}
\begin{tabular}{ | p{3.0cm} | p{4.5cm}| p{4.5cm} |p{4.2cm}|}
  \hline
  Step & Reduction B & Reduction C & Reduction D \\ 
  \hline

  Meaning of posterior flux density, $p(S_\mathrm{i}|S_\mathrm{o},\sigma_\mathrm{o})$ &

  Flux density probability for best fit of flux density to the underlying
  `zero-footprint' map distribution. &

  Flux density probability for brightest individual source in a measurement 
  aperture. &

  Total flux density in a measurement aperture. \\
  \hline

  Posterior flux density expression &

  \[ \frac{p(S_{\mathrm{i}})p(S_{\mathrm{o}},\sigma_{\mathrm{o}}|S_{\mathrm{i}})}{p(S_{\mathrm{o}},\sigma_{\mathrm{o}})} \]&

  \[ \frac{p(S_{\mathrm{i}})p_\mathrm{d}(S_{\mathrm{o}}|S_{\mathrm{i}},\sigma_{\mathrm{o}})p_\mathrm{d}(\sigma_{\mathrm{o}}|S_{\mathrm{i}})}
       {p_\mathrm{d}(S_{\mathrm{o}},\sigma_{\mathrm{o}})}\]&
  
  \[ \frac{p(S_{\mathrm{i}})p(S_{\mathrm{o}},\sigma_{\mathrm{o}}|S_{\mathrm{i}})}{p(S_{\mathrm{o}},\sigma_{\mathrm{o}})}\] \\ 
  \hline

  Prior information used &

  Actual S/N map for the given field. & 

  Simulated noisy flux density maps, assuming a form of the number counts,
  and completeness for given field. &
  
  Simulated noiseless flux density map, assuming a form of the number counts,
  and Gaussian photometric errors. \\
  \hline

  Source list selection criteria &

  $S_{\mathrm{o}}/\sigma_{\mathrm{o}}>3.5$ &

  $S_{\mathrm{o}}/\sigma_{\mathrm{o}}>2.5$&

  $S_{\mathrm{o}}/\sigma_{\mathrm{o}}>2.5$, $p(S_{\mathrm{i}}\leq0|S_{\mathrm{o}},\sigma_{\mathrm{o}})<5$ per cent \\
  \hline

  Number counts &

  Place sources in bins at peak posterior probability. &

  Fit prior $p(S_{\mathrm{i}})$ by comparing modelled $p(S_{\mathrm{o}},\sigma_{\mathrm{o}})$ with
  data. &

  Place sources in bins by integrating posterior probability. \\
  \hline

  Completeness &

  Add individual sources to real maps, compare input to output catalogue. &

  Completely simulate maps, compare input to output catalogue. &

  Add individual sources to real map, detect nearest peak and compare 
to input position. \\
  \hline

  Spurious Detections &

  Completely simulate maps, detect nearest peak. &

  Completely simulate maps, compare input to output catalogue.&

  See source list selection criteria. \\
  \hline

  Counts Uncertainties &

  Analytic propagation of errors. &

  Monte Carlo, using realizations of completely simulated data. &

  Monte Carlo, using bootstraps of real data. \\
  \hline
  \hline
	
\end{tabular}
\end{center}
\label{tab:bias}
\end{table*}

\subsection{Direct estimate of the differential source counts}\label{redD}

The direct estimate works with Reduction D's source list and calculates 
the differential counts directly using the posterior flux
density distributions for individual sources in the list 
following \citet{Coppin}.  

The source list used to calculate the number counts
is constructed by identifying all $2.5\,\sigma$ peaks in the map, 
and keeping all of the peaks likely to be real (i.e.~having $<5$ per cent deboosted 
probability of having $S_\mathrm{i}<0$).  For the purposes of measuring the counts the deboosting criterion could be relaxed, but with the added complication of statistically taking 
into account the FDR in the counts (see Section~\ref{redC}).  
For this reason, the same criterion that is used to construct the SHADES catalogue is applied -- the difference being that here only Reduction D's data are used.

An `effective area' is calculated.  It is the area times the completeness, and these are estimated together as a function of intrinsic source flux density, $\Omega(S_\mathrm{i})$, in order to correct the source counts for incompleteness.  Fake sources of known flux density are injected one at a time into the real maps (without worrying if the sources fall entirely within the region that was measured) and then they are extracted using the source extraction method.  This procedure is repeated 2000 times each at flux density levels of 4, 6, 8, 10, 12, 14, 16, 40, 60, and $100\,\mathrm{mJy}$.  A source is considered recovered if it is found within $7.5\,\mathrm{arcsec}$ of its input position \textit{and} survived the flux density deboosting.  $\Omega(S_\mathrm{i})$ is the ratio of the number of sources found to the number put in {\it per square degree\/} (see Fig.~\ref{fig:completeness}).  A smooth best-fitting function of the form $(S^{a})/(b + c\,S^{a})$ is fitted to the data points and is used in correcting the raw source counts.  Here $S$ is the flux density and $a,b,c$ are constants. 

\begin{figure}
\psfig{file=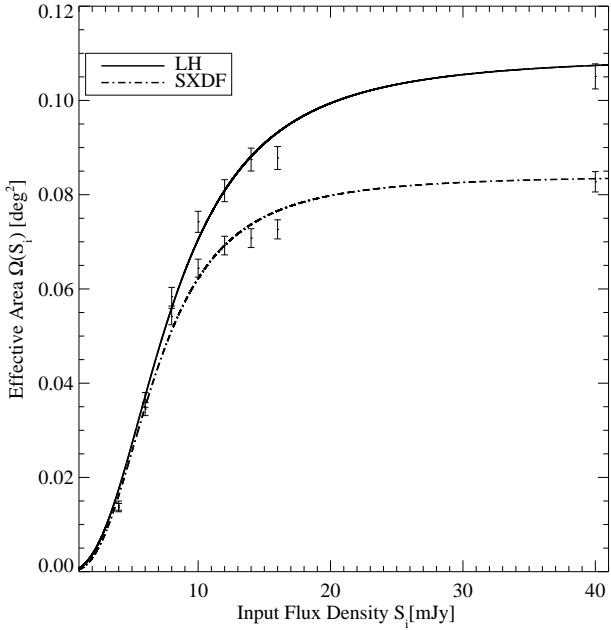,width=0.5\textwidth}
\caption{A fit to $\Omega(S_\mathrm{i})$, the effective area 
(survey area times completeness) of the $850\,\mu$m source 
recovery at each level of input 
flux density for $3\,\sigma$ map detections which have 
$P(S_\mathrm{i}\leq0\,\mathrm{mJy})<5$ per cent, as determined from Monte 
Carlo simulations of individual sources added to the SHADES
 maps of Reduction D (see Section~\ref{redD}).  The curves are fits to the form $(S^{a})/(b + c\,S^{a})$.}
\label{fig:completeness}
\end{figure}

The values of $N(S_\mathrm{i})$ in the bins are estimated using Monte Carlos, which are also used to estimate the errors.  In the past, submillimetre survey groups have placed error bars on 
the counts by simply accounting for the simple $1\,\sigma$ Poisson 
counting errors (the square root of the raw counts in each bin, 
scaled to unit solid angle).  The estimated 
number of spurious and confused sources are sometimes added in quadrature 
to the lower error bar in the corrected number counts 
(e.g.~\citealt{Scott}).  Since the deboosting procedure provides a distribution for each $S_\mathrm{i}$, rather than just a single value, a modified boot-strapping simulation is used to estimate the differential source counts and uncertainties in bins of width $2\,\mathrm{mJy}$.  This simultaneously accounts for the Poisson error, as now described.  First, a number of sources is chosen from a Gaussian distribution centred on the number of sources included in the real source list, $N_{\mathrm{true}}$, with a standard deviation $\sigma =\sqrt{N_{\mathrm{true}}}$ to account for counting errors.  This number of sources is then randomly selected from the actual source list and their probability distributions sampled {\it with  replacement\/} (i.e.~boot-strapping; see Section~6.6 of \citealt{Wall});  once a flux density is determined at random (in proportion to the source's posterior probability distribution), one source per effective area is added into the appropriate flux density bin.  This procedure is repeated 10,000 times, in order to make well-sampled histograms of the count distributions for each bin.  These histograms are used to estimate the mean counts and the frequentist 68 per cent confidence intervals in each flux density bin and are given in Table~\ref{tab:count_num}.  Simultaneously, the linear Pearson covariance matrix of the bootstraps across the flux density bins can be calculated to assess the correlation between bins and this can then be used in model fitting procedures; the covariance matrix is given in Table~\ref{tab:corr_num} for the counts of the combined LH and SXDF fields in $2\,\mathrm{mJy}$-wide bins.  The integral counts are obtained by directly summing over the differential counts and are tabulated in Table~\ref{tab:count_num} and shown in Fig.~\ref{lh_prior_counts_fig}.


\subsubsection{Tests of deboosting on source count recovery}\label{deboosttest2}

A test for bias in the recovery of the number counts was carried out in the following way.  A fake sky populated with the source counts of \citet{Borys2003} was created.  This was observed using the actual SXDF observing scheme and a map was made in the same way as for the real data, while simultaneously injecting random Gaussian noise with an RMS similar to the real map ($\sim 2\,\mathrm{mJy}$).  Sources were then extracted and deboosted according to the prescription described in Section~\ref{deboost}.  The recovered cumulative number counts (scaled by the effective area) were found to be consistent with the input number count realisation in all except the lowest flux density bin, where the uncertainty in the completeness estimates dominates (see Fig.~\ref{fig:fakects}).  We have therefore corrected the differential (integral) counts in each of our lowest flux density bins by a factor of 1.6 (1.33).  The input source counts were also recovered when different source count models were used as input to the simulated skies, while keeping the form of the prior distribution of pixel flux densities fixed.  We found that the correction factor hardly changed when we used different input models.

\begin{figure}
\psfig{file=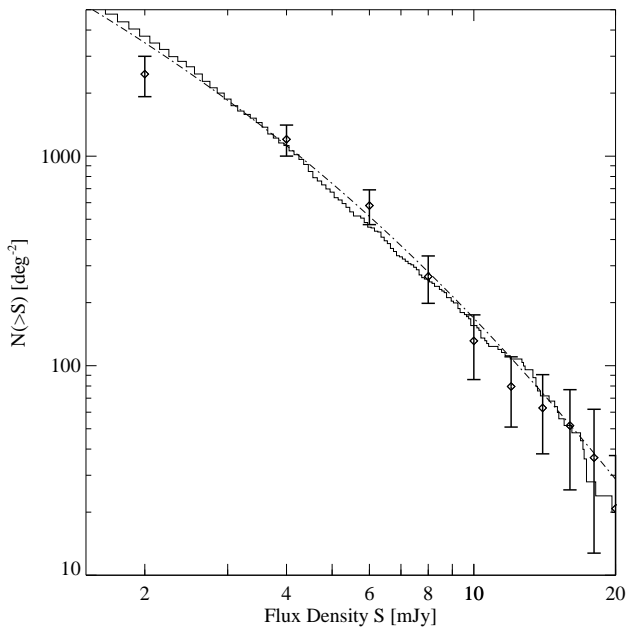,width=0.5\textwidth}
\caption{$850\,\mathrm{\mu m}$ cumulative source counts recovered from a fake sky populated with a known source count model (explicitly for Reduction D).  The diamonds and error bars are the recovered source counts for these simulated data, observed using the same scheme as for the real SXDF data.  The overplotted histogram (solid line) is the actual realisation of the \citet{Borys2003} source count model that was used (dot-dashed line).  This same source count model was used in creating the prior in the Bayesian flux density deboosting method described in Section~\ref{membership}.}
\label{fig:fakects}
\end{figure}

We have also attempted to quantify an overall FDR in a different manner to that described in Section~\ref{deboost}.  A noiseless fake sky was populated with the \citet{Borys2003} source counts, and the region was sampled using the real observing scheme and map reduction steps for each field, while simultaneously adding Gaussian random noise into the timestream to get the same RMS as the real maps.  Sources were extracted in the usual way and then deboosted using the \citet{Coppin} prescription to create a final refined catalogue of flux density deboosted sources.  We found that more than 90 per cent of sources detected in the simulated maps correspond with input sources above the faintest deboosted flux densities of the actual SHADES catalogue.  The interpretation of the remainder is complicated, particularly as one approaches the confusion regime.  We are thus confident that the overall FDR lies below 10 per cent, but determining the precise FDR from simulations is complicated by source confusion, i.e.~interpretation of precisely what \textit{the} source means.

\subsubsection{Another direct estimate of the differential source counts}\label{redB}

Direct estimation of the source count density was carried out independently working from a catalogue derived from Reduction B.  The main differences in the approach are listed below.

\begin{list}{}{}
\item Instead of calculating an effective area, an explicit coverage area, $A$, corresponding to the portion of a given map with observed noise below $\sigma_\mathrm{o}= 10\,\mathrm{mJy}$ is used.   Source candidates are rejected outside of this region.

\item  Deboosted source posterior flux probability density  functions are obtained using the normalised histogram of S/N in the coverage area of a given map,  $H(S_\mathrm{o}/\sigma_\mathrm{o})$ to estimate $P(S_\mathrm{i}|S_\mathrm{o}, \sigma_\mathrm{o})$ in Equation~\ref{bayes} instead of the Gaussian distribution used in \citet{Coppin}.  This is a small difference, since the noise is very well described by a Gaussian distribution.  However, the posterior probability density functions are truncated outside the region $(S_\mathrm{o}/2 \leq S_\mathrm{i} \leq 2S_\mathrm{o})$.  Examples of these deboosted flux density distributions are plotted in Fig.~\ref{fig:deboosted_compare}.

\item Source completeness is estimated by Monte Carlo techniques.  This is defined to be $C(S_\mathrm{i})$, the detection probability for a source of actual flux density $S_\mathrm{i}$ which is located within the coverage area.  Sources are inserted into a given map and are counted as detected if  they are found by the source detection algorithm, if their recovered location is within $7\,\mathrm{arcsec}$ of the insertion position, and if the recovered flux density is within a factor of two of the input flux density.   Simulated sources recovered within $7\,\mathrm{arcsec}$ of a genuine catalogue source in this process are discarded.  $C(S_\mathrm{i})$ is the ratio of recovered sources to simulated sources, calculated in half mJy bins.

\item Source reliability, $R(S_\mathrm{o}/\sigma_\mathrm{o})$, is calculated as a function of recovered S/N.  For each of the six chop-maps, an artificial sky is generated consistent with the source count models of \citet{Scott} and random noise is added, consistent with the actual noise maps.  Source extraction is performed just as it is on the actual survey data.  Recovered sources are identified with input sources if they lie within $7\,\mathrm{arcsec}$ in position and their recovered flux densities are within a factor of two of each other.  $R$ is calculated as the ratio of the number of identified sources at a given S/N to the total number of recovered sources at that S/N.      
\end{list}

A selection catalogue is formed from Reduction B containing all sources with S/N $\geq3.5$ and with $\sigma_\mathrm{o}\leq 10\,\mathrm{mJy}$.  The mode of the posterior flux density probability distribution is found for each source, $\hat S = \mathrm{mode}[P_\mathrm{d}(S_\mathrm{i}| S_\mathrm{o}, \sigma_\mathrm{o})]$, and the contribution of each source in this catalogue to the number of sources per square degree is calculated as
\begin{equation}
\Delta N(\hat S) = {R(\hat S/\sigma_\mathrm{o}) \over A\times C(\hat S)} .
\end{equation}
The total source number density in a given flux density bin is the sum of $\Delta N(\hat S)$ over all members of the catalogue whose mode, $\hat S$, lies in the  flux  bin. The uncertainty is calculated as the quadrature sum of uncertainties estimated for $R$, $A$, and $C$, added to the Poisson uncertainties calculated for the number of members of the catalogue with $\hat S$ in the flux bin.

\subsection{Parametric model fits to estimate the differential source counts}\label{redC}

Number counts are calculated from Reduction C by fitting models to
observed source catalogues.  This technique is similar to the methods
used by \citet{Borys2003} and \citet{Laurent} for the analyses of
SCUBA and BOLOCAM data, respectively. Source catalogues were first
generated for each field by identifying all 2.5\,$\sigma$ peaks in the
maps. The area of the maps analysed are defined as the regions having
a photometric error $\sigma_\mathrm{o} < 5$\,mJy\,beam$^{-1}$.  The
model is developed by first expressing the probability distribution
of the source catalogue $p(S_\mathrm{o},\sigma_\mathrm{o})$, where
$S_\mathrm{o}$ and $\sigma_\mathrm{o}$ are the observed flux densities
and photometric uncertainties respectively, as the sum of the
probabilities of the source being a real detection of an object with
intrinsic flux density $S_\mathrm{i}$,
$p_\mathrm{d}(S_\mathrm{i},S_\mathrm{o},\sigma_\mathrm{o})$, and being
a false detection, $p_\mathrm{f}(S_\mathrm{o},\sigma_\mathrm{o})$,

\begin{equation}
  p(S_\mathrm{o},\sigma_\mathrm{o}) =
  \int p_\mathrm{d}(S_\mathrm{i},S_\mathrm{o},\sigma_\mathrm{o})dS_\mathrm{i}
  + p_\mathrm{f}(S_\mathrm{o},\sigma_\mathrm{o}).
  \label{eq:probdata}
\end{equation}


\noindent The subscript d (f) is shorthand for the conditional probability that the source is a true (false) detection.  Also, note that $p(S_\mathrm{o},\sigma_\mathrm{o})$ integrates to 1.  The joint probability distribution for all of the true detections can be further factored:

\begin{equation}
  p_\mathrm{d}(S_\mathrm{i},S_\mathrm{o},\sigma_\mathrm{o}) =
  p_\mathrm{d}(S_\mathrm{o}|S_\mathrm{i},\sigma_\mathrm{o})
  p_\mathrm{d}(\sigma_\mathrm{o}|S_\mathrm{i}) p(S_\mathrm{i}).
  \label{eq:factordet}
\end{equation}

\noindent The scattering function
$p_\mathrm{d}(S_\mathrm{o}|S_\mathrm{i},\sigma_\mathrm{o})$ is the
probability distribution of observed flux densities given an intrinsic
flux density and measurement error, and hence contains information
about the flux bias due to source blending.  The function
$p_\mathrm{d}(\sigma_\mathrm{o}|S_\mathrm{i})$ is the differential
completeness, since integrating over $\sigma_\mathrm{o}$ is $C(S_\mathrm{i})$, the
probability of detecting a source with intrinsic flux density
$S_\mathrm{i}$ (see Figs.~\ref{fig:diff_compl} and \ref{fig:scatter}). 
The final factor $p(S_\mathrm{i})$ is the
underlying probability distribution of sources with intrinsic flux
densities $S_\mathrm{i}$.

The total number of sources detected in the catalogue is the number
density of sources per square degree $N$ multiplied by the survey area
$A$.  Multiplying each side of Equation~\ref{eq:probdata} by $AN$, and
applying the factorisation in Equation~\ref{eq:factordet}, gives the
observed number density of sources in the catalogue $\cal{N}$ as
a function of $S_\mathrm{o}$ and $\sigma_\mathrm{o}$:

\begin{eqnarray}\label{eq:comexpectation}
&  {\cal N}(S_\mathrm{o},\sigma_\mathrm{o}) = A \int p_\mathrm{d}(S_\mathrm{o}|S_\mathrm{i},\sigma_\mathrm{o}) p_\mathrm{d}(\sigma_\mathrm{o}|S_\mathrm{i}) N(S_\mathrm{i}) dS_\mathrm{i} \\ \nonumber
&    + A N_\mathrm{f}(S_\mathrm{o},\sigma_\mathrm{o}).
\end{eqnarray}

\noindent Here ${\cal N}(S_\mathrm{o},\sigma_\mathrm{o}) \equiv
ANp(S_\mathrm{o},\sigma_\mathrm{o})$, $N(S_\mathrm{i}) \equiv
Np(S_\mathrm{i})$ is the underlying differential source counts per
square degree, and $N_\mathrm{f} \equiv
Np_\mathrm{f}(S_\mathrm{o},\sigma_\mathrm{o})$ is the spurious
detection rate per square degree. The left hand side of this equation 
and the area $A$ are measured directly from the survey.  Both the
scattering and differential completeness functions are calculated
using Monte Carlo simulations, leaving the differential source counts
$N(S_\mathrm{i})$ as the only free parameter to be solved for. It is
trivial to extend this model to a combined source catalogue
${\cal{N}}^{\mathrm{com}}(S_\mathrm{o},\sigma_\mathrm{o})$
from $M$ independent surveys $i$ taken with the same
instrument, provided the different
$p^i_\mathrm{d}(\sigma_\mathrm{o}|S_\mathrm{i})$,
$N^i_\mathrm{f}(S_\mathrm{o},\sigma_\mathrm{o})$, and $A^i$ are known:

\begin{eqnarray}\label{eq:expectation}
{\cal N}^{\mathrm{com}}(S_\mathrm{o},\sigma_\mathrm{o}) = \int p_\mathrm{d}(S_\mathrm{o}|S_\mathrm{i},\sigma_\mathrm{o}) N(S_\mathrm{i}) \sum_i \left[A^i p_\mathrm{d}^i(\sigma_\mathrm{o}|S_\mathrm{i}) \right] dS_\mathrm{i} \\ \nonumber
 + \sum_i A^i N^i_\mathrm{f}(S_\mathrm{o},\sigma_\mathrm{o}).
\end{eqnarray}

To calculate $p_\mathrm{d}(S_\mathrm{o}|S_\mathrm{i},\sigma_\mathrm{o})$ and
$p_\mathrm{d}(\sigma_\mathrm{o}|S_\mathrm{i})$ mock bolometer data
were generated using realisations of Gaussian noise with the same
variance as the real data. To these data were added the effect of a
population of spatially uniformly distributed point sources with a
flux density distribution following the number counts measured by
\citet{Borys2003}.  Source catalogues of all $2.5\,\sigma$ peaks were
created in the same way as the catalogue for the real data. An attempt
was then made to identify each observed point source with objects in
the input catalogue within a $6\,\mathrm{arcsec}$ radius.  If there
were intrinsic sources associated with the peak, the brightest was
considered the match, $S_\mathrm{i}$. In this way the observed flux
density in an aperture was related to the flux density for the single
brightest source that fell within the measurement aperture; other
fainter sources simply contribute to the upward flux bias of this one
source through blending.  To avoid sensitivity to extremely faint
counts that were not sampled by the survey (whose effect is highly
model dependent) only sources with $S_\mathrm{i} \ge 2$\,mJy were
allowed to be matched (the survey was found to be approximately 10 per
cent complete at this level).  The survey is therefore defined to have
a completeness of 0 for $S_\mathrm{i} < 2$\,mJy with this model.  

The
rate of detection of sources $S_\mathrm{i}$, and the distribution of
$S_\mathrm{i}$, $S_\mathrm{o}$ and $\sigma_o$ using 500 simulated maps
was used to estimate
$p_\mathrm{d}(S_\mathrm{o}|S_\mathrm{i},\sigma_\mathrm{o})$ (see
Fig.~\ref{fig:scatter}) and
$p_\mathrm{d}(\sigma_\mathrm{o}|S_\mathrm{i})$ (see
Fig.~\ref{fig:diff_compl}) in bins of width 1\,mJy for $S_\mathrm{i}$
and 0.125\,mJy for $S_\mathrm{o}$ and $\sigma_\mathrm{o}$.  The bin
sizes for $S_\mathrm{o}$ and $\sigma_\mathrm{o}$ were chosen so that
the probability of having more than one source in a bin is small, so
that we can use simple Poisson statistics.  The coarser bin size for
$S_\mathrm{i}$ was adopted so that the Monte Carlo simulations would
converge more quickly, and since $S_\mathrm{i}$ does not require high
resolution because of the width of the posterior flux density
distribution (see Fig.~\ref{fig:deboosted_compare}).  At 
flux densities $>10$\,mJy
$p_\mathrm{d}(S_\mathrm{o}|S_\mathrm{i},\sigma_\mathrm{o})$ did not
fully converge after the 500 simulations since the number density 
of sources is so low ($\ll1$ source per bin).  At fainter flux 
densities it was compared with a
Gaussian model truncated appropriately for the $2.5\,\sigma$ source
selection criteria (see Fig.~\ref{fig:scatter}) and was found to be
indistinguishable.  Rather than run a much larger number of Monte 
Carlo simulations, the scattering function was instead replaced 
by the smooth theoretical model.  Peaks in the map
with no intrinsic counterparts were considered spurious detections
(either pure noise or in extremely rare cases blended sources with
flux densities $< 2$\,mJy) and were used to estimate the false
detection rate $N_\mathrm{f}(S_\mathrm{o},\sigma_\mathrm{o})$ per
square degree using the same bins.

To solve for $N(S_\mathrm{i})$ in Equation~\ref{eq:expectation}, a
discrete non-parametric binned model was first adopted,
$\widetilde{N}(j)$.  Each flux density bin $j$ was chosen to be 2\,mJy
wide (comparable to the photometric uncertainty). Rather than assuming
a constant density of sources across the bin, the counts were modelled
by the product of a free scale parameter, $a_j$, with an exponential
template function similar to the counts spectrum measured in previous
surveys:

\begin{eqnarray}
  \widetilde{N}(j) & = & a_jT_j \\
  T_j & = & \int_{S^j_{\mathrm{low}}}^{S^j_{\mathrm{high}}} S^{-3.8} dS,
\end{eqnarray}

\noindent where $S^j_{\mathrm{low}}$ and $S^j_{\mathrm{high}}$ are the
flux density limits for the $j$th bin.  A downhill simplex optimiser
was used to solve for the $a_j$ by maximizing the joint Poisson
likelihood, ${\cal L}$, of observing the true number of detected
objects in each bin ${\cal N}(j,k)$ given the expected
distribution $\widetilde{\cal N}(j,k)$ produced by the model $\widetilde{\cal N}(j)$ in Equation~\ref{eq:comexpectation}, where $j$
and $k$ denote bins of $S_\mathrm{o}$ and $\sigma_\mathrm{o}$
respectively:

\begin{equation}
  {\cal L} = \prod_{j,k}
  \frac{\widetilde{\cal N}(j,k)^{{\cal N}(j,k)}
    e^{-\widetilde{\cal N}(j,k)}}
  {{\cal N}(j,k)!} .
\end{equation}

\noindent In order to prevent non-physical answers the $a_j$ were
constrained to be positive. In addition, it was discovered that the
solutions were highly unstable and adjacent bins would frequently
oscillate between 0 and very large values. To remedy this type of
problem the fits were further constrained such that the
$\widetilde{N}(j)$ were monotonically decreasing with $S_\mathrm{i}$.

To calculate the uncertainty in the model, maximum likelihood
solutions were re-calculated for 500 realisations of mock data. These
data were generated by drawing the same number of sources as in the
real list from the maximum likelihood distribution
$\widetilde{\cal{N}}(S_\mathrm{o},\sigma_\mathrm{o})$ for the
real data.  In Figs.~\ref{n(s)_lh_fig} and \ref{n(s)_sxdf_fig}, 
the error bars represent the frequentist 68 per
cent confidence intervals for the distribution in each bin. The
integral source count spectrum (and uncertainty) was obtained by
directly integrating each model $\widetilde{N}(j)$.  Finally, the 500 fits of
$\widetilde{N}(j)$ allowed us to directly calculate the sample
covariance matrix $\left<\widetilde{N}(j),\widetilde{N}(k)\right>$.

Despite the non-negative and monotonically decreasing constraints
placed on the binned source counts model, the binned differential
number counts have a significantly larger scatter than was observed in
the other groups' estimates.  This behaviour in the model fitting
process is probably due to the bin size being inappropriately small
given the uncertainty in the posterior flux density distributions for
individual sources, and also the fact that less prior information was
used as a constraint (see discussion in Section~\ref{diffs}).  This
problem is analagous to the increased noise one obtains in an
astronomical image when trying to deconvolve the point spread function
with small pixels compared with the FWHM.

Finally, the model fitting procedure was constrained by replacing the
binned representation of $\widetilde{N}(S_\mathrm{i})$ in
Equation~\ref{eq:expectation} with a smooth parametric model following
Equation~\ref{doublepl} (for consistent notation replace $dN/dS$ with
$\widetilde{N}(S_\mathrm{i})$).  As with the binned model, maximum likelihood 
solutions were found for
the model parameters $N', S', \alpha$ and $\beta$. The parameter
covariance matrix was also obtained using 500 sets of data generated
from Monte Carlo simulations. The 68 per cent confidence envelope for
these models is clearly smaller than the uncertainties of the
individual count bins (see Figs.~\ref{n(s)_lh_fig} and \ref{n(s)_sxdf_fig}).

All of the analysis undertaken for each field separately was repeated
using Equation~\ref{eq:comexpectation} to calculate joint fits of the
differential source counts to both fields simultaneously.

\begin{figure*}
\epsfig{file=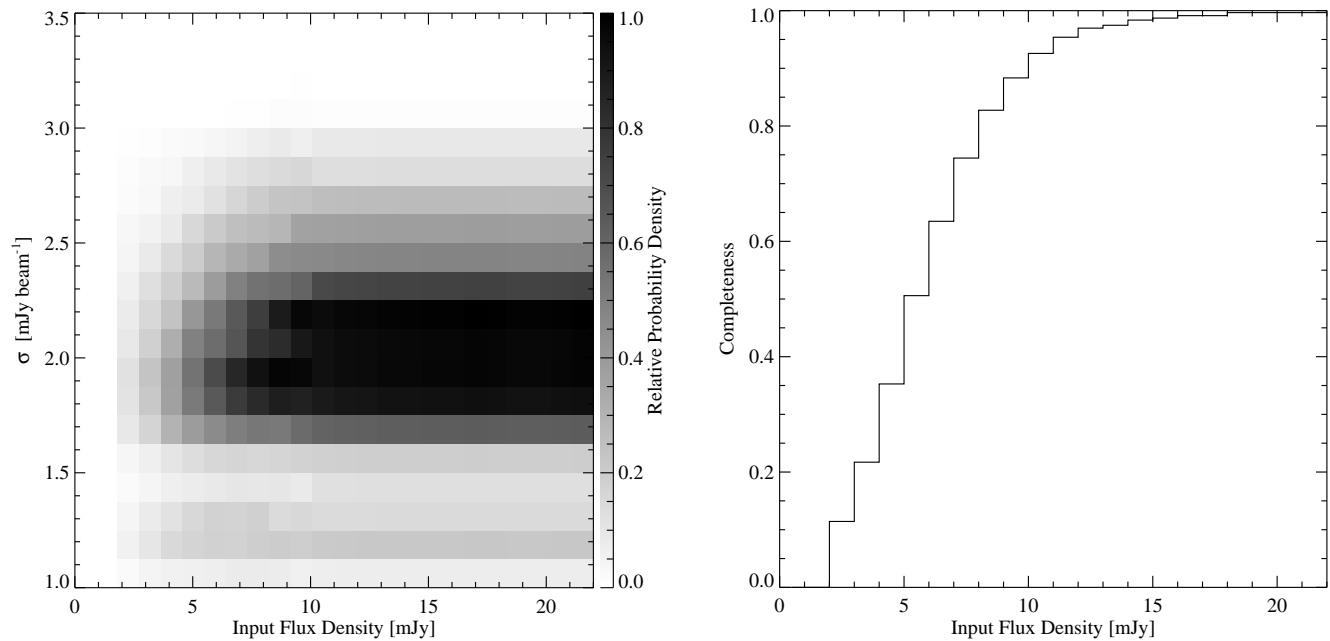,width=1.\textwidth}
\caption{The completeness for LH as calculated by Reduction C.  The
  left panel shows the differential completeness
  $p_\mathrm{d}(\sigma_\mathrm{o}|S_\mathrm{i})$, the probability of a
  source being detected at a noise level $\sigma_\mathrm{o}$ given an
  intrinsic flux density $S_\mathrm{i}$. The variations in the
  vertical direction in this plot reflect the relative areas of the
  map that reached a given noise level. The LH map has a small deep
  portion with a mean noise $\sim1.2$\,mJy, and a large shallower
  region with a mean noise $\sim2$\,mJy, corresponding to the lower and
  upper horizontal bands, respectively. Marginalization over $\sigma_\mathrm{o}$
  (right panel) gives the more typical definition of completeness, the
  probability that a source is detected as a function of instrinsic
  flux density, $p_\mathrm{d}(S_\mathrm{i})$.}
\label{fig:diff_compl}
\end{figure*}

\begin{figure}
\centering
\epsfig{file=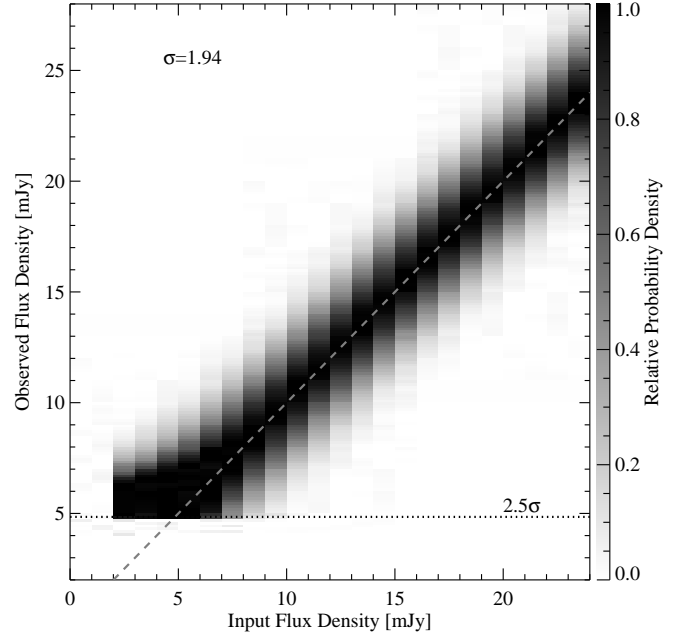,width=0.5\textwidth}
\caption{The scattering function
  $p_\mathrm{d}(S_\mathrm{o}|S_\mathrm{i},\sigma_\mathrm{o})$ at a
  fixed noise level $\sigma_\mathrm{o}=1.94$\,mJy for Reduction C.
  Vertical bands in this plot are the distribution of observed flux
  densities that could be measured for a fixed intrinsic flux density
  and photometric error. In the absence of bias and with Gaussian
  photometric uncertainties this distribution is simply a Gaussian
  with standard deviation $\sigma_\mathrm{o}$ and a mean
  $S_\mathrm{o}$ equal to $S_\mathrm{i}$ (the dashed line).  However,
  at flux densities $S_\mathrm{i} < 10$\,mJy the scattering function
  changes shape. The decision to impose a $2.5\,\sigma$ cut in the
  source list makes it impossible to detect a source at lower S/N and
  is shown by the dotted line. In addition, confusion will tend to
  cause the faintest sources to appear brighter than they really are.
  Since the 850\,$\mu$m extra-galactic confusion limit is at
  $\sim1$\,mJy, however, this effect is negligible.}
\label{fig:scatter}
\end{figure}

\begin{figure}
\centering
\epsfig{file=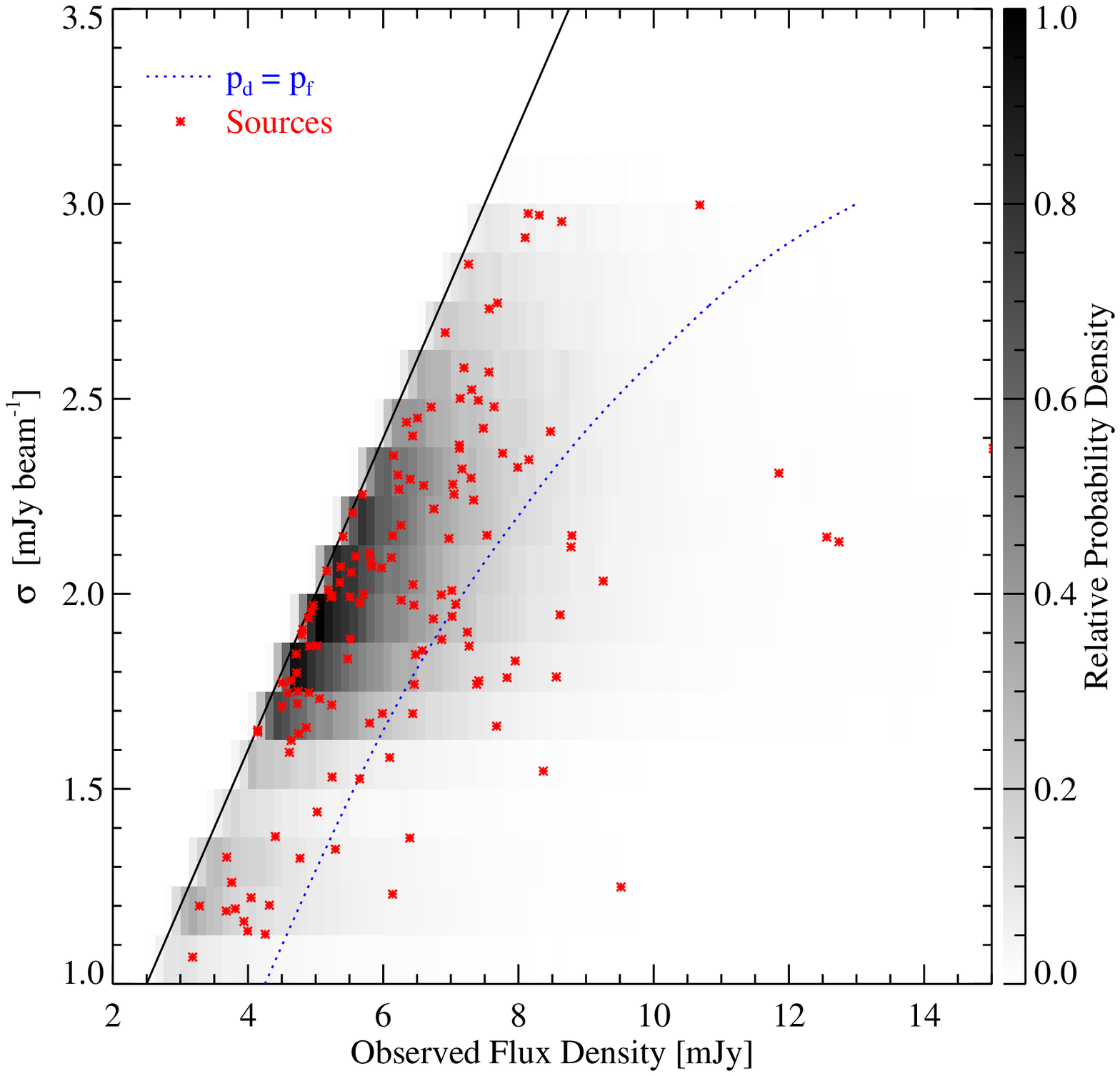,width=0.5\textwidth}
\caption{Reduction C's observed source distribution (red stars)
  compared with the expected distribution
  $\widetilde{N}(S_\mathrm{o},\sigma_\mathrm{o})$ (shaded region) in
  LH for the maximum likelihood differential counts distribution
  $\widetilde{N}(S_\mathrm{i})$.  The diagonal black line shows the
  detection threshold of $2.5\,\sigma_\mathrm{o}$ that was used to
  construct the source list for determining the counts. The dashed 
  contour indicates where the real and spurious parts of 
  Equation~\ref{eq:expectation} are equal. Therefore sources detected
  along this contour have equal chances of being true detections
  of objects with intrinsic flux densities greater than 2\,mJy, or 
  spurious detections (i.e.~fainter than 2\,mJy). Similar to the detection 
  threshold line in Fig.~\ref{fig:flux_noise} this contour is not parallel to the line of 
  constant S/N. Sources are more likely to be true detections in the 
  bottom right region of this plot. 
}
\end{figure}

Reduction C tested for bias in the recovered number counts by simulating 10 data sets with a range of reasonable source count models.  Source catalogues were produced from these data in the same manner as for the real data. Using the methods of Section~\ref{redC} (with the same fixed estimates of $p_\mathrm{d}(S_\mathrm{o}|S_\mathrm{i},\sigma_\mathrm{o})$ and $p_\mathrm{d}(\sigma_\mathrm{o}|S_\mathrm{i})$) the recovered binned counts were in each case consistent with the input counts with insignificant systematic bias.  The correction factor in the lowest bin is consistent with what was found by Reduction D's direct estimate of the source counts (cf.~\ref{deboosttest2}).

\begin{table*}
\begin{tabular}{llll}
\hline 
\multicolumn{1}{l}{Flux density} & \multicolumn{1}{c}{$850\,\mathrm{\mu m}$ differential counts} & \multicolumn{1}{l}{Flux density} & \multicolumn{1}{c}{$850\,\mathrm{\mu m}$ integral counts} \\ 
\multicolumn{1}{l}{(mJy)} & \multicolumn{1}{c}{$dN/dS\,\mathrm{mJy}^{-1}\,\mathrm{deg}^{-2}$} & \multicolumn{1}{l}{(mJy)} & \multicolumn{1}{c}{$N(>S)\,\mathrm{deg}^{-2}$}\\
\hline
2.77 & $831^{+230}_{-227}$ & 2.0 & $2506^{+407}_{-407}$ \\ 
4.87 & $240^{+51}_{-51}$ & 4.0 & $844^{+117}_{-117}$\\
6.90 & $106^{+24}_{-24}$ & 6.0 & $362^{+61}_{-59}$ \\
8.93 & $41^{+13}_{-13}$ & 8.0 & $150^{+35}_{-35}$\\
10.94 & $17^{+7.6}_{-7.0}$ & 10.0 & $68^{+21}_{-21}$ \\
12.95 & $8.8^{+4.2}_{-5.6}$ & 12.0 & $33^{+16}_{-15}$\\
14.96 & $3.9^{+2.2}_{-3.8}$ & 14.0 & $15^{+8.5}_{-9.4}$ \\
16.96 & $1.8^{+1.1}_{-1.7}$ & 16.0 & $7.4^{+4.2}_{-7.3}$ \\
18.96 & $1.0^{+1.8}_{-0.9}$ & 18.0 & $3.9^{+1.9}_{-3.7}$ \\
20.97 & $0.6^{+2.2}_{-0.4}$ & 20.0 & $2.0^{+3.6}_{-1.8}$ \\
\hline
\end{tabular}
\caption{$850\,\mathrm{\mu m}$ SHADES differential (in 2\,mJy wide bins; quoted per mJy) and integral (in 2\,mJy bins) source counts of Reduction D.  The error bars represent the frequentist 68 per cent confidence intervals of the boot-strapped count distribution in each bin (see text).  Each differential count flux density bin is indicated by the midpoint of each bin weighted by $S^{-3}$, whereas the integral count flux density bins are indicated by the lower flux density bound of each bin.  The counts and errors in each of the lowest flux density bins of the differential (integral) estimates have been corrected by a factor of 1.6 (1.33), for the known undercounting in this bin seen in simulations (see Section~\ref{deboosttest2}).  Note that a Gaussian approximation to the error bars becomes invalid for high flux densities.}\label{tab:count_num}
\end{table*}

\begin{table*}
\begin{tabular}{lrrrrrrrrrr}
\hline 
\multicolumn{1}{l}{Flux density (mJy)} & \multicolumn{1}{r}{2.77} & \multicolumn{1}{r}{4.87} & \multicolumn{1}{r}{6.90} & \multicolumn{1}{r}{8.93} & \multicolumn{1}{r}{10.94} & \multicolumn{1}{r}{12.95} & \multicolumn{1}{r}{14.96} & \multicolumn{1}{r}{16.96} & \multicolumn{1}{r}{18.96} & \multicolumn{1}{r}{20.97}\\
\hline
2.77     &   19926.5	 & 109.1	 & $-29.3$	& $-1.5$	& $-12.5$	& $-14.5$ & $-1.8$	& $-2.5$	& $-0.1$ &	1.5\\
4.87     &   109.1	 & 2511.9	 & $-9.0$	& $-3.8$	& 0.2		& 2.4 & 1.2&	 0.1	&$-1.2$	& 0.5 \\
6.90     &   $-29.3$	 & $-9.0$	 & 599.4	& 3.8		& $-1.2$	& 1.1 & $-0.1$	& 0.5	& 0.1 &	 0.6	\\
8.93     &   $-1.5$	 & $-3.8$	 & 3.8		& 176.5		& 2.2		& 0.57 & 0.5 &	 0.0 &	 0.2 &	$-0.1$\\
10.94	 &  $-12.5$	& 0.2		& $-1.2$	& 2.3		& 62.5		& 0.27 & $-0.2$	 &0.1 &	 0.1 &	0.0 \\
12.95	 &  $-14.5$	& 2.4		& 1.1		& 0.6		& 0.3		& 28.6 & 0.0	 &0.1 &	$-0.1$	& 0.1 \\
14.96	&  $-1.8$	& 1.2		&$-0.1$	 	&0.5	&	$-0.2$	 	& 0.0	& 11.5	& 0.0 &	 0.0 &	 0.0\\
16.96   & $-2.5$	& 0.1		&0.5		&0.0	&	0.1	 & 0.1	& 0.0 &	 5.2	& 0.0 &	0.0 \\
18.96   & $-0.1$	&$-1.2$		&0.1	&0.2	& 0.1&	$-0.1$	 &0.0	& 0.0 &	 2.7&	0.0 \\
20.97   & 1.5	 & 0.5	& 0.6	&$-0.1$	& 0.0 &	 0.1 &	 0.0 &	0.0	&0.0	& 1.5	\\
\hline
\end{tabular}
\caption{Covariance matrix for the $850\,\mathrm{\mu m}$ SHADES combined differential source counts.  This can be used along with Table~\ref{tab:count_num} to fit models to the counts using the figure-of-merit $\chi^{2}=(d-m)^{\mathrm{T}}\,C^{-1}\,(d-m)$, where $d$ is the data, $m$ is the model, and $C^{-1}$ is the inverse of the covariance matrix.}\label{tab:corr_num}
\end{table*}

\subsection{Summary of differences between the three methods}\label{diffs}

Having described each group's techniques for calculating posterior flux
density distributions and number counts, it is now appropriate to
discuss several key differences:  the amount of information used; the
interpretation of posterior flux densities; and the problems inherent
in calculating the FDR and completeness. 

The philosophies adopted by each group to calculate posterior flux
density distributions vary in a number of subtle ways. First, what
flux density was calculated?  In the case of Reduction C the 
method is attempting to estimate the flux density of the brightest 
source in the beam, while Reductions B and D calculate the 
posterior distribution 
for the beam-smoothed flux density map.  The former
may be more desirable for determining source counts, since it directly ties
measurements to individual objects.  In practice,
correcting for the confusion is a difficult procedure, which
depends not only on knowledge of the source counts, but also
their spatial clustering \citep{Takeuchi}.  However, since the bulk of the
sources detected in SHADES have flux densities well above the
extra-galactic confusion limit, simply taking the total posterior flux
density in a beam as in the case of Reductions B and D is acceptable for the
purpose of calculating source counts.
Furthermore, this deboosting technique avoids the possibility of
introducing further model-dependent errors into the posterior flux densities
in the SHADES catalogue; for this reason the selection procedure for the SHADES catalogue
uses the \citet{Coppin} algorithm.

How much information is derived completely from the data, and how
much information is assumed? Each of the reductions uses prior
knowledge of the 850\,$\mu$m extra-galactic source counts measured in
previous blank-field and lensing cluster surveys to create simulated
source catalogues, and to simulate maps exhibiting the chop pattern.  
Reductions B and
C use such maps (including noise) to Monte Carlo the source detection
procedure, and to cross-correlate detected sources with the input
catalogue in order to determine completeness.  Sources in the output catalogue
with no corresponding sources in the input catalogue are used to
measure the FDR based on the selection criteria
for the source list. Reduction D uses the simulated maps
(with no instrumental noise added) strictly as a prior for the
posterior flux density distributions.  Spurious sources are handled
by constructing the catalogue in such a way that it is nearly
free of spurious sources. Completeness is determined by introducing
fake sources into the real map and attempting to recover them over a
range of input flux densities. An additional completeness correction is 
calculated in a manner similar to Reductions B and C by ensuring that the 
recovered source density matches the input source density in a separate 
simulation that includes both sources and noise.

Reductions B and D both choose to calculate the differential
source counts in bins directly from the posterior flux density
distributions of sources in their catalogues. There is clearly a
possiblity for the prior used in the posterior flux density
calculation to bias the estimate of the source counts. Reduction
B handles this problem by using the pixel distribution in the observed
map itself as a prior for individual sources. Reduction D tests the
procedure on maps with sources drawn from several different source
count models, while keeping the prior for de-boosting fixed to test
for bias (see Section~\ref{deboosttest2}). Rather than assuming a prior, 
Reduction C leaves the prior as a free parameter and attempts to fit it to the 
observed catalogue.

The simulations undertaken by all reductions indicate that the uncertainty in 
the completeness introduced by different source count models is dominated by 
other uncertainties in the counting procedure (map noise and Poisson counting 
uncertainties). In the case that sources are completely isolated, the 
probability that they are detected (i.e.~completeness) is 
only a function of the noise. On the other 
hand, source blending near the detection threshold may affect completeness as 
a function of the underlying source density. Sources that blend together and 
are detected as a single bright source increase the completeness if 
individually they would not have been detected, but decrease the completeness 
if individually they could have been detected. In this paper, the faintest 
sources that we claim to count are $>2\,\mathrm{mJy}$. The full range of 
integral source counts at this flux density measured here, and in previous 
work, is conservatively between 1000 and 10000 sources per $\mathrm{deg}^{2}$.
 Therefore, for spatially uniformly distributed objects there are on average 
$\simeq0.02$--0.2 sources that land within a beam (the solid angle of the 
SCUBA beam is $\simeq2\times10^{-5}\,\mathrm{deg}^{2}$; the 
$\mathrm{FWHM}^{2}$). 
Given the Monte Carlo simulations undertaken in 
Sections~\ref{redD} and ~\ref{redC}, using a plausible range of different 
input source counts models, it is therefore not surprising that the 
variations in the completeness corrections down to $2\,\mathrm{mJy}$ are dominated by 
uncertainties other than the variations in the counts model.

Which method is best? Since the 850$\mu$m source counts are
known well enough to construct a useful prior, it makes sense 
to use this information
in the analysis of SHADES data. It is not surprising that Reductions B and
D quote smaller uncertainties than Reduction C in the binned differential
counts, since more information is being used. On the other hand,
for surveys at a wavelength for which little information is known, the
technique for binned counts described in Section~\ref{redC} is
more conservative. The technique adopted by Reduction C is more useful
once a smooth parametric model is assumed to describe the source
counts. Constraining the counts in this way produces a range
of models, with a spread generally consistent with the smaller error
bars quoted by the other groups for the flux density bins. 
Furthermore it offers the
cleanest way for constraining the model; the model 
parameters are varied directly to calculate the likelihood of observing
the source catalogue.  With Reduction D, on the other hand, parametric
models are fit to the binned counts calculated previously. However,
the binned differential counts from Reduction D can more easily be
combined with counts from other surveys to constrain models over
wider ranges in flux density, because the data products that come out 
of the procedure are binned counts and a bin-to-bin covariance matrix.

For this last reason we proceed with the number counts and bin-to-bin covariance matrix of Reduction D.  It is reasonable to select one group's reduction, since we cannot easily combine the three sets of counts like we did for flux densities or positions, and moreover all of the counts across reductions appear to be consistent with each other within the error bar estimation, for all but the lowest bin (where Reduction B appears to be low).  In model tests we quote the best fit parameters using the counts of Reduction D, since that reduction provides the smallest error bars (cf.~Section~\ref{dependent}), while using the 68 per cent confidence intervals of Reduction C as a consistency check.  The differential counts for LH and SXDF are shown in Figs.~\ref{n(s)_lh_fig} and \ref{n(s)_sxdf_fig}, respectively.

\begin{figure}
\psfig{file=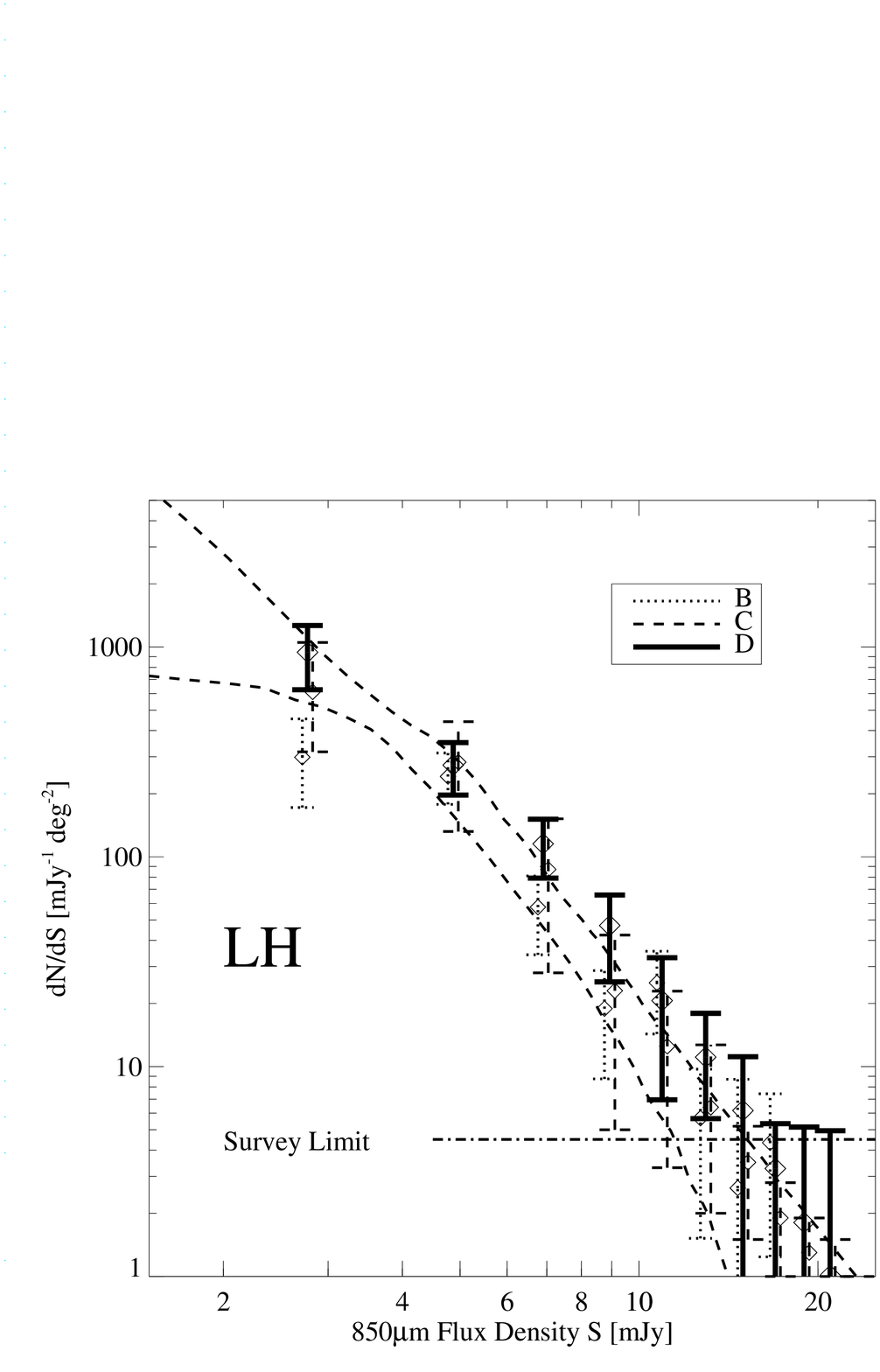,width=0.5\textwidth}
 \caption{Differential source count densities in the LH region.  The number of sources $\mathrm{deg}^{-2}\,\mathrm{mJy}^{-1}$ at a given flux density is plotted against flux density (with different reductions offset horizontally for clarity).  All error bars are estimates of $1\,\sigma$ uncertainties of the full error distribution including completeness estimates (see text).   The horizontal line marked `Survey Limit' corresponds to the 68 per cent confidence limit which can be drawn from finding zero sources in $1/8\,\mathrm{deg}^{2}$ (in a $2\,\mathrm{mJy}$-wide bin).  The 68 per cent confidence interval on the set of best-fitting double power-laws of the form of Equation~\ref{doublepl} (fit to the differential counts of Reduction C) are overplotted as dashed lines; the error bars show that there is little constraint above this level.}
 \label{n(s)_lh_fig}
\end{figure}
 
\begin{figure}
\psfig{file=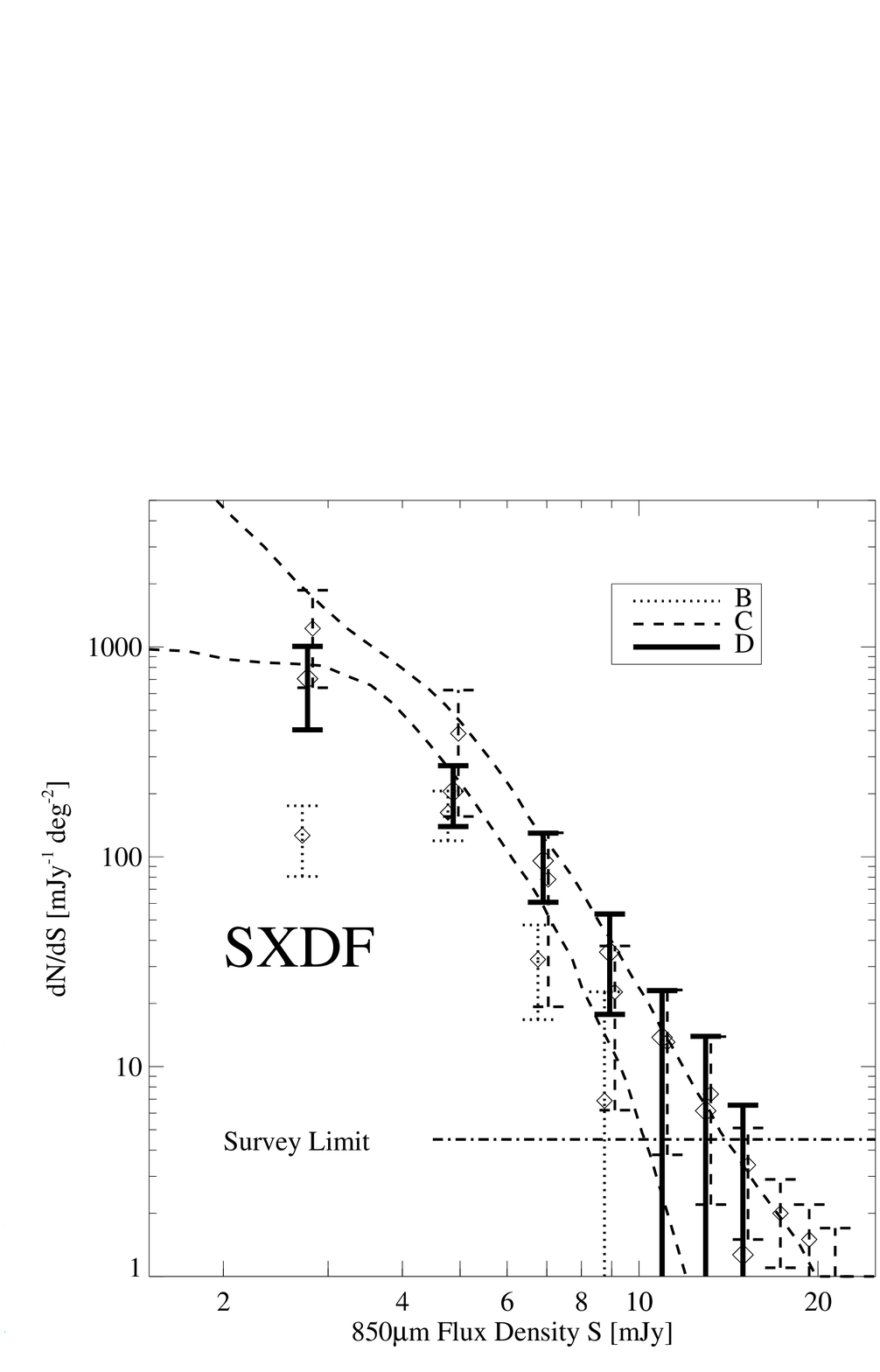,width=0.5\textwidth}
 \caption{Differential source count densities in the SXDF region, as for 
Fig.~\ref{n(s)_lh_fig}.  Notice that in all except the lowest two flux density bins for Reduction C, the density of sources inferred in the SXDF is approximately
   $1/4\,\sigma$ lower than for the LH data; see Section~\ref{cluster} for a discussion of possible field-to-field variations.}
 \label{n(s)_sxdf_fig} 
 \end{figure}

\section{Results:  Models and Cumulative Source Counts}\label{numcts}

\subsection{Fits to differential counts}\label{diffcounts}

In fitting models of the numbers of sources at different flux densities to our data there are different approaches which could be taken.  The first choice is whether to fit to the list of sources directly or to count the sources in a series of discrete flux density bins and fit to those numbers.  In principle these methods could each be applied to the results of each reduction, but there is little to be gained from generating eight different fits in this way.  We have tried one approach each on the data of two separate reductions.  

Starting with the catalogue from Reduction C, we use Monte Carlo techniques to compute the maximum-likelihood distribution of models that fit the source catalogue, given the completeness and FDR.   The dashed curves in Figs.~\ref{n(s)_lh_fig} and \ref{n(s)_sxdf_fig}, indicate the 68 per cent frequentist interval on the best-fitting models.  This procedure requires detailed knowledge of the FDR and completeness as a function of flux density and therefore it is difficult to compare directly with published counts.  Note that the 68 per cent range on the models fit to Reduction C is about the size of the individual error bars in Reduction D, and that the reductions are all consistent with each other within the error bars in all but the lowest flux density bin.

Starting with the counts and covariance matrix, based on Reduction D (found in Tables~\ref{tab:count_num} and \ref{tab:corr_num}), we fit a simple power-law of the form

\begin{equation}\label{single}
\frac{dN}{dS}=N'\left(\frac{S}{S'}\right)^{-\alpha},
\end{equation}

\noindent using a minimum $\chi^{2}$ parametric approach.  Here, $N'$ is the normalisation at $S'=5\,\mathrm{mJy}$ (the approximate `pivot' point for our data).  

The `Survey Limit' (as indicated by a dot-dashed line in the figures) corresponds to the 68 per cent confidence limit which can be drawn from finding zero sources in $1/8\,\mathrm{deg}^{2}$ (in a $2\,\mathrm{mJy}$-wide bin) in each of the two fields.  This provides a rule-of-thumb counting limit to which the SHADES fields are sensitive, given the observed areas and depths achieved.  Note that the Gaussian error approximation breaks down for the highest flux density bins but these carry little weight in the fits in any case.  The approach taken by Reduction C also makes full use of the range of flux densities in which there are no detected objects, and thus provides a consistency check. 

The best fitting line has parameters:  $\alpha=2.9\,\pm{0.2}$ and $N'=189\,\pm{26}$ in LH; and $\alpha=3.0\,\pm{0.3}$ and $N'=136\,\pm{24}$ in SXDF.  Note that $\alpha$ and $N'$ are essentially uncorrelated with each other, because we choose the normalisation to be around $S'= 5\,\mathrm{mJy}$, near the centroid of our data.  This result, that $\alpha$ is virtually the same in both fields while $N'$ is lower in SXDF, is consistent with the data in Figs.~\ref{n(s)_lh_fig} and \ref{n(s)_sxdf_fig}, where every bin in SXDF appears lower than the corresponding bin in LH by about $1/4\,\sigma$ uniformly across source brightness.  The slope result agrees with previous estimates obtained by other groups (we quote $1\,\sigma$ limits from the literature): $\alpha=2.8\pm{0.7}$ \citet{Blain99}; $\alpha=2.9\pm{0.25}$ \citet{Borys2003}; $\alpha=3.2^{+0.35}_{-0.3}$ \citet{Barger99}; and $\alpha=3.25\pm{0.7}$ \citet{Eales}.

The total number of sources per square degree in LH versus SXDF differ by less than $2\,\sigma$ (for purely Poisson scatter) and there is no convincing evidence to suggest that the ratio of the counts between the fields is different from 1.  In the rest of our analysis we combine the data for the two fields to improve the statistical power slightly.  The combined counts (i.e.~for both fields) are plotted in Fig.~\ref{fig:singlepl}. 

\begin{figure}
\psfig{file=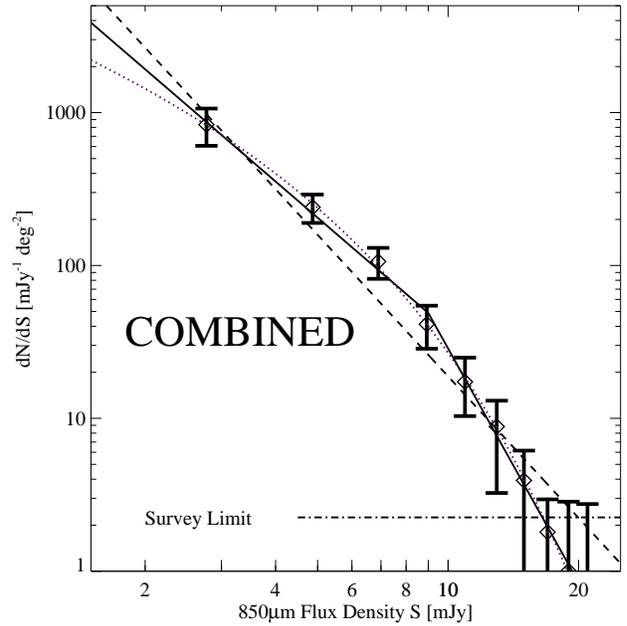,width=0.5\textwidth}
 \caption{Differential source counts in the combined fields (LH + SXDF).  The number of sources $\mathrm{mJy}^{-1}\,\mathrm{deg}^{-2}$ at a given flux density is plotted against flux density.  All error bars are estimates of $1\,\sigma$ uncertainties of the full error distribution, including completeness estimates (see text).  The horizontal line marked `Survey Limit' corresponds to the 68 per cent confidence limit which can be drawn from finding zero sources in $1/4\,\mathrm{deg}^{2}$ (in a $2\,\mathrm{mJy}$-wide bin).  The best-fitting single power-law (Equation~\ref{single}; dashed line), Schechter function (Equation~\ref{schech}; dotted curve), and broken power-law (Equation~\ref{realpl}; solid lines) to the differential counts of Reduction D are plotted.  It is clear that the Schechter function or the broken power-law are a better fit to the data than the single power-law, indicating a break somewhere in the range $\simeq\,5$--$13\,\mathrm{mJy}$.}
 \label{fig:singlepl}
 \end{figure}

It is clear that a single power-law model for the combined data, shown as the dashed line in Fig.~\ref{fig:singlepl}, is a poor fit to the data.  The fit remains poor even if the counts in the lowest bin are arbitrarily doubled, and we do not believe we have mis-estimated completeness by this much.  We have thus explored fits to broken power-laws, and to a Schechter function \citep{Schechter}.  

A full fit to Equation~\ref{doublepl} involves solving for four parameters: $N'$, $S'$, $\alpha$, and $\beta$, which are not simultaneously well-constrained by the few data points.  The parameter error bars are therefore large and correlated (see Table~\ref{tab:fits}).  If instead, we hold the break flux density fixed at $S'=9$, a visually plausible value (and close to the best-fitting value of $S'$), the resulting uncertainties in $N'$, $\alpha$, and $\beta$ are reduced, as also shown in Table~\ref{tab:fits}, and the errors become almost uncorrelated.  The $\chi^{2}$ of the broken power-law fit is reduced from the single power-law by about 10 for one new parameter ($N'$, $\alpha$ $\rightarrow$ $N'$, $\alpha$, $\beta$).  However, we note that the resulting reduced $\chi^{2}$ values are unrealistically small (i.e.~less than 1.0) in all but the single power-law fits.  This probably reflects the fact that the errors are non-Gaussian or that there are some correlations not taken into account properly in the covariance matrix.  But the fact remains that fits to a single power-law do not describe the data well.

We also fit a broken power-law of the form
\begin{eqnarray}\label{realpl}
&\frac{dN}{dS}=N'\left(\frac{S}{S'}\right)^{-\beta}\,\mathrm{for}\,S>S'\,\,\mathrm{and} \\\nonumber
&\frac{dN}{dS}=N'\left(\frac{S}{S'}\right)^{-\alpha}\,\mathrm{for}\,S<S',
\end{eqnarray}
\noindent using the same minimum $\chi^{2}$ parametric approach.  The best-fitting form of Equation~\ref{realpl}, holding $S'$ fixed at $9\,\mathrm{mJy}$, is plotted in Fig.~\ref{fig:singlepl} as 2 solid lines.

A Schechter function \citep{Schechter} was also fit to the counts, since it is at least a physically-motivated functional form:

\begin{equation}\label{schech}
\frac{dN}{dS} = \frac{N'}{S'}S\left(\frac{S}{S'}\right)^{\alpha}\mathrm{exp}(-S/S').
\end{equation}

\noindent  The parameters $N'$, $S'$ and $\alpha$ of the best-fitting form of Equation~\ref{schech} are tabulated in Table~\ref{tab:fits} (see Fig.~\ref{fig:singlepl}, dotted curve).  The Schechter function above fits the data as well as the other 2 power-laws we fit here.  This may not be surprising given that the relationship between $850\,\mathrm{\mu m}$ flux density and rest-frame FIR luminosity is nearly constant and independent of the redshift for $1\lesssim z \lesssim8$ (see \citealt{Blain}).  But there is no evidence to favour an exponential fall off over a steeper power-law at large flux densities.

The data clearly favour a change in the slope for the differential counts, although the precise position of this break is not well constrained.  This change in slope of the differential counts should be helpful in breaking degeneracies in fitting models of luminosity function evolution.

If we use additional information from low flux density counts (particularly from cluster lens fields) the evidence for a break becomes stronger.  Also, a shallower slope at low flux densities is required in order not to overproduce the submillimetre background.  However, neither of these arguments requires that the break be at a large enough flux density to be seen within the SHADES data.  It is our direct fits to the differential counts which, for the first time, constrain the power-law break to be at a flux density of several mJy.

\begin{table}
\begin{tabular}{lllll}
\hline
\multicolumn{1}{c}{Equation}
& \multicolumn{1}{c}{$S'$} & \multicolumn{1}{c}{$N'$} & \multicolumn{1}{c}{$\alpha$} & \multicolumn{1}{c}{$\beta$} \\
\hline
\ref{doublepl} & $9.4\,\pm{4.0}$ &  $647\,\pm{739}$ & $2.0\,\pm{0.6}$ & $6.0\,\pm{2.2}$ \\
\ref{doublepl} &  9    & $735\,\pm{123}$ & $2.0\,\pm{0.2}$ & $5.8\,\pm{0.9}$ \\
\hline
\ref{realpl} &  9   & $49\,\pm{9.3}$ & $2.4\,\pm{0.2}$ & $5.1\,\pm{0.9}$ \\
\ref{schech} & $3.3\,\pm{1.2}$ & $1599\,\pm{1183}$ & $-2.0\,\pm{0.7}$ & -- \\
\hline
\end{tabular}
\caption{Best-fitting parameters of Equations~\ref{doublepl} (smooth double power-law), \ref{realpl} (a broken power-law) and \ref{schech} (Schechter function) fit to the combined counts.  Fixing the break location in the source counts, $S'$, has the effect of reducing the error bars on the other parameters (cf. top 2 rows).}\label{tab:fits}
\end{table}

\subsection{Background estimate}

One can use any of the best-fitting models to estimate the total flux density which has been resolved into discrete sources.  Adopting Equation~\ref{realpl} and using the best-fitting parameters in Table~\ref{tab:fits}, with $S'$ fixed at 9\,mJy, we can calculate the total $850\,\mathrm{\mu m}$ flux density by integrating $S\,N(S)\,dS$.  The estimated background of sources brighter than $2\,\mathrm{mJy}$ is $9.7^{+4.6}_{-3.2}\,\times 10^{3}\,\mathrm{mJy}\,\mathrm{deg}^{-2}$.  We can also estimate the background directly from the SHADES counts by performing the sum $\sum S N(S)$ over the bins, which gives $1.0^{+0.29}_{-0.28}\,\times\,10^{4}\,\mathrm{mJy}\,\mathrm{deg}^{-2}$.

By comparison, the total FIR $850\,\mathrm{\mu m}$ background inferred from \textit{COBE}-FIRAS  is 3.1--$4.4\,\times\,10^{4}\,\mathrm{mJy}\,\mathrm{deg}^{-2}$ (\citealt{Puget}; \citealt{Fixsen}; \citealt{Lagache_rev}).  We find that a survey complete down to $2\,\mathrm{mJy}$ at $850\,\mathrm{\mu m}$ would resolve between 20 and 30 per cent of the FIR background into point sources.  The uncertainty in this fraction is dominated by uncertainty in the unresolved background, rather than uncertainty in the SHADES sources.  Our result is consistent with values quoted by other groups (e.g.~\citealt{Barger99}, \citealt{Eales}, \citealt{Borys2003}).  A significant fraction of the submillimetre emission lies below the detection limit of blank-field SCUBA surveys at $850\,\mathrm{\mu m}$.  Therefore, knowing the number counts accurately down to much fainter flux density limits ($\sim 0.1\,\mathrm{mJy}$) is essential in order to constrain models that predict the evolution of luminous IR galaxies.

\subsection{Cumulative source counts}\label{cumul}

Previous SCUBA surveys with fewer sources have reported source densities in the form of cumulative counts rather than differential counts, since the latter are somewhat sensitive to the bin choice for small source catalogues.  The SHADES integral counts are obtained by directly summing over the differential counts of Reduction D.  The best-fitting double power-law model and Schechter function (cf. Equations~\ref{realpl} and \ref{schech} and Table~\ref{tab:fits}) of the differential counts have been integrated to produce models of the cumulative counts.  The cumulative counts and these models are plotted in Fig.~\ref{lh_prior_counts_fig} and given in Table~\ref{tab:count_num}.  Note that the models are \textit{not} fit in the cumulative counts domain.  These should be the most accurate $850\,\mathrm{\mu m}$ number counts in the flux density range 2--$15\,\mathrm{mJy}$ achieved so far by any single survey. 

\subsection{Comparison of cumulative counts to previous estimates}\label{compare_with_previous}

\begin{figure*}
\psfig{file=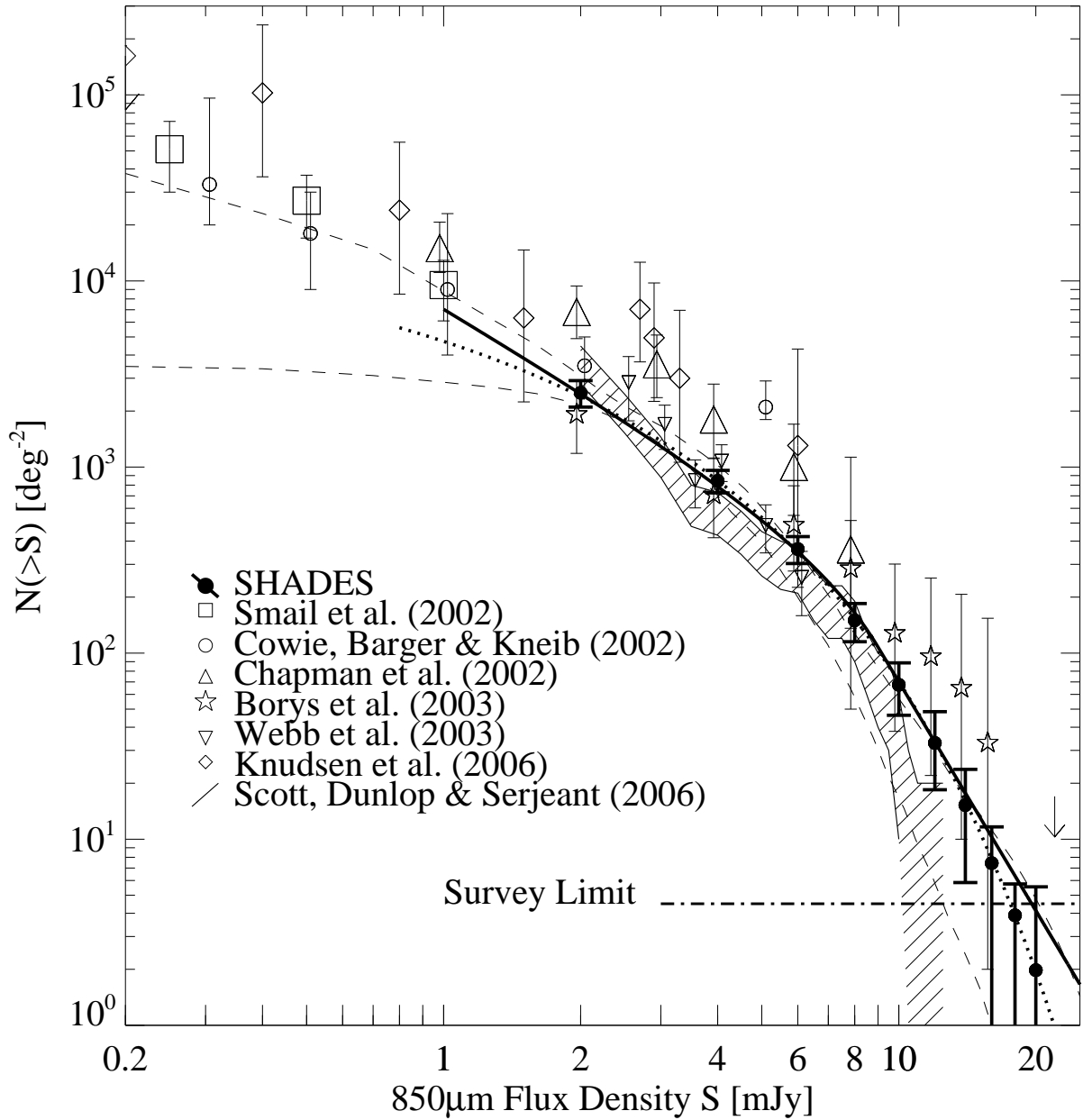,width=1.\textwidth}
\caption{Cumulative combined SHADES source counts (solid points) compared to
 previous estimates.  The 68 per cent confidence interval in Reduction C's model fits is indicated by the dashed lines.  The 95 per cent Poisson upper confidence limit to the surface density of sources brighter than 22\,mJy in SHADES from Section~\ref{bright} is shown by the downward arrow.  The source counts determined from various other blank-field and cluster lensing surveys (corrected for lensing) are indicated using different symbols, with the \citet{SDS} compiled counts being represented by the hatched region.  The best fitting form of Equation~\ref{realpl} (dark solid curve), holding $S'$ fixed at $9\,\mathrm{mJy}$, and Equation~\ref{schech} (dark dotted curve) to the differential counts are integrated and plotted here.  See Table~\ref{tab:fits} for the best-fitting parameter values for these functions.  The data confirm a break in the power-law somewhere in the middle of the flux density range.  The lensing data are consistently high, which could have a number of explanations (see text).}
\label{lh_prior_counts_fig}
\end{figure*}

Previous measurements of the cumulative number counts have differed by factors of $\simeq\,2.5$--3 among groups observing various small (10s to $\sim200\,\mathrm{arcmin}^{2}$) patches of sky, especially at flux densities between 2--$6\,\mathrm{mJy}$ (e.g.~\citealt{Barger99}, \citealt{Blain99}, \citealt{Eales}, \citealt{Scott}, \citealt{Borys2003}, \citealt{Webb}; see discussion in \citealt{Scott}).  The culprits are most likely sampling variance, clustering and/or low number statistics, due to the small-area surveys, as well as different ways of treating flux density boosting in the low S/N regime.

We might expect most agreement at the bright end of the source counts (among previous wide-area surveys to varying depths), since only the brightest sources would have prevailed.  However, we might also expect most disagreement between small area blank-field surveys, since very bright sources are rare and a large degree of sampling variance may therefore come into play.  In Fig.~\ref{lh_prior_counts_fig}, the SHADES counts are plotted in comparison with some previous work in order to assess the level of agreement.

Gravitational lensing by clusters of galaxies has been used as a tool to study the faintest SMGs (e.g.~\citealt{Blain99}; \citealt{Cowie}; \citealt{Smail2002}; \citealt{Chapman2002}; \citealt{Webb_lens}; and \citealt{Knudsen}).  The agreement between these independent surveys is very good below about $2\,\mathrm{mJy}$.  We note that there is a smooth transition also at $3\,\mathrm{mJy}$ between the cluster surveys and this work.  However, we agree with \citet{Borys2003}, \citet{Webb_lens} and others who noted that the lensing number counts above 4\,mJy (where the cluster and blank-field counts overlap) appear higher than the combined blank-field survey counts.  This is most likely due to the fact that the quoted error bars in the literature do not contain a variance component and that the flux density boosting bias has been treated differently by different groups.

At the bright end, the \citet{Borys2003} blank-field counts follow an approximately power-law decline with increasing flux density, whereas ours appear to steepen more dramatically, but the results agree well at low flux densities (2--$6\,\mathrm{mJy}$).  The \citet{Borys2003} counts appear to be higher on average than the SHADES counts and those of \cite{SDS}, but are within the error bars.  Sampling variance may be responsible for these differences, since no survey so far has enough area to overcome this, particularly if these objects cluster on arcminute scales.  SHADES agrees well, particularly at the higher flux densities, with the counts of \citet{SDS}, which is a compilation of data covering a similar area to each of the two SHADES fields (the `band' in Fig.~\ref{lh_prior_counts_fig}).  For flux densities $\gtrsim10\,\mathrm{mJy}$ the SHADES counts appear to be intermediate between those of \citet{Borys2003} and \citet{SDS}.

\subsection{Bright source constraint}\label{bright}
We can also estimate a limit to the surface density of the brightest SMGs.  There are various ways to do this, but the simplest is just to take the fact that SHADES contains no sources brighter than 22\,mJy in the entire surveyed area to constrain the bright counts using Poisson statistics.  We find $N({>}22\,\mathrm{mJy})<17\,\mathrm{deg}^{-2}$ at 95 per cent confidence.  This estimate is probably more robust than previous upper limits at similar flux densities (e.g.~\citealt{Borys2003}; \citealt{Scott}) and can be compared with the even brighter \citet{Barnard} limit of $N({>}100\,\mathrm{mJy})<2.9\,\mathrm{deg}^{-2}$.

\subsection{Comparison of the two fields -- evidence for sampling variance?}\label{cluster}

We can estimate the expected sampling variance between the fields in the
following way.  The variance of counts in cells is generally given by
(\citealt{Peebles}, eqn.~45.6)

\begin{equation}
$$
\sigma^{2}({\cal N}) = {\cal N} \Omega + {\cal N}^2 \int \int w(\theta_{12}) d\Omega_1 d\Omega_2,
$$
\end{equation}

\noindent where ${\cal N}$ is the number per unit area on the sky, $\Omega$ is the solid
angle of the cell, $w(\theta)$ is the angular 2-point correlation
function and the integrals are over all separations $\theta_{12}$
between 2 positions.  If the size of the fields is
much bigger than the correlation length, then one of the integrals can
be done trivially, giving

\begin{equation}
$$
\sigma^{2}({\cal N}) ={\cal N} \Omega \left\{1 + {\cal N} \int w(\theta) d\Omega\right\},
$$
\end{equation}

\noindent where $\theta$ can now be considered as the angle from the centre, say.  The term with the integral is the excess variance over Poisson.  If we are
in the power-law regime for $w(\theta)$, then
$w(\theta) = (\theta/\theta_0)^{-\gamma}$, with $\gamma\simeq0.8$ typically.
For simplicity we can assume that the area is a circle of radius
$\theta_{\rm max}$.  In that case the
integral gives $(2\pi \theta_0^{0.8}/1.2) \theta_{\rm max}^{1.2}$ (with any
cut-off at some $\theta_{\rm min}$ being relatively unimportant).
Eventually this will break down, since $w(\theta)$ goes negative, but it is
a reasonable approximation in the power-law regime.
Putting in $\Omega\simeq 400\,{\rm arcmin}^2$ and ${\cal N}\Omega\simeq60$ for
SHADES, one obtains an RMS scatter in the counts of

\begin{equation}
$$
\sigma({\cal N}) \simeq ({\cal N}\Omega)^{1/2}\times
 (1 + 0.5 (\theta_0/1^{\prime\prime})^{0.8})^{1/2}.
$$
\end{equation}

Measuring the value of the clustering angle, $\theta_0$ is
one of the key goals of SHADES, and a careful analysis will be made once the forthcoming redshift distributions (Itziar et al. and Clements et al., in preparation) are included.  One might ask what constraints on $\theta_0$ can be obtained from comparing the number of sources per square degree found in the two SHADES fields.  In Section~\ref{diffcounts} we found $189\pm\,26$ sources $\mathrm{deg}^{-2}\,\mathrm{mJy}^{-1}$ at $S=5\,\mathrm{mJy}$ in the LH and $136\pm\,24$ in the SXDF.  The uncertainties are essentially Poisson, and are therefore underestimated by a factor
$$
{\cal F}(\theta_0) =
 (1 + 0.5 (\theta_0/1^{\prime\prime})^{0.8})^{1/2}
$$
if clustering is important.  In Fig.~\ref{fig:ratio} we have plotted the likelihood to observe a difference, $\Delta N$ between the number counts in the two fields which is larger than the one we have observed as a function of clustering angle, $\theta_0$. \citet{vanKampen} estimate that for surveys like SHADES, $\theta_0$ might range from $5\,\mathrm{arcsec}$ (for a hydrodynamic model) to perhaps $20\,\mathrm{arcsec}$ (for a high mass merger model).  Notice that the likelihoods in Fig.~\ref{fig:ratio} remain within the central $68\,\mathrm{per cent}$ region even over a broader range of $\theta_0$ than is anticipated, corresponding to values of ${\cal F}$ running from 1 to $\simeq4$.  Thus, for this particular comparison we do not provide even an interesting $1\,\sigma$ constraint on the clustering angle.

\begin{figure}
\psfig{file=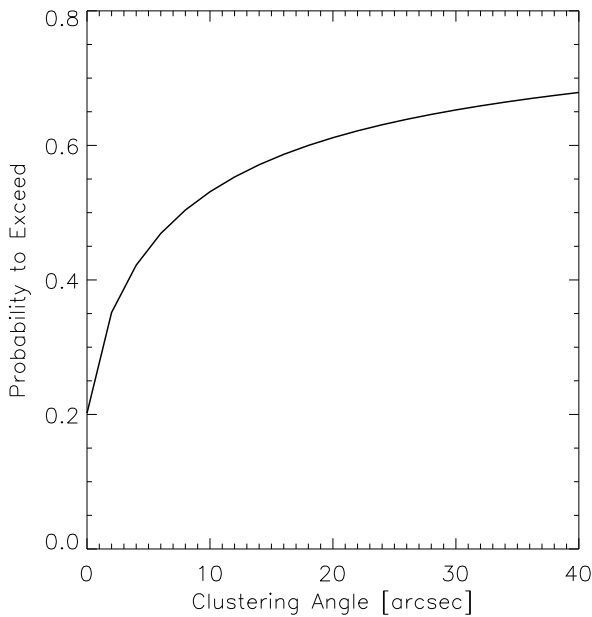,width=0.5\textwidth}
 \caption{The likelihood that fluctuations would lead to a larger difference in the density of $5\,\mathrm{mJy}$ sources than has been observed between the LH and SXDF regions is plotted against clustering angle, $\theta_0$.  Even though the run of $\theta_0$ which is plotted is broader than the range anticipated from physical models, the likelihood values all lie in the central 68\,per cent of the distribution.  Mild clustering is a slightly better fit to the data than no clustering ($\theta_0=0$), but these data do not provide an interesting constraint on clustering amplitude, even at the $1\,\sigma$ level.}
 \label{fig:ratio}
 \end{figure}

\section{Conclusions}\label{conc}

This is the first instance that a SCUBA data set has been subjected to a detailed comparison of differing independent reductions.  As a result, we believe we have produced the most robust and reliable submillimetre-selected source catalogue to date.  We have learned that using monthly average FCFs is just as good as using nightly calibration observations, but with the advantage of less time lost to calibration observations.  Also, using larger pixels (3\,arcsec as opposed to 1\,arcsec pixels) to make maps of these data does not seem to increase the positional uncertainty in the centroiding of detected sources.

The SHADES survey is the final legacy of SCUBA, being the largest extragalactic survey undertaken to investigate the nature and redshift distribution of a complete and consistent sample of $>100$ SMGs.  Here, we have presented $850\,\mathrm{\mu m}$ source catalogues and number counts (differential and integral) using information from four independent reductions of the data, complete to about $720\,\mathrm{arcmin}^{2}$, down to an RMS noise of $\simeq2.2\,\mathrm{mJy}$.  We also provide $3\,\sigma$ upper limits to the $450\,\mathrm{\mu m}$ flux densities of each SHADES source.  With approximately double the area of all blank-field surveys combined, this work has produced the most robust submillimetre source catalogue for multi-wavelength studies, as well as the most accurate measurement of SMG number counts.  

We find that the differential counts are better fit with a broken power-law than a single power-law, the location of the break in the source counts being at several mJy.  Using the ratio of the counts between the two SHADES fields, we are unable to provide a useful constraint on SMG clustering.

Overall, the increased accuracy of the source counts due to the new measurement by SHADES will allow us to study the evolutionary nature of SMGs in much more detail than has previously been possible.  SMG number counts have been able to put severe constraints on galaxy evolution models, where new ingredients or assumptions had to be adopted in order to reproduce them (e.g.~\citealt{Kaviani}, \citealt{Granato}, \citealt{Baugh}).

The maps and catalogues that have resulted from this work, in combination with multi-wavelength measurements from deep radio, far-, mid-, and near-infrared, optical and X-ray imaging, will yield the photometric redshifts necessary to investigate the SMG contribution to the cosmic star-formation history of the Universe and to discover the true nature of the SMG population.  

Further work on the SHADES data will include a $P(D)$ fluctuation analysis (which should be relatively straightforward because of the fairly uniform noise), and clustering estimates (again, significantly easier for SHADES than for previous surveys), as well as follow-up studies, focussed on estimating redshifts and determining luminosities and other physical properties for the sample.  A detailed analysis and modelling of the number counts will be presented in future paper (Rowan-Robinson et al. in preparation).  Together these will provide much tighter constraints on galaxy evolution models.  We note that wider surveys, such as those planned for SCUBA-2, the LMT, \textit{Planck} or \textit{Herschel}, will be able to place better constraints on the bright end of the number counts, since these SHADES maps are insufficiently wide to detect rare bright SMGs.  On the other hand, deeper surveys, such as those also planned for SCUBA-2, are needed to push into the JCMT's confusion regime, and eventually ALMA will allow us to study these SMGs with the desired angular resolution.

\section{Acknowledgments}\label{ack}
We thank an anonymous referee for helpful comments and suggestions which improved the paper.  We also thank Jasper Wall for useful discussions on boot-strapping and model fitting to correlated data.  The SHADES consortium would like to acknowledge the following people, who are not members of SHADES, for help with the SHADES observations:  numerous JAC staff members, Alex van Engelen and Payam Davoodi.  The paper working group acknowledges support from the Natural Sciences and Engineering Research Council of Canada (NSERC), the Particle Physics and Astronomy Research Council of the United Kingdom (PPARC), and partial support through CONACyT grants 39953-F and 39548-F.  The JCMT is operated on behalf of the PPARC, the Netherlands Organisation for Scientific Research, and the National Research Council of Canada.  

\setlength{\bibhang}{2.0em}

\begin{appendix}

\section{Noise Spike Removal Tests}\label{noisespike}
In order to assess its effect on the data, an aggressive attempt was made to remove the noise spike (see Section~\ref{dr}) as follows.  First, the array average was removed as a function of time, using only the least-affected bolometers to estimate the sky at each time, so as not to introduce the spike into bolometers which do not already exhibit the excess power.  Next, another function of time (essentially a template of the spike effect in the timestreams) was subtracted, for which each bolometer sees a different fraction.  This fraction was found by minimising the spread in each bolometer over time (i.e.~using a minimum variance estimator) and is expected to be close to 0 for the bolometers which exhibit little evidence of any excess power and close to 1 (after being appropriately normalised) for the bolometers exhibiting the highest degree of excess power.  This approach was applied to Reduction D, but the results should be applicable to all reductions.

Approximately 20 per cent of the SXDF data files appear to be contaminated by the noise spike.  While all 37 bolometers may contain the effect to some extent, about 30 per cent of them exhibit a significant excess of power in the power spectrum at a period of 16 samples. So overall, about 8 per cent of the data going into the SXDF map is significantly affected by the noise spike.  As a result of the spike removal, the RMS of Reduction D's SXDF map was decreased by a mere 1 per cent, as compared to the reduction in which the spike's presence was ignored.

We find that the LH map is slightly more affected than the SXDF map:  30 per cent of the Lockman data files are affected and once the spike is removed the map RMS is improved by 3 per cent.  For the 10/14 Reduction D-detected LH SHADES catalogue sources lying in regions with more than 50 per cent of their data affected by the spike, no positional offsets are detected from the corresponding spike-removed map sources.  Positional offsets of up to the $3\,\mathrm{arcsec}$ pixel size occur in the spike-removed Reduction D map for only 4 of the LH SHADES catalogue sources.  

In general the contaminated files are distributed fairly uniformly around the maps, although the worst examples are localised to specific regions.  Hence even although the overall map RMS changed very little, it may still be that the flux densities or positions of particular sources could be affected significantly.  We checked for signal, noise and S/N variations and found that small differences do occur for sources detected in the noise-spike removed maps versus the regular maps, as expected.  The flux densities and S/N values differ from the regular map values by about 5 per cent on average, and never by more than 10 per cent.  Overall 28 per cent of the SHADES sources have been affected to any discernible extent by the presence of the noise spike, in terms of flux density, S/N or position.  Approximately 25 per cent of these sources show positional offsets of $\lesssim3$--4 arcsec from their original positions in the map, while the rest show no discernable positional differences.  Therefore, we conclude that the noise spike mainly contributes additional random noise to the source flux densities and centroids of the data, and thus can be safely ignored in our data reduction treatment.  However, for a few of the most-severely affected sources we have included a discussion of the effects in the individual source notes in Appendix~\ref{notes}.


\section{Tests of the Data}\label{tests}

Several different checks were performed by the reduction groups in order to examine the noise properties of the maps and test the individual source robustness by examining spatial and temporal fits to point sources in the maps.

\subsection{Tests of Gaussianity in the maps}

Reduction C's photometric error distributions were tested for Gaussianity by performing Monte Carlo simulations.  A realisation of Gaussian noise was created from the measured detector signal variances for SXDF.  The simulated noise signals were reduced and rebinned into pixels using the same procedures as for the real data.  Sources were placed into the map one at a time at random locations and their recovery was attempted.  The joint probability distribution from the simulations was calculated using the same technique that was used for the real data.  We found that the photometric uncertainty is very close to Gaussian, as expected, and the parameter, $\sigma$ that comes out of the source extraction routine is entirely consistent with what is expected.

\subsection{General source robustness}\label{robust}

A quick test of source reality is to create the negative of the map and search for sources using the same triple-beam template.  Aside from pixels associated with the off-beams of positive detections, we find 3 and 4 $\geq3.5\,\sigma$ `detections' and 21 and 25 $\geq3.0\,\sigma$ `detections' in the inverted reduction D SXDF and LH maps, respectively.  This is consistent with the expected number of false positive detections in noisy data.  Note that only one of these negative sources ($3.9\,\sigma$, in LH) has high enough S/N and/or low enough noise to survive the deboosting procedure of Section~\ref{membership}.

We have also performed spatial and temporal $\chi^{2}$ tests in order to determine how well the raw timestream data fit the final PSF-fitted maps.  A candidate source may be `detected' in the map, but it may not necessarily be well-fit by the PSF, or it may be a poor fit to the set of difference data, or both.  See \citet{Pope} and \citet{Coppin} for more discussion of such tests in other SCUBA surveys.  Note that these two tests also check for Gaussianity of the noise in the maps, but that they are not a strong test of this distribution.

The `spatial $\chi^{2}$' test provides a gauge across the map of the goodness-of-fit of the triple-beam differential PSF to the data, and thus indicates if a source is poorly fit by the assumed PSF.  We have picked out $>\!2.5\,\sigma$ sources with $\chi^{2}$ values lying outside of the $\pm2\,\sigma\!=\!2.16$ (LH), or 2.05 (SXDF) regions of the complete spatial $\chi^{2}$ distribution of the smoothed LH and SXDF maps.  We find 6 such sources in Reduction D's LH map, 3 of which correspond to SHADES catalogue sources, and we find 16 such sources in Reduction D's SXDF map, 5 of which correspond to SHADES catalogue sources.

The `temporal $\chi^{2}$' provides a measure of the self-consistency of the raw timestream data which contribute flux density to each map pixel.  We have selected $>\!2.5\,\sigma$ sources in the maps lying outside the $\pm 2\,\sigma\!=\!1.60$ (LH and SXDF) regions of the temporal $\chi^{2}$ distribution.  In Reduction D's LH map, we find 17 such sources, 5 of which correspond to SHADES catalogue sources.  In Reduction D's SXDF maps, we find 16 such sources, 6 of which correspond to SHADES catalogue sources.  

We have noted those source candidates with relatively poor spatial and/or temporal $\chi^{2}$ values in Appendix~\ref{notes}.  All of these flagged sources have either temporal or spatial $\chi^{2}$ values \textit{just} outside the $2\,\sigma$ regions of the $\chi^{2}$ distributions, except for one source that has a spatial $\chi^{2}$ value $>\!3\,\sigma$ (LOCK850.15).

\subsection{Analysis of the $450\,\mathrm{\mu m}$ data}\label{450umfluxes}

$450\,\mathrm{\mu m}$ data were taken simultaneously with those at 
$850\,\mathrm{\mu m}$, providing complementary short-wavelength 
photometry (or flux density limits) over a poorly sampled wavelength regime of the 
spectral energy distribution of SMGs. For objects in the
 SHADES catalogue, these data can provide useful constraints for
 Far-IR-to-mm wavelength
 photometric redshift estimates, a powerful tool for measuring the 
star formation history of the submillimetre galaxy population 
(e.g.~\citealt{Aretxaga2003}). However, $450\,\mathrm{\mu m}$ observations
of faint objects with SCUBA are of limited use, as: (1) the JCMT beamshape is non-Gaussian at 
$450\,\mathrm{\mu m}$, due to the non-optimization of the telescope surface 
for short-wavelength observations; (2) the atmosphere is 
more opaque at this wavelength and therefore the noise will be more
 sensitive to
 small variations in the sky transmission than at longer wavelengths; and (3) atmospheric emission fluctuations are much more severe at $450\,\mathrm{\mu m}$, and are not sufficiently reduced by removing the $1\,\mathrm{Hz}$ array average (which is so successful at $850\,\mathrm{\mu m}$).  As the SHADES survey was conducted throughout a range of weather conditions ($\tau_{\mathrm{CSO}}>0.05$; mostly unsuitable for sensitive $450\,\mathrm{\mu m}$ 
observations), these data are not expected to add much to our
understanding of the $850\,\mathrm{\mu m}$ source sample.
 Nevertheless, for completeness we describe the main results derived from our analysis of these data.  

\subsubsection{$450\,\mathrm{\mu m}$ reduction strategies}

Reduction methods applied to the $450\,\mathrm{\mu m}$ data are similar 
to those described in Section~\ref{dr}, with the following minor 
exceptions. Reduction C uses smaller, $1\,\mathrm{arcsec}$ pixels when 
rebinning, and adopts a $7.5\,\mathrm{arcsec}$ Gaussian rather than the full, chopped
 PSF to extract flux densities. 
Reduction D also ignores the off-beams in the data and thus does not fold 
them in when rebinning. The results of the Reduction A analysis are not 
included in the $450\,\mathrm{\mu m}$ comparison.

The spatial variation in $450\,\mathrm{\mu m}$ flux densities across the rebinned maps
 is more correlated between Reductions B and C than between either of 
these and Reduction D. This is not surprising, given that Reductions B and C 
follow a similar calibration 
strategy, i.e.~applying FCFs derived from calibration observations 
taken on the same night as the data, rather than adopting a monthly average
 FCF as was the strategy for Reduction D (cf.~Table~\ref{procedures}). 
There are plausible reasons why using frequent measurements 
of the FCF can either be beneficial, 
or detrimental, and our data cannot distinguish between these possibilities,
 as we do not know the true amplitude of the underlying signal. Based on 
 the consistently lower noise in reduction B, one might conclude that
 the longer, monthly average calibration strategy of Reduction D is introducing 
noise, and that at $450\,\mathrm{\mu m}$ the more frequent calibration 
strategy adopted by Reductions B and C is optimal. However, within the large 
uncertainties of the $450\,\mathrm{\mu m}$ photometry (described below), 
the reductions are broadly consistent. 

\subsubsection{A search for $450\,\mathrm{\mu m}$ blank-field sources}

As the beamsize at $450\mathrm{\mu m}$ is roughly half that at 
$850\,\mathrm{\mu m}$, we expect to uncover many more spurious sources 
in each map above a given S/N threshold.
This translates to a greater number of blank-field source candidates 
(i.e.~those found at random positions in the map),
 so care must be taken to assess the likelihood of finding spurious 
sources in each map.  

As an initial test of the data, each group focused on the SXDF 
data, extracting a list of blank-field, 
$3\,\sigma$ source candidates. From these source lists, a preliminary 
cross-identification candidate source list was created using a similar 
grading scheme to that described in Section~\ref{prelim}. To 
determine the likely number of spurious objects in this list of 
`positive' sources, 
each group then applied the same source extraction methods to 
inverted, or `negative' versions of these same SXDF maps 
(cf.~Section~\ref{robust}). The result was that
in the combined SXDF lists \textit{more} `negative' than `positive' source 
candidates were identified (142 versus 131). 
From these lists, we then removed those source candidates found in  
noisier regions of the maps 
(see Figs.~1 and 2),
 as these should be less reliable.
 This resulted in an equal number of `positive' and `negative' source 
candidates in the combined SXDF list (68). Higher S/N threshold 
cuts (up to $5\sigma$) did not yield an excess of `positive' over 
`negative' source candidates.  Spatial and temporal 
$\chi^{2}$ tests (see Appendix~\ref{robust}) were performed on Reduction D's
 $>3\,\sigma$ source candidates, and we found that the majority of the 
sources fit within the allowed $2\,\sigma$ area of the map's $\chi^{2}$ 
distribution and therefore could not be rejected on these grounds. 
Based on these analyses, no single $450\,\mathrm{\mu m}$ blank-field
 source candidate can be claimed as a reliable detection in the SHADES data.

Although Reductions B and C do
 agree on the detection of a few $450\,\mathrm{\mu m}$ counterparts to the 
$850\,\mathrm{\mu m}$ sources in 
the LH field (some of which was observed under excellent weather conditions
as part of the SCUBA 8-mJy survey), in no case do all three reductions 
agree on a detection with a 
consistent position and significance level. As such, 
we have chosen to adopt the $3\,\sigma$ limits on the $450\,\mathrm{\mu m}$ 
flux densities for subsequent analyses, such as the photometric redshift estimates 
(Aretxaga et al.\ in preparation), which benefit from the shorter wavelength 
data.

\subsubsection{$450\,\mathrm{\mu m}$ photometry}\label{survival}

As the reductions show broad agreement in the $450\,\mathrm{\mu m}$
flux densities within the uncertainties, and there are no large systematic
 differences, we adopt the $450\,\mathrm{\mu m}$ flux densities and photometric errors of Reduction B.  
We do not claim any individual $450\,\mathrm{\mu m}$ detections.  Equivalent $3\,\sigma$ upper 
limits are calculated for each source in the following way.  We construct an error distribution from a 
histogram of pixel S/N.  This error function is nearly Gaussian but has slightly larger wings.  For each source, 
the error distribution is scaled to match the peak and $\sigma$ of the source.  We quote in Table~\ref{tab:lock} as upper limits the flux densities bounding 99.73 per cent of the area of the error function (corresponding to the percentage area between the tails of the $3\,\sigma$ region under a Gaussian distribution).

\subsubsection{$450\,\mathrm{\mu m}$ stacking analyses}

Although we are unable to claim a $450\,\mathrm{\mu m}$ detection of any $850\,\mathrm{\mu m}$ 
SHADES source, we attempt to determine if the population as a whole 
is detected in these data. To do this, we perform a series of stacking
 analyses on the $450\,\mu$m data at the positions of the $850\,\mu$m sources, and also 
their proposed $1.4\,\mathrm{GHz}$ radio counterparts (Ivison et al.\ in preparation). These analyses were
 performed independently by each group following 2 strategies:  (1) pointed $450\,\mathrm{\mu m}$ photometry
 at the precise $850\,\mathrm{\mu m}$-selected source positions (this result 
will be biased low since the peak of emission may be offset by an amount 
as great as the JCMT pointing error of 2--3 arcsec); and (2) a search 
for the nearest $450\,\mathrm{\mu m}$ peak
within a $7\,\mathrm{arcsec}$ radius of the $850\,\mathrm{\mu m}$ source position, 
if one exists (an estimate which should be biased high).  A $7\,\mathrm{arcsec}$ search radius was chosen since the probability of a spurious $450\,\mathrm{\mu m}$ source lying within this distance of a known $850\,\mathrm{\mu m}$ is very low (see \citealt{Fox}).  The results from each reduction are given in columns 2 and 3 of Table~\ref{tab:450stacked}.

\begin{table*}
\caption{$450\,\mathrm{\mu m}$ stacked flux densities (in mJy) for the B, C and D
 reductions. The results are for average flux densities measured at the 
indicated positions, related to those of the $850\,\mathrm{\mu m}$ SHADES catalogue sources.  Reduction B has calculated unweighted mean stacked flux densities, while Reductions C and D have calculated mean stacked flux densities with inverse noise variance weighting.}
\begin{tabular}{lrrrrrr}
\hline 
\multicolumn{1}{l}{Reduction} & \multicolumn{1}{r}{$850\,\mathrm{\mu m}$ position} & \multicolumn{1}{r}{highest S/N detection} & \multicolumn{1}{r}{$850\,\mathrm{\mu m}$ position} & \multicolumn{1}{r}{$850\,\mathrm{\mu m}$ position} & \multicolumn{1}{r}{radio position}\\
\multicolumn{1}{l}{} & \multicolumn{1}{r}{} & \multicolumn{1}{r}{within $7\,\mathrm{arcsec}$ radius of} & \multicolumn{1}{r}{} & \multicolumn{1}{r}{} & \multicolumn{1}{r}{}\\
\multicolumn{1}{l}{} & \multicolumn{1}{r}{(all)} & \multicolumn{1}{r}{$850\,\mathrm{\mu m}$ position (all)} & \multicolumn{1}{r}{(radio ID subset)} & \multicolumn{1}{r}{(non-radio ID subset)} & \multicolumn{1}{r}{(radio ID subset)}\\

\hline
LH & & & & &\\
\hline
B & $10.6\pm3.8$ & $38.3\pm4.0$ & $16.9\pm4.3$ & $10.6\pm4.8$ & $19.2\pm4.2$ \\ 
C & $12.4\pm2.1$ & $30.6\pm2.1$ & $20.8\pm3.3$ & $7.4\pm2.7$ & $22.0\pm3.3$ \\ 
D & $10.5\pm2.3$ & $28.8\pm2.9$ & $19.6\pm3.8$ & $4.9\pm2.9$ & $18.8\pm3.8$ \\
\hline
SXDF & & & & & \\
\hline
B & $5.7\pm4.6$ & $24.3\pm3.7$ & $11.1\pm4.2$ & $14.6\pm5.4$ & $16.6\pm4.2$ \\
C & $12.5\pm2.4$ & $37.5\pm2.4$ & $14.1\pm3.9$ & $11.4\pm3.1$ & $17.9\pm3.8$ \\
D & $9.9\pm2.7$ & $36.7\pm3.4$ & $12.1\pm4.3$ & $7.9\pm3.5$ & $14.0\pm4.2$ \\
\hline
\label{tab:450stacked}
\end{tabular}
\end{table*}

To determine how frequently the stacked flux density measurements would occur 
given the number of $850\,\mathrm{\mu m}$ sources in each field, 
10,000 Monte Carlo simulations were performed with uniformly-selected 
random positions in the Reduction D maps. A stacked flux density greater 
than or equal to the measured value at the 
$850\,\mathrm{\mu m}$-selected positions (strategy 1) occurs 
$\lesssim\!0.1\,\mathrm{per cent}$ of the time at random 
in both fields. The simulations
 therefore indicate a significant detection of the SHADES catalogue of 
$850\,\mathrm{\mu m}$-selected galaxies at $450\,\mathrm{\mu m}$.  
Another indication that we have marginally detected the $850\,\mathrm{\mu m}$
sources at $450\,\mathrm{\mu m}$, is shown by the slight 
positive skewness in the distribution of S/N values at the  
$850\,\mathrm{\mu m}$ SHADES catalogue positions (as shown in Fig.~\ref{fig:450hist}).
The simulations were then repeated for strategy 2, searching for the nearest 
$450\,\mathrm{\mu m}$ peak within $7\,\mathrm{arcsec}$ of the SHADES catalogue 
$850\,\mathrm{\mu m}$ positions. We found that a value above the measured stacked value 
occurred 90 (22) per cent of the time in the LH (SXDF) field. 
 The stacked $450\,\mathrm{\mu m}$ flux density obtained using strategy 2 is 
therefore not high enough compared with the Monte Carlos (which pick up 
$450\,\mathrm{\mu m}$ noise peaks within the $7\,\mathrm{arcsec}$ search radius) 
to be regarded as statistically significant.

\begin{figure}
\epsfig{file=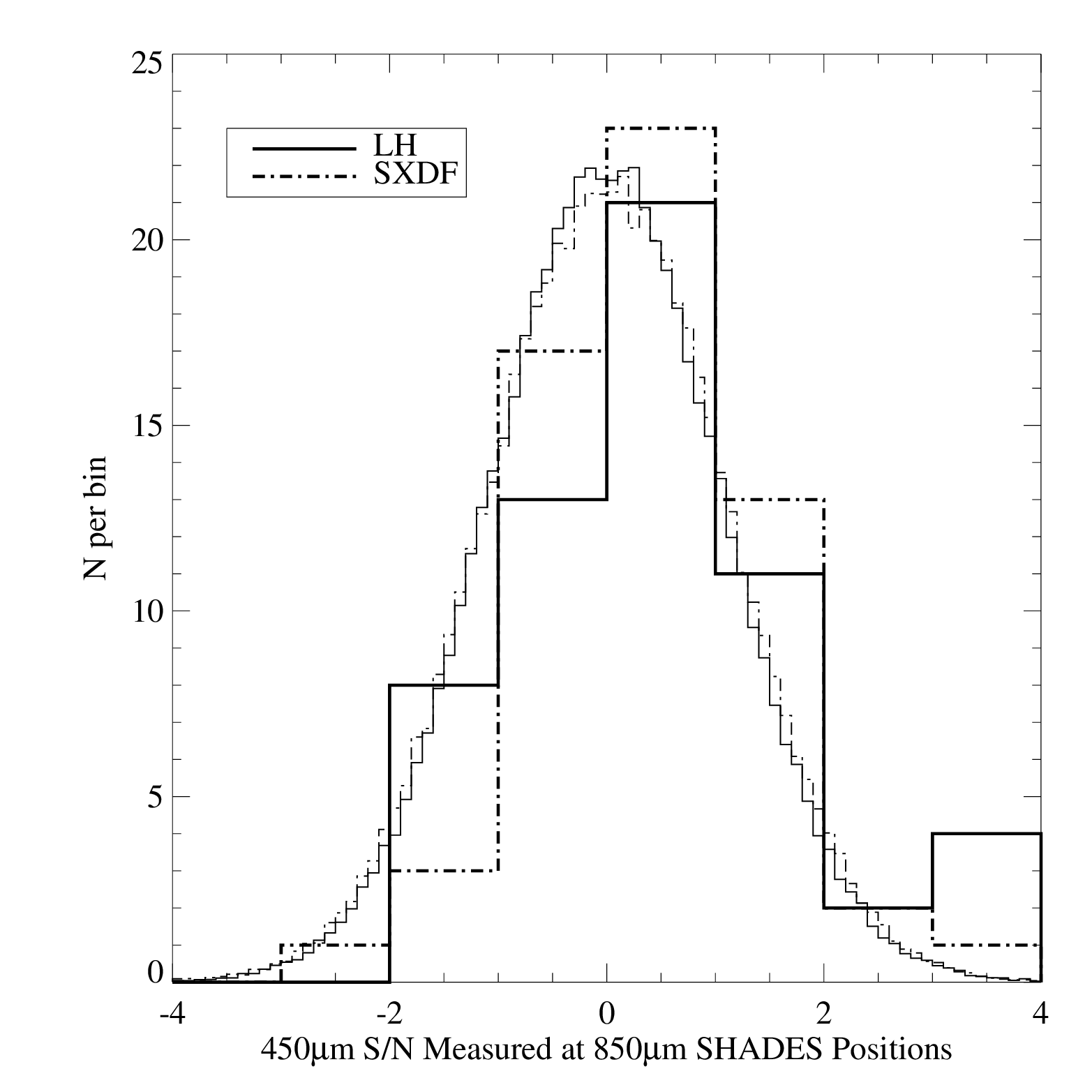,width=0.5\textwidth}
\caption{Histogram of stacked flux densities in the $450\,\mathrm{\mu m}$ S/N map at
 the positions of the $850\,\mathrm{\mu m}$-selected 
SHADES catalogue positions (see Table~\ref{tab:lock}).
 The stacked flux densities for the LH and SXDF are shown by the large-binned thick 
histograms.  The finer-binned histograms represent the $450\,\mathrm{\mu m}$
 pixel values from the S/N LH and SXDF maps. Notice the appearance 
of excess positive $450\,\mathrm{\mu m}$ S/N when stacked at the 
$850\,\mathrm{\mu m}$ positions, indicating that we \textit{do} detect an overlap 
between these populations.  While any single $850\,\mathrm{\mu m}$ object 
is not reliably detected, the $850\,\mathrm{\mu m}$ population as a whole is detected 
statistically. This plot is specifically for Reduction D; the 
results based on the other two reductions are similar.}
\label{fig:450hist}
\end{figure}

The ratio of stacked $450\,\mathrm{\mu m}$ flux density to $850\,\mathrm{\mu m}$ 
flux density of SHADES catalogue sources is around 2--2.5, depending on the precise 
choice of data and reduction (and including the effect of $850\,\mathrm{\mu m}$ flux 
deboosting).  This ratio is low compared to what one would expect, independent of any 
reasonable choice for a redshifted sample of local SED templates (which typically yield 
$S_{450}/S_{850}\gsim4$).  We found that fitting a range of template galaxies from 
\citet{Aretxaga2003} yielded redshifts $\gsim3$ for the population, which is higher 
than that found for SMGs.  This result of low stacked $450\,\mathrm{\mu m}$ flux 
density is consistent with that found for SMGs in the GOODS-N field \citep{Pope} and 
suggests a systematic bias when pushing SCUBA data to these faint stacked values.

The $450\,\mathrm{\mu m}$ stacking analysis was also carried out at the
positions of the preliminary sub-sample of radio-identified sources
 (see Section~\ref{compareposition}), and compared with the results of
an analysis at the positions of the non-radio-identified sub-sample.  Note that the radio-detected SMGs are more likely to be detected at $450\,\mathrm{\mu m}$ (e.g.~\citealt{Chapman2005}).  The stacked $450\,\mathrm{\mu m}$ flux densities at the $850\,\mathrm{\mu m}$ positions of the sources with radio IDs are similar to 
the stacked $450\,\mathrm{\mu m}$ flux densities at the radio positions.
 This implies that moving to the positions of the proposed 
radio identifications (which one might expect would 
provide a more accurate measure 
of the true source positions) does not actually raise the average 
$450\,\mathrm{\mu m}$ flux density significantly for the same subset of sources. 
However, a marked difference was found between the stacked 
$450\,\mathrm{\mu m}$ flux densities of the radio-identified subset and the 
non-radio-identified subset.  This result may be explained by
 one of the following:
(1) the non-radio-identified subset lies 
at a higher redshift on average; or (2) there are spurious sources in 
the non-radio-identified subset, diluting the $450\,\mathrm{\mu m}$ 
stacked flux density measurement (\citealt{Ivison}; \citealt{Greve}). Since there is no evidence that 
the radio-identified SHADES sources are more likely to be real,
 the radio-identified subset is probably biased to lower redshifts \citep{Chapman2005}, yielding a higher $S_{450}/S_{850}$ ratio. 

\section{Notes on individual sources}\label{notes}

Here we give notes for some of the individual SHADES sources.  

\textit{LOCK850.1}: This source has a relatively poor spatial $\chi^{2}$ (3.13, where the $+2\,\sigma$ range of the distribution extends to 2.16), which could perhaps be explained by its proximity to LOCK850.41 (22\,arcsec away). 

\textit{LOCK850.2}:  More than 50 per cent of data in this region of the map were affected by the noise spike.  Treating the noise spike as discussed in Appendix~\ref{noisespike} resulted in a 2 per cent decrease in the flux and S/N and no discernible positional offset.

\textit{LOCK850.5}:  Treating the noise spike as discussed in Appendix~\ref{noisespike} resulted in an apparent offset in RA of $3\,\mathrm{arcsec}$, an offset of $3\,\mathrm{arcsec}$ in Dec., and approximate differences in flux, noise and S/N of 2, 4, and 2 per cent, respectively.

\textit{LOCK850.8}:  This source has a relatively poor temporal $\chi^{2}$ ($-1.72$, where the $-2\,\sigma$ range of the distribution extends to $-1.60$).

\textit{LOCK850.10}:  This source was detected in the noisier edge region of the map ($\sigma >3\,\mathrm{mJy}$).

\textit{LOCK850.11}:  Treating the noise spike as discussed in Appendix~\ref{noisespike} resulted in an apparent offset of $-3\,\mathrm{arcsec}$ in Dec. and approximate differences in flux, noise and S/N of 1, 2, and 0 per cent, respectively.

\textit{LOCK850.12}:  This source has a relatively poor temporal $\chi^{2}$ ($-1.70$, where the $-2\,\sigma$ range of the distribution extends to $-1.60$).

\textit{LOCK850.14}:  LOCK850.18 is a nearby neighbour, 20\,arcsec away.  

\textit{LOCK850.15}:  This source has a relatively poor temporal $\chi^{2}$ ($-2.12$, where the $2\,\sigma$ of the distribution extends to $-1.60$).  This source also has a relatively poor spatial $\chi^{2}$ (7.82, where the $+2\,\sigma$ range of distribution extends to 2.16).  This source was detected in the noisier edge region of the map ($\sigma >4\,\mathrm{mJy}$).  

\textit{LOCK850.18}:  LOCK850.14 is a near neighbour, 20\,arcsec away.

\textit{LOCK850.21}:  Treating the noise spike as discussed in Appendix~\ref{noisespike} resulted in an apparent offset of $2.9\,\mathrm{arcsec}$ in RA and $-3\,\mathrm{arcsec}$ in Dec., and approximate differences in flux, noise and S/N of 1, 1, and 2 per cent, respectively.

\textit{LOCK850.22}:  Treating the noise spike as discussed in Appendix~\ref{noisespike} resulted in an apparent offset of $-3\,\mathrm{arcsec}$ in RA and approximate differences in flux, noise and S/N of 5, 2, and 2 per cent, respectively.

\textit{LOCK850.33}:  Treating the noise spike as discussed in Appendix~\ref{noisespike} resulted in an apparent offset of $3\,\mathrm{arcsec}$ in RA and $-3\,\mathrm{arcsec}$ in Dec., and approximate differences in flux, noise and S/N of 3, 2, and 0 per cent, respectively.

\textit{LOCK850.34}:  This source was detected in the noisier edge region of the map ($\sigma >3\,\mathrm{mJy}$).   

\textit{LOCK850.35}:  More than 50 per cent of data in this region of the map were affected by the noise spike.  Treating the noise spike as discussed in Appendix~\ref{noisespike} resulted in a 4 per cent decrease in the flux, 2 per cent decrease in S/N and no noticeable positional offset.  This source has a relatively poor temporal $\chi^{2}$ ($-1.94$, where the $-2\,\sigma$ range of the distribution extends to $-1.60$).

\textit{LOCK850.39}:  More than 50 per cent of data in this region of the map were affected by the noise spike.  Treating the noise spike as discussed in Appendix~\ref{noisespike} resulted in a 7 per cent decrease in the flux, 5 per cent decrease in S/N and no noticeable positional offset. 

\textit{LOCK850.40}:  This source has a relatively poor temporal $\chi^{2}$ ($-2.05$, where the $-2\,\sigma$ range of the distribution extends to $-1.60$).

\textit{LOCK850.41}:  This source has a relatively poor spatial $\chi^{2}$ (5.98, where the $+2\,\sigma$ range of the distribution extends to 2.16), which could be explained by its proximity to LOCK850.1 (22\,arcsec away).

\textit{LOCK850.43}:  Treating the noise spike as discussed in Appendix~\ref{noisespike} resulted in an apparent offset of $-3\,\mathrm{arcsec}$ in Dec. and approximate differences in flux, noise and S/N of 12, 0, and 12 per cent, respectively.

\textit{LOCK850.47}:  Treating the noise spike as discussed in Appendix~\ref{noisespike} resulted in an apparent offset of $-3\,\mathrm{arcsec}$ in RA and approximate differences in flux, noise and S/N of 6, 2, and 3 per cent, respectively.

\textit{LOCK850.48}:  Treating the noise spike as discussed in Appendix~\ref{noisespike} resulted in an apparent offset of $-3\,\mathrm{arcsec}$ in Dec.. and approximate differences in flux, noise and S/N of 4, 3, and 3 per cent, respectively.

\textit{LOCK850.60}:  Treating the noise spike as discussed in Appendix~\ref{noisespike} resulted in an apparent offset of $3\,\mathrm{arcsec}$ in Dec. and approximate differences in flux, noise and S/N of 6, 2, and 3 per cent, respectively. 

\textit{LOCK850.66}:  Treating the noise spike as discussed in Appendix~\ref{noisespike} resulted in an apparent offset of $3\,\mathrm{arcsec}$ in Dec. and approximate differences in flux, noise and S/N of 2, 1, and 5 per cent, respectively. 

\textit{LOCK850.67}:  Treating the noise spike as discussed in Appendix~\ref{noisespike} resulted in an apparent offset of $3\,\mathrm{arcsec}$ in Dec. and approximate differences in flux, noise and S/N of 3, 1, and 3 per cent, respectively. 

\textit{LOCK850.76}:  More than 50 per cent of data in this region of the map were affected by the noise spike.  Treating the noise spike as discussed in Appendix~\ref{noisespike} resulted in a 13 per cent decrease in the flux, a 9 per cent decrease in the S/N and no discernible positional offset.

\textit{LOCK850.77}:  Treating the noise spike as discussed in Appendix~\ref{noisespike} resulted in an apparent offset of $-3\,\mathrm{arcsec}$ in RA and approximate differences in flux, noise and S/N of 1, 0, and 3 per cent, respectively. 

\textit{LOCK850.79}:  Treating the noise spike as discussed in Appendix~\ref{noisespike} resulted in an apparent offset of $-3\,\mathrm{arcsec}$ in RA, $3\,\mathrm{arcsec}$ in Dec., and approximate differences in flux, noise and S/N of 4, 4, and 9 per cent, respectively. 

\textit{LOCK850.100}:  This source was detected in the noisier edge region of the map ($\sigma >3\,\mathrm{mJy}$).

\textit{SXDF850.1}:  This source appears to be extended in the map in the NS direction.

\textit{SXDF850.5}:  More than 50 per cent of data in this region of the map were affected by the noise spike.  Treating the noise spike as discussed in Appendix~\ref{noisespike} resulted in a 6 per cent decrease in the flux, a 2 per cent decrease in S/N and no discernible positional offset.

\textit{SXDF850.6}:  This source has a relatively poor spatial $\chi^{2}$ ($-2.28$, where the $-2\,\sigma$ range of the distribution extends to $-2.05$).

\textit{SXDF850.7}:  This source has a relatively poor spatial $\chi^{2}$ (2.03, where the $+2\,\sigma$ range of the distribution extends to 2.05).  The poor fit could be a result of it being separated by less than $22\,\mathrm{arcsec}$ from a lower significance rejected source (SXDF850.110).

\textit{SXDF850.9}:  More than 50 per cent of data in this region of the map were affected by the noise spike.  Treating the noise spike as discussed in Appendix~\ref{noisespike} resulted in a 1 per cent decrease in the flux, and no discernible change in S/N or position.

\textit{SXDF850.10}: This source has a relatively poor temporal $\chi^{2}$ (2.27, where the $+2\,\sigma$ range of the distribution extends to 1.60).

\textit{SXDF850.18}:  More than 50 per cent of data in this region of the map were affected by the noise spike.  Treating the noise spike as discussed in Appendix~\ref{noisespike} resulted in a 6 per cent decrease in the flux, a 0 per cent change in S/N and a positional offset of $-3\,\mathrm{arcsec}$ in Dec.

\textit{SXDF850.19}:  Treating the noise spike as discussed in Appendix~\ref{noisespike} resulted in an apparent offset of $3\,\mathrm{arcsec}$ in RA and approximate differences in flux, noise and S/N of 1, 2, and 0 per cent, respectively.  This source has a relatively poor spatial $\chi^{2}$ (2.75, where the $+2\,\sigma$ range of the distribution extends to 2.05).

\textit{SXDF850.21}:  This source has a relatively poor spatial $\chi^{2}$ (2.23, where the $+2\,\sigma$ range of the distribution extends to 2.05).  This source also has a relatively poor temporal $\chi^{2}$ (1.84, where the $+2\,\sigma$ range of the distribution extends to 1.60).

\textit{SXDF850.22}:  Treating the noise spike as discussed in Appendix~\ref{noisespike} resulted in an apparent offset of $3\,\mathrm{arcsec}$ in RA, $-3\,\mathrm{arcsec}$ in Dec., and approximate differences in flux, noise and S/N of 11, 2, and 7 per cent, respectively.

\textit{SXDF850.27}:  More than 50 per cent of data in this region of the map were affected by the noise spike.  Treating the noise spike as discussed in Appendix~\ref{noisespike} resulted in a 5 per cent decrease in the flux, a 3 per cent decrease in S/N and no noticeable positional offset.  This source has a relatively poor spatial $\chi^{2}$ (2.15, where the $+2\,\sigma$ range of the distribution extends to 2.05).

\textit{SXDF850.28}:  More than 50 per cent of data in this region of the map were affected by the noise spike.  Treating the noise spike as discussed in Appendix~\ref{noisespike} resulted in a 10 per cent increase in the flux, a 13 per cent increase in S/N and a positional offset of $3\,\mathrm{arcsec}$ in RA.  This source has a relatively poor temporal $\chi^{2}$ ($-1.75$, where the $-2\,\sigma$ range of the distribution extends to $-1.60$).  This source was detected in the noisier edge region of the map ($\sigma >3\,\mathrm{mJy}$).  

\textit{SXDF850.37}:  Treating the noise spike as discussed in Appendix~\ref{noisespike} resulted in an apparent offset of $-3\,\mathrm{arcsec}$ in RA, and approximate differences in flux, noise and S/N of 1, 1, and 0 per cent, respectively.

\textit{SXDF850.45}:  This source has a relatively poor temporal $\chi^{2}$ ($-1.85$, where the $-2\,\sigma$ range of the distribution extends to $-1.60$).  This source was detected in the noisier edge region of the map ($\sigma >4\,\mathrm{mJy}$).

\textit{SXDF850.47}:  This source has a relatively poor temporal $\chi^{2}$ (1.80, where the $+2\,\sigma$ range of the distribution extends to 1.60).

\textit{SXDF850.50}:  More than 50 per cent of data in this region of the map were affected by the noise spike.  Treating the noise spike as discussed in Appendix~\ref{noisespike} resulted in a 10 per cent decrease in the flux, a 10 per cent decrease in S/N and no discernible positional offset.

\textit{SXDF850.56}:  Treating the noise spike as discussed in Appendix~\ref{noisespike} resulted in an apparent offset of $3\,\mathrm{arcsec}$ in Dec., and approximate  differences in flux, noise and S/N of 4, 1, and 3 per cent, respectively.

\textit{SXDF850.63}:  More than 50 per cent of data in this region of the map were affected by the noise spike.  Treating the noise spike as discussed in Appendix~\ref{noisespike} resulted in a 5 per cent decrease in the flux, no discernible change in S/N and a positional offset of $-3\,\mathrm{arcsec}$ in RA.

\textit{SXDF850.69}:  More than 50 per cent of data in this region of the map were affected by the noise spike.  Treating the noise spike as discussed in Appendix~\ref{noisespike} resulted in a 15 per cent increase in the flux, a 15 per cent increase in S/N and a positional offset of $-3\,\mathrm{arcsec}$ in RA.

\textit{SXDF850.76}:  Treating the noise spike as discussed in Appendix~\ref{noisespike} resulted in an apparent offset of $-3\,\mathrm{arcsec}$ in RA, $-3\,\mathrm{arcsec}$ in Dec., and approximate differences in flux, noise and S/N of 1, 3, and 3 per cent, respectively.

\textit{SXDF850.86}:  This source has a relatively poor temporal $\chi^{2}$ (2.21, where the $+2\,\sigma$ range of the distribution extends to 1.60).

\textit{SXDF850.93}:  Treating the noise spike as discussed in Appendix~\ref{noisespike} resulted in an apparent offset of $3\,\mathrm{arcsec}$ in Dec. and approximate differences in flux, noise and S/N of 3, 1, and 0 per cent, respectively.

\textit{SXDF850.95}:   More than 50 per cent of data in this region of the map were affected by the noise spike.  Treating the noise spike as discussed in Appendix~\ref{noisespike} resulted in a 8 per cent decrease in the flux, a 6 per cent decrease in S/N and positional offsets of $-3\,\mathrm{arcsec}$ in RA and $-3\,\mathrm{arcsec}$ in Dec.

\textit{SXDF850.96}:   More than 50 per cent of data in this region of the map were affected by the noise spike.  Treating the noise spike as discussed in Appendix~\ref{noisespike} resulted in a 4 per cent decrease in the flux and a 3 per cent decrease in S/N, with no noticeable positional offset.

\end{appendix}

\end{twocolumn}
\end{document}